%% file: arXiv_v3/arXiv_v3.tex
\documentclass[prx,aps,amsfonts,amsmath,amssymb,twocolumn,superscriptaddress,10pt]{revtex4-2}

\usepackage[version=4]{mhchem}
\usepackage{natbib}
\usepackage{siunitx}
\usepackage{amsmath}
\usepackage{amssymb}
\usepackage{amsfonts}
\usepackage{dsfont}
\usepackage{bbold}
\usepackage{float}
\newcommand{\dashfill}{%
  \leavevmode
  \leaders\hbox{\rule[0.5ex]{3pt}{0.4pt}\hskip2pt}\hfill\kern0pt
}
\usepackage{booktabs}

\usepackage{graphicx}
\usepackage{dcolumn}

\usepackage{bm}

\usepackage{multirow}
\usepackage{makecell} 
\usepackage{xcolor}

\usepackage{hyperref}                       

\begin{document}

\makeatletter
\let\oldaddcontentsline\addcontentsline
\renewcommand{\addcontentsline}[3]{}
\input{arXiv_v3/main_body}
\let\addcontentsline\oldaddcontentsline
\makeatother

\clearpage
\onecolumngrid
\begin{center}
{\Large\bfseries Supplementary Materials for ``Plasmon-driven electron-hole pair formation in a non-equilibrium Fermi liquid''}
\end{center}

\setcounter{section}{0}
\setcounter{subsection}{0}
\setcounter{figure}{0}
\setcounter{table}{0}
\setcounter{equation}{0}

\renewcommand{\thefigure}{S\arabic{figure}}
\renewcommand{\thetable}{S\arabic{table}}

\makeatletter
\makeatother

\tableofcontents
\newpage
\input{arXiv_v3/SI_body}
\end{document}

%% file: arXiv_v3/main_body.tex
\title{Plasmon-driven electron-hole pair formation in a non-equilibrium Fermi liquid}

\author{Rishi Acharya} 
\affiliation{Department of Physics, The Grainger College of Engineering, University of Illinois at Urbana-Champaign, Urbana, 61801 IL, USA}
\affiliation{Materials Research Laboratory, The Grainger College of Engineering, University of Illinois at Urbana-Champaign, Urbana, 61801 IL, USA}

\author{Eli Gerber}
\affiliation{Institut de Physique Th\'eorique, Universit\'e Paris-Saclay, CEA, CNRS, F-91191 Gif-sur-Yvette, France}

\author{Nina Bielinski}
\affiliation{Department of Physics, The Grainger College of Engineering, University of Illinois at Urbana-Champaign, Urbana, 61801 IL, USA}
\affiliation{Materials Research Laboratory, The Grainger College of Engineering, University of Illinois at Urbana-Champaign, Urbana, 61801 IL, USA}

\author{Hannah E. Aguirre}
\affiliation{Department of Physics, The Grainger College of Engineering, University of Illinois at Urbana-Champaign, Urbana, 61801 IL, USA}
\affiliation{Materials Research Laboratory, The Grainger College of Engineering, University of Illinois at Urbana-Champaign, Urbana, 61801 IL, USA}

\author{Younsik Kim}
\affiliation{Department of Physics, The Grainger College of Engineering, University of Illinois at Urbana-Champaign, Urbana, 61801 IL, USA}
\affiliation{Materials Research Laboratory, The Grainger College of Engineering, University of Illinois at Urbana-Champaign, Urbana, 61801 IL, USA}

\author{Camille Bernal-Choban}
\affiliation{Department of Physics, The Grainger College of Engineering, University of Illinois at Urbana-Champaign, Urbana, 61801 IL, USA}
\affiliation{Materials Research Laboratory, The Grainger College of Engineering, University of Illinois at Urbana-Champaign, Urbana, 61801 IL, USA}

\author{Gaurav Tenkila}
\affiliation{Department of Physics, The Grainger College of Engineering, University of Illinois at Urbana-Champaign, Urbana, 61801 IL, USA}
\affiliation{Institute of Condensed Matter Theory,
University of Illinois at Urbana-Champaign, Urbana, 61801 IL, USA}

\author{Suhas Sheikh}
\affiliation{Department of Physics, The Grainger College of Engineering, University of Illinois at Urbana-Champaign, Urbana, 61801 IL, USA}
\affiliation{Institute of Condensed Matter Theory,
University of Illinois at Urbana-Champaign, Urbana, 61801 IL, USA}

\author{Pranav Mahaadev}
\affiliation{Department of Physics, The Grainger College of Engineering, University of Illinois at Urbana-Champaign, Urbana, 61801 IL, USA}
\affiliation{Materials Research Laboratory, The Grainger College of Engineering, University of Illinois at Urbana-Champaign, Urbana, 61801 IL, USA}

\author{Faren Hoveyda-Marashi}
\affiliation{Department of Physics, The Grainger College of Engineering, University of Illinois at Urbana-Champaign, Urbana, 61801 IL, USA}
\affiliation{Materials Research Laboratory, The Grainger College of Engineering, University of Illinois at Urbana-Champaign, Urbana, 61801 IL, USA}

\author{Subhajit Roychowdhury}
\affiliation{Max Planck Institute for Chemical Physics of Solids, Dresden 01187, Germany}
\affiliation{Department of Chemistry, Indian Institute of Science Education and Research Bhopal, Bhopal 462066, India}

\author{Chandra Shekhar}
\affiliation{Max Planck Institute for Chemical Physics of Solids, Dresden 01187, Germany}

\author{Claudia Felser}
\affiliation{Max Planck Institute for Chemical Physics of Solids, Dresden 01187, Germany}

\author{Peter Abbamonte}
\affiliation{Department of Physics, The Grainger College of Engineering, University of Illinois at Urbana-Champaign, Urbana, 61801 IL, USA}
\affiliation{Materials Research Laboratory, The Grainger College of Engineering, University of Illinois at Urbana-Champaign, Urbana, 61801 IL, USA}

\author{Benjamin J. Wieder}
\affiliation{Institut de Physique Th\'eorique, Universit\'e Paris-Saclay, CEA, CNRS, F-91191 Gif-sur-Yvette, France}

\author{Fahad Mahmood} 
\email{fahad@illinois.edu}
\affiliation{Department of Physics, The Grainger College of Engineering, University of Illinois at Urbana-Champaign, Urbana, 61801 IL, USA}
\affiliation{Materials Research Laboratory, The Grainger College of Engineering, University of Illinois at Urbana-Champaign, Urbana, 61801 IL, USA}

\date{\today}

\begin{abstract}
Collective modes in Fermi liquids are usually regarded as dissipation channels that relax electronic excitations through Landau damping. Whether such modes can instead mediate the formation of correlated electronic states under non-equilibrium conditions remains an open question. 
Here, we show that under optical photo-doping, a bulk plasmon can drive correlated inter-band transfer within a transient electronic continuum. Using time- and angle-resolved photoemission spectroscopy (Tr-ARPES) on \ce{EuCd2As2} supported by electronic structure calculations, we observe that at high excitation density, plasmons transfer energy from a weakly dispersing bulk band into unoccupied surface states. This bulk-to-surface redistribution stabilizes a long-lived, energy-localized spectral feature consistent with a Mahan exciton. Our results reveal a non-equilibrium regime of Fermi-liquid physics in which collective modes do not merely dissipate energy, but also stabilize correlated bound states.
\end{abstract}

\maketitle

\begin{figure}[t]
    \includegraphics[width=\linewidth]{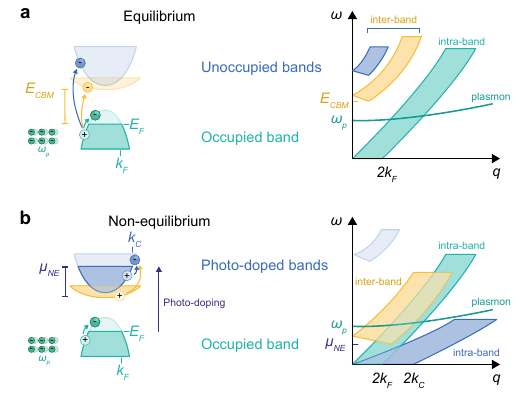}
    \caption{\label{fig:excitation_spectrum}\textbf{Excitations of a Fermi liquid at and away from equilibrium:} \textbf{(a)} Schematic depiction of the excitations of a Fermi liquid at equilibrium, represented as scattering processes along the occupied and unoccupied band dispersions ($E$ vs $k$) (left) and as a spectrum showing the excitation energy $\omega$ as a function of the excitation wave vector $q$ (right). 
    $E_F$ and $k_F$ indicate the Fermi level and Fermi momentum, respectively. $E_{CBM}$ indicates the energy minimum of the lower conduction band, specified relative to $E_F$, and $\omega_p$ denotes the energy of the equilibrium plasmon.
    \textbf{(b)} Excitations of a Fermi liquid with a secondary chemical potential ($\mu_{NE}$) photo-doped into a conduction band. 
    In (b), $\mu_{NE}$ is specified relative to the bottom of the subset of photo-doped bands.}
\end{figure}

Collective excitations in interacting Fermi systems are traditionally regarded as decay channels. In a conventional Fermi liquid, the plasmon branch lies outside the particle–hole continuum at long wavelengths, but enters it at finite momentum, where Landau damping transfers plasmon energy into electron–hole pairs (Fig.~\ref{fig:excitation_spectrum}a)~\cite{Nozieres2018TheLiquids, Wehling2014DiracMaterials}.
This paradigm underlies the standard description of relaxation in metals, where collective modes redistribute electronic energy into particle–hole continua.

Such relaxation also extends to non-equilibrium settings: in optically excited metals and semiconductors, bosonic-mode-assisted inter- and intra-band scattering channels have been directly observed transferring high-energy carriers into lower-energy electronic continua and bound states~\cite{Na2019DirectDomain,Schmitt2022FormationTime, Madeo2020}. Moreover, transient excitonic states are suppressed in metals when the dephasing of plasmons screens the electron-hole Coulomb interaction~\cite{Cui2014TransientSurfaces}.
In these cases, collective modes function as an efficient pathway toward equilibration. Yet this established framework leaves open the fundamental question of whether collective modes, rather than merely dissipating energy, can also actively drive excitations of a particle-hole continuum, and generate correlated electronic states under non-equilibrium conditions.

In general, inter-band optical excitation generates a non-equilibrium carrier population that rapidly thermalizes to a secondary chemical potential within transiently filled bands~\cite{Gierz2013SnapshotsGraphene}. The resulting inter- and intra-band excitation continua can extend to $\omega=0$, overlapping with the plasmon branch (Fig.~\ref{fig:excitation_spectrum}b). In this regime, collective modes need not act solely as sinks of energy; they may instead mediate inter-band transfer within the transient continuum.

Observing such an interaction channel in a bulk metal, however, is challenging. In 3D systems, strong screening and the large phase space tend to smear interaction-enhanced features. Plasmons are heavily damped once they overlap with bulk continua, and competing scattering channels redistribute spectral weight broadly in momentum and energy. A more favorable setting would involve 2D surface electronic bands interacting with the bulk plasmon of a 3D system. The reduced dimensionality of the surface suppresses phase space, weakens screening, and therefore enhances interaction effects.
Moreover, the near-field interaction between the dipole moment of a bulk collective mode in a low-carrier-density material and surface-localized conduction electrons is exponentially enhanced at zero-momentum transfer~\cite{Lee2014InterfacialSrTiO3}.
In this configuration, a bulk plasmon could hence transfer energy to the more confined surface electron bands, giving rise to correlated features that can be spectrally resolved.

In this work, we experimentally and theoretically demonstrate that \ce{EuCd2As2} precisely provides the latter setting -- we observe that the electronic structure of \ce{EuCd2As2} hosts a bulk plasmon at an energy scale comparable to the separation between an unoccupied, weakly dispersing bulk band and an unoccupied surface state. Our density-functional theory (DFT) calculations establish the existence of these unoccupied surface states, motivating the experimental investigation of plasmon-mediated transfer between the bulk and surface in the non-equilibrium continuum. 
Using time- and angle-resolved photoemission spectroscopy (Tr-ARPES), we not only directly demonstrate this plasmon-mediated transfer, but also find that the reduced screening and constrained phase space of the surface potentially stabilizes a long-lived, energy-localized spectral feature consistent with a Mahan exciton~\cite{Mahan1967ExcitonsSemiconductors, Mahan1967ExcitonsMass, Asnin1983CoulombGermanium, Palmieri2020MahanPerovskites}. 
Our results reveal a non-equilibrium regime of Fermi-liquid physics in which collective modes do not simply dissipate energy, but also actively reshape electronic distributions by driving bulk-to-surface transfer and generating electronic bound states.

\begin{figure}[t!]
\includegraphics[width=\linewidth]{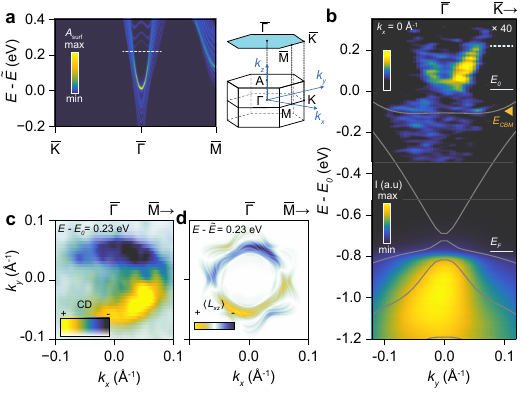}
\caption{\textbf{Unoccupied band structure of \ce{EuCd2As2}:} 
\textbf{(a)} $(001)$-surface spectral function, obtained from density-functional theory (DFT) calculations in the paramagnetic (PM) state (SM Sec.~\ref{siDFT}~\cite{2026SeeURL}), plotted relative to the energy minimum of the surface band $\widetilde{E}$.
\textbf{(b)} Time-integrated ($t-t_0\in[\qty{0}{ps},\qty{2}{ps}]$) ARPES spectrum showing the occupied and unoccupied band structure. Light gray lines indicate the bulk band structure along $\Gamma-\mathrm{K}$ obtained from DFT. Dark gray horizontal rules indicate where different scans were truncated to render the composite figure.
Above the Fermi level $E_{F}$, we observe surface states with an energy minimum $E_0\approx E_F+\qty{0.8}{eV}$ (SM Sec.~\ref{sislabgreens}~\cite{2026SeeURL}).
A weakly dispersing bulk conduction band (yellow triangle) lies at $E_{CBM}$, below $E_{0}$.
\textbf{(c)} Constant-energy cut (dashed line in (b)) of circular dichroism (CD-) ARPES measurement at $t_0$, where the hue (saturation) represents the CD signal (polarization-integrated intensity). 
\textbf{(d)} Constant-energy cut (dashed line in (a)) of the local atomic orbital angular momentum (OAM) of the DFT-obtained surface spectrum in (a), projected along the incident axis of the probe (see SM Sec.~\ref{sispintext}~\cite{2026SeeURL} and End Matter).
Tr-ARPES measurements were taken at \qty{20}{K}, below $T_C\approx\qty{25}{K}$, using \qty{1.2}{eV} excitation at low fluence (\qty{15}{\micro J/cm^2}); we obtained qualitatively identical Tr-ARPES spectra above $T_{C}$ (\qty{42}{K}, SM Sec.~\ref{CD_temp_SI}).}
\label{fig:overview}
\end{figure}

\ce{EuCd2As2} has previously been characterized as semiconducting~\cite{Santos-Cottin2023EuCd2As2:Semiconductor, Nelson2024RevealingLa-Doping, Shi2024AbsenceEffect} or semimetallic~\cite{Ma2019SpinEuCd2As2, Roychowdhury2023AnomalousEuCd2As2} throughout its magnetic phase diagram (Supplementary Material (SM) Sec.~\ref{Semimetal_SI}~\cite{2026SeeURL}), with an antiferromagnetic phase reported below $T_N=\qty{9}{K}$. In the samples used in this work, ferromagnetic (FM) order has been stabilized below $T_C=\qty{25}{K}$ through \ce{Eu} vacancies~\cite{Roychowdhury2023AnomalousEuCd2As2}.
Above $T_{C}$, \ce{EuCd2As2} lies in a paramagnetic (PM) state with bulk inversion ($\mathcal{I}$) and time-reversal ($\mathcal{T}$) symmetries. To analyze the bulk and surface spectra of \ce{EuCd2As2}, we performed DFT and Wannier-based tight-binding calculations of the electronic structure (SM Sec.~\ref{siDFT}~\cite{2026SeeURL}).
Because our ARPES measurements produced qualitatively identical spectra both above and below $T_{C}$ over the energy and momentum range probed in this study (SM Sec.~\ref{CD_temp_SI}~\cite{2026SeeURL}), we performed our DFT calculations in the PM state.
Our DFT calculations reveal unoccupied $(001)$-surface bands that coexist with the projected continuum of unoccupied bulk states (the projected bulk conduction manifold) and exhibit an approximately hyperbolic dispersion with an energy minimum that we denote as $\widetilde{E}$ (Fig.~\ref{fig:overview}a, see SM Sec.~\ref{sislabgreens}~\cite{2026SeeURL}).
Through calculations of the $(001)$-surface spin texture in SM Sec.~\ref{sispintext}~\cite{2026SeeURL}, we find that the $(001)$-surface states form a 2D Dirac cone with weak spin-orbit-coupling- (SOC-) driven splitting away from ${\bf k}={\bf 0}$.

To understand the origin of the hyperbolic $(001)$-surface Dirac cone, we applied the group-theoretic methodology in Ref.~\cite{Vergniory2022AllMaterials} to compute the band topology at and away from $E_{F}$ (SM Sec.~\ref{sirepeattopo}~\cite{2026SeeURL}).
We identified a large range of electronic fillings -- spanning from 5 electrons below intrinsic filling to 7 electrons above -- at which \ce{EuCd2As2} in the PM state hosts 3D topological band-insulating gaps.
We may therefore associate the unoccupied $(001)$-surface states in Fig.~\ref{fig:overview}a to a 3D topological band insulator phase at an electronic filling above $E_{F}$, which can be accessed by photo-doping experiments.
In this sense, \ce{EuCd2As2} is reminiscent of the repeat-topological materials introduced in Ref.~\cite{Vergniory2022AllMaterials} (\emph{e.g.} \ce{Bi2Mg3}~\cite{Chang2019RealizationMg3Bi2,Zhou2019ExperimentalEpitaxy}), which host topological gaps at and just below $E_{F}$.
The unoccupied $(001)$-surface states in \ce{EuCd2As2} and the associated topological band gaps above $E_{F}$ can hence be viewed as a generalization of the concept of repeat-topology to topological surface conduction bands that are accessible by doping and Tr-ARPES experiments like those in the present study.

\begin{figure}[t!]
\includegraphics[width=\linewidth]{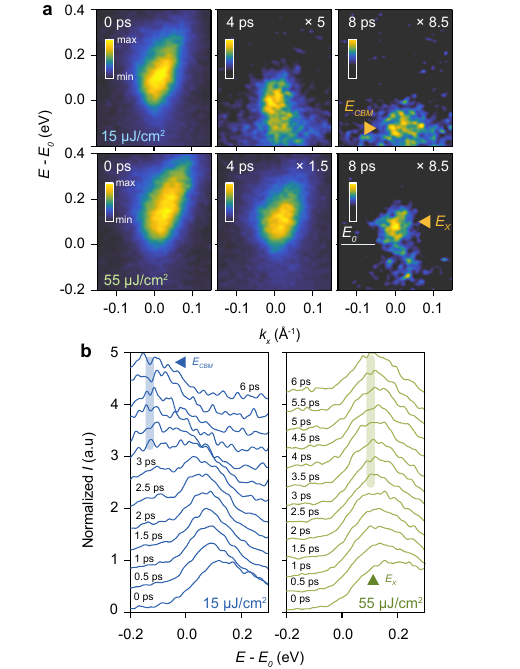}
\caption{\label{fig:fluence_dependence}\textbf{Population dynamics of the unoccupied surface states at representative excitation densities: }\textbf{(a)} ARPES spectra of the unoccupied band structure at selected temporal delays after low fluence (top row) and high fluence (bottom row) excitation.
\textbf{(b)} Normalized energy distribution curves from the spectra in (a) at various temporal delays after photo-excitation for low (\qty{15}{\micro J/cm^2}, left) and high (\qty{55}{\micro J/cm^2}, right) fluence photo-excitation, where the shaded region at $E_{X}$ in the right panel is a guide to the eye.
The Tr-ARPES measurements in (a, b) were taken at \qty{9.5}{K} using \qty{1.5}{eV} excitation.}
\end{figure}

Having established the presence and energies of unoccupied surface states in \ce{EuCd2As2}, we now use Tr-ARPES to dynamically populate and resolve the surface bands.
In our measurements, \textit{p}-polarized IR pulses at either \qty{1.2}{eV} or \qty{1.5}{eV} excite inter-band transitions, after which we collect ARPES spectra using \qty{6}{eV} photons with a variable delay from photo-excitation. In momentum space, the photon scattering plane coincides with a crystallographic mirror plane whose normal vector lies along $k_{y}$ and is related to the Brillouin zone line $\overline{\Gamma}-\overline{\mathrm{K}}$ in Fig.~\ref{fig:overview}a by crystal symmetry.
In Fig.~\ref{fig:overview}b, we show an ARPES spectrum ($E$ vs $k_y$) of the occupied and unoccupied bulk and surface band structure of \ce{EuCd2As2}.
Below $E_F$ (\emph{i.e.} at energies $E-E_0<\qty{-0.8}{eV}$, where $E_0$ is assigned to the bottom of a bright electron-like unoccupied band), a broad valence band is visible, consistent with equilibrium synchrotron ARPES measurements~\cite{Roychowdhury2023AnomalousEuCd2As2} and the bulk DFT calculations performed in the present study (gray lines, see SM Sec.~\ref{siDFT}~\cite{2026SeeURL}).

Around \qty{0.8}{eV} above $E_F$ in Fig.~\ref{fig:overview}b ($\emph{i.e.}$ above $E_0$), we observe a sharp, nearly-linear (hyperbolic) electron-like feature, in strong agreement with the Wannier model surface spectral function in Fig.~\ref{fig:overview}a. 
Next, at $E_{CBM}\equiv E_0-\qty{0.12}{eV}$ in Fig.~\ref{fig:overview}b, we observe a faint incoherent spectral weight that we attribute to a weakly dispersing bulk conduction band \qty{0.7}{eV} above $E_F$, consistent with optical spectroscopy~\cite{Santos-Cottin2023EuCd2As2:Semiconductor, Wu2024TheSingularity} and high-energy electron energy loss spectroscopy (EELS) measurements (SM Sec.~\ref{bulk_gap_SI}~\cite{2026SeeURL}), as well as our bulk DFT calculations (SM Sec.~\ref{siDFT}~\cite{2026SeeURL}).
In Fig.~\ref{fig:overview}c, we show circular dichroism (CD-) ARPES measurements of the $(001)$-surface states, which provide further details of their wavefunction textures, and exhibit excellent agreement with theoretical calculations of the surface-projected local (in real space) atomic-orbital contribution to the orbital angular momentum (Fig.~\ref{fig:overview}d, see SM Sec.~\ref{sispintext}~\cite{2026SeeURL}). 

We next investigate the dynamics of the photo-doped surface and bulk conduction bands under different photo-excitation fluences.
In Fig.~\ref{fig:fluence_dependence}a, we plot ARPES spectra at several different time delays after excitation for representative cases of low (\qty{15}{\micro J/cm^2}, top) and high (\qty{55}{\micro J/cm^2}, bottom) fluences (SM Sec.~\ref{sec:Clarity}~\cite{2026SeeURL}). 
Immediately after excitation, a broad energy range is populated at both low and high fluence (left panels of the top and bottom rows of Fig.~\ref{fig:fluence_dependence}a, respectively).
The photo-doped population above $E_{F}$ (here, in the vicinity of the bottom of the surface band $E_{0}$, see Fig.~\ref{fig:overview}b) corresponds to the non-equilibrium phase depicted in Fig.~\ref{fig:excitation_spectrum}b. 
At low fluence, only states in the nondispersing bulk conduction band ($E\approx E_{CBM}<E_0$) remain occupied several picoseconds after excitation (Fig.~\ref{fig:fluence_dependence}a, top right). 
This is consistent with previous studies that mapped the unoccupied band structure in solids~\cite{Santos-Cottin2023EuCd2As2:Semiconductor, Gierz2013SnapshotsGraphene, Puppin2022Excited-stateMapping, Bao2022PopulationCd3As2, Fragkos2024Excited1T-ZrTe2}, in which long-lived populations above $E_F$ were associated with the energy minima of unoccupied bulk bands. At negative delays, we faintly resolve a non-equilibrium energy-localized population distribution, induced by photon-energy-dependent direct inter-band excitations (SM Sec.~\ref{sec:si_rise}~\cite{2026SeeURL}).
At and after $t_0$, the momentum-integrated population at different times (Fig.~\ref{fig:fluence_dependence}b, left panel) can be fit using Fermi-Dirac distributions multiplied by linear densities of states (SM Sec.~\ref{fitting_SI}~\cite{2026SeeURL}).
We find that the resulting chemical potential at later delay times $\mu(t\gg t_0)$ is largely independent of fluence, and that $\mu(t\gg t_0)\approx\qty{100}{meV} + E_{0}$ for \qty{1.5}{eV} photon excitation. 
Such cooling is largely consistent with a two-temperature model in which photo-excitation raises the temperature of electrons, which then cool by transferring energy to a bath of collective bosonic modes, such as phonons, magnons, or plasmons (see SM Sec.~\ref{sec:si_req_model}~\cite{2026SeeURL} for further details).

Surprisingly, at high fluence (Fig.~\ref{fig:fluence_dependence}a, bottom row), a strong spectral weight persists at $E_X\equiv E_0+\qty{100}{meV}$, significantly above the surface band minimum $E_{0}$ (SM Sec.~\ref{charging_SI}~\cite{2026SeeURL}), even after the populations at lower energies have largely decayed (Fig.~\ref{fig:fluence_dependence}a, bottom right). 
The strong energy localization of this long-lived spectral weight, and its qualitative absence for low fluence excitation, are shown in Fig.~\ref{fig:fluence_dependence}b. 
A clear transfer of spectral weight to lower energies over several picoseconds is visible for low fluence excitation (Fig.~\ref{fig:fluence_dependence}b, left), whereas for high fluence excitation (Fig.~\ref{fig:fluence_dependence}b, right), the spectrum after $t_0$ remains distinctly sharpened at $E_X$ relative to the initial hot electron distribution.

\begin{figure}[t]
\includegraphics[width=\linewidth]{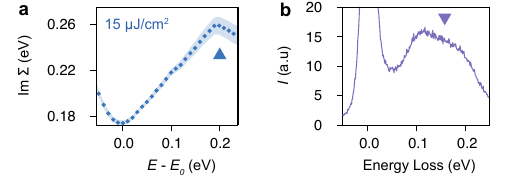}
\caption{\label{fig:electron-plasmon_coupling}\textbf{Electron-plasmon coupling in the unoccupied conduction states:}
\textbf{(a)} The imaginary part of the self energy $\text{Im }\Sigma(E)$ of the unoccupied band structure, measured at various lattice temperatures under low fluence excitation (\qty{15}{\micro J/cm^2}) by \qty{1.5}{eV} photons. The shaded colored regions indicate a 95\% confidence interval from Lorentzian fits. 
In (a), the ARPES cuts were integrated over $t-t_0\in[0,2]\ \qty{}{ps}$ to provide sufficient spectral weight, and the blue triangle is a guide to the eye to highlight the temperature dependence of the kink (derivative discontinuity).
\textbf{(b)} The momentum-integrated spectrum obtained from equilibrium electron energy loss spectroscopy (EELS) at room temperature, showing a broad plasmon feature between \qty{0.1}{eV} and \qty{0.2}{eV} (purple triangle).}
\end{figure}

To identify the origin of the long-lived population at $E_X$ in Fig.~\ref{fig:fluence_dependence}, we next examine the role of many-body interactions in the unoccupied bands.
Under low fluence excitation, the imaginary part of the self energy $\text{Im }\Sigma(E)$ extracted from ARPES linewidths~\cite{Kaminski2005OnSpectroscopy} (SM Secs.~\ref{fitting_SI} and~\ref{charging_SI}~\cite{2026SeeURL}) shows a strong discontinuity in its derivative (a ``kink") at $E-E_0\approx\qty{200}{meV}$ (Fig.~\ref{fig:electron-plasmon_coupling}a). 
Under non-equilibrium distributions in weakly interacting systems, such kinks can arise from changes in quasiparticle lifetimes due to crossovers in phase-space-allowed scattering mechanisms~\cite{duvel}, and therefore here indicate strong coupling of the surface states in \ce{EuCd2As2} to a collective mode with the corresponding energy. 
Previous optical spectroscopy experiments~\cite{Wang2016AnisotropicProperties} and EELS measurements at equilibrium performed in this work (Fig.~\ref{fig:electron-plasmon_coupling}b) identify the relevant mode between \qty{0.1}{eV} and \qty{0.2}{eV} above $E_{0}$ as the bulk plasmon. 
This suggests that, although the surface conduction band dynamics in \ce{EuCd2As2} differ sharply from previous Tr-ARPES studies of conduction band dynamics in a broad set of materials~\cite{Sobota2013DirectBi2Se3, Sobota2014UltrafastSpectroscopy, Tanimura2015UltrafastSpectroscopy, Puppin2022Excited-stateMapping} -- in which electron-electron scattering is the dominant interaction far from $E_F$ -- the low fluence conduction band dynamics in \ce{EuCd2As2} can still be understood with a conventional two-temperature model in which a cold bath of bosons (here plasmons) provides a relaxation channel for hot electrons (see SM Sec.~\ref{sec:si_req_model}~\cite{2026SeeURL} and Refs.~\cite{Allen1987TheoryMetals,Sun2012DynamicsGraphene,Gierz2014Non-equilibriumSpectroscopy, Bao2022PopulationCd3As2}).

\begin{figure}[t]
\includegraphics[width=\linewidth]{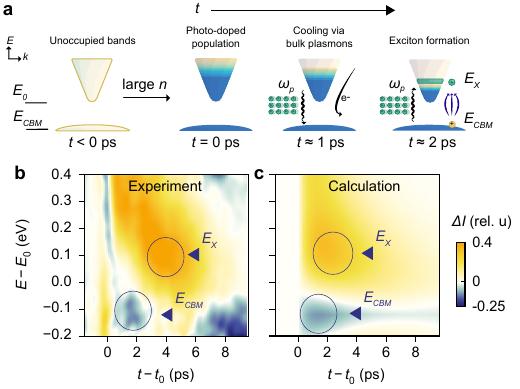}    \caption{\label{fig:mahan_exciton}\textbf{Plasmon-driven exciton formation: }
\textbf{(a)} Schematic of the dynamics of electrons excited into unoccupied surface states, showing the plasmon-driven formation of Mahan excitons. 
\textbf{(b)} Difference between time traces for high (\qty{55}{\micro J/cm^2}) and low (\qty{15}{\micro J/cm^2}) excitation fluence, with the maximum intensity at each energy level for both fluences (reached approximately at $t_0$) normalized to 1 after subtracting a pre-$t_0$ background signal (see SM Sec.~\ref{RateEq_SI}~\cite{2026SeeURL} for detailed descriptions of the background subtraction and normalization procedures).
\textbf{(c)} Theoretical calculation of a normalized difference time trace like that in (b) in the presence of plasmon-driven exciton formation, obtained from a modified rate equation model detailed in SM Sec.~\ref{sec:si_req_model}~\cite{2026SeeURL}.}
\end{figure}

However at high fluence excitation, we posit that the role of plasmons is qualitatively different.
We specifically instead propose that in the high-fluence regime, the persistence of localized spectral weight at $E_X$ arises due to a hot plasmon-driven inter-band transition that drives electrons from the bulk conduction band just below $E_0$ (yellow triangle in Fig.~\ref{fig:overview}b) to states at $E_X$ in the surface conduction band.
This process corresponds precisely to the interaction between the plasmon branch and the non-equilibrium particle-hole continuum in Fig.~\ref{fig:excitation_spectrum}b.
Under high excitation fluence, a Mahan exciton may then form from the Coulomb interaction of the electron-hole pairs (schematically depicted in Fig.~\ref{fig:mahan_exciton}a).

Mahan excitons specifically form with electrons at the effective Fermi level of an \textit{n}-doped system when hole densities are negligible due to valence degeneracy~\cite{Mahan1967ExcitonsSemiconductors, Mahan1967ExcitonsMass}.
In support of this scenario in \ce{EuCd2As2}, we previously noted that the chemical potential $\mu$ of the population in the unoccupied surface bands cools to $\mu(t\gg t_0)\approx E_0 + \qty{100}{meV}$ (SM Sec.~\ref{fitting_SI}~\cite{2026SeeURL}), which is equal to $E_X$.
Moreover, our assessment of high valence degeneracy is supported by optical spectroscopy studies, which identify a bulk conduction band with a large mass at $\Gamma$ in the energy range of the yellow triangle at $E_{CBM}$ in Fig.~\ref{fig:overview}b~\cite{Wu2024TheSingularity, Santos-Cottin2023EuCd2As2:Semiconductor}, as well as our PM-state DFT calculations, which show a doubly degenerate bulk band at approximately the same energy that is flat to leading order in 3D $k$-space near $\Gamma$ (SM Sec.~\ref{siDFT}~\cite{2026SeeURL}).

To test the interpretation of a Mahan exciton at high fluence excitation, we first search for signatures of holes in the bulk conduction band after photo-excitation. 
This is challenging, because at equilibrium, all of the bulk and surface states participating in this scenario are unoccupied. 
To identify holes in the bulk conduction band, we therefore compare population dynamics between the excitonic and non-excitonic conditions, which in our experiments respectively correspond to high (\qty{55}{\micro J/cm^2}) and low (\qty{15}{\micro J/cm^2}) fluence excitation.
In Fig.~\ref{fig:mahan_exciton}b, we show the difference between time traces taken at high and low fluence, with the maximum spectral weight at each energy for both fluences normalized to one after subtracting a pre-$t_0$ background signal (see SM Sec.~\ref{RateEq_SI}~\cite{2026SeeURL} for a detailed description and justification of the background subtraction and normalization procedure).
Strikingly, a relative depletion of spectral weight for high fluence accumulates at $E_{CBM}\equiv E_0-\qty{100}{meV}$ during the first \qty{2}{ps} (dark blue region in Fig.~\ref{fig:mahan_exciton}b). 
This indicates that the initial population decay at $E_{CBM}$ is faster for high fluence excitation than at low fluence. 
Simultaneously, a substantially larger relative spectral weight at high fluence accumulates at $E_X$ within a few picoseconds (gold region in Fig.~\ref{fig:mahan_exciton}b), consistent with the longer lifetime of an excitonic population.

We interpret these observations as evidence for an additional process -- apparent only at high fluence -- that transfers spectral weight across the inter-band transition from $E_{CBM}$ to $E_X$ (blue arrow in Fig.~\ref{fig:mahan_exciton}b) over the corresponding \qty{2}{ps} timescale. 
The energy difference $E_X-E_{CBM}$ is approximately equal to the plasmon energy previously identified in Fig.~\ref{fig:electron-plasmon_coupling}, strongly suggesting that the transfer is plasmon-induced. 
In turn, the rate of such plasmon absorption should scale quadratically with the excited carrier density, because the plasmons must first be excited by the relaxation of electrons that were themselves previously excited into the unoccupied bands in a system that started at low temperature.
Hence, the plasmon-induced transition from $E_{CBM}$ to $E_{X}$ is only dominant at high fluence, and excitons should not form in the low-fluence regime, consistent with our observations in Fig.~\ref{fig:fluence_dependence}a.

The onset of such a second-order process should also be continuous: at low fluences (including \qty{15}{uJ/cm^2}), plasmon-induced transitions would be negligible, but not strictly forbidden.
Consistent with this expectation, we observe a crossover from the low-fluence to high-fluence regime at approximately \qty{46}{uJ/cm^2} (SM Secs.~\ref{fitting_SI} and \ref{sec:Clarity}~\cite{2026SeeURL}).
Specifically, at this intermediate fluence, we can resolve the initial thermalization of the hot electron distribution -- due to a relatively small population of plasmon-driven electron-hole pairs compared to hot electrons (SM Sec.~\ref{fitting_SI}~\cite{2026SeeURL}) -- while simultaneously resolving a relative dearth of electron accumulation near $E_{CBM}$ immediately after excitation (SM Sec.~\ref{sec:Clarity}~\cite{2026SeeURL}).
For completeness, we additionally note that in Fig.~\ref{fig:mahan_exciton}b, a relative dearth of electrons for high fluence excitation is also present around $E_{CBM}$ at later delay times ($t-t_0>\qty{6}{ps}$). 
We attribute this long-lived dearth of electrons at $E_{CBM}$ -- the relative intensity of which shows fluence dependence (SM Sec.~\ref{sec:si_beyondREQ}~\cite{2026SeeURL}) -- to a persisting population in the conduction band that is larger at low fluence.
Importantly, however, this long-lived conduction band population evolves over slower timescales, and hence does not influence our proposed mechanism of plasmon-driven electron-hole pair formation.

The contrast between the dynamics at low and high excitation fluence can be captured using a modified rate equation that models the photo-excited population as relaxing via plasmon-mediated scattering through an additional channel that at high fluence transfers spectral weight between electron and hole energies upon the formation of excitons.
A detailed description of the model is provided in SM Sec.~\ref{sec:si_req_model}~\cite{2026SeeURL}.
In Fig.~\ref{fig:mahan_exciton}c, we show numerical data generated using this theoretical model, which are in close agreement with the experimental data in Fig.~\ref{fig:mahan_exciton}b.
We obtained the input parameters for the theoretical model through independent fits of Tr-ARPES time traces at low and high fluence.
The model parameters respectively indicate an exciton formation time and lifetime of approximately \qty{1.4}{ps} and \qty{4.4}{ps}, though with moderate correlation between the two values.
Lastly, in SM Sec.~\ref{SI_alternativeExplanations}~\cite{2026SeeURL}, we discuss in detail alternative, non-excitonic interpretations of the fluence-dependent electron dynamics in our experimental data, and ultimately find these alternative scenarios to be inconsistent with our complete dataset.

To conclude, the signatures of Mahan excitons in our measurements motivate future studies to probe their spin polarization~\cite{Mori2023Spin-polarizedInsulator}, electronic chirality~\cite{Kung2019ObservationBi2Se3}, condensation~\cite{Pertsova2018ExcitonicMaterials, Mori2025PossibleInsulator}, and tunable binding energies~\cite{Pertsova2018ExcitonicMaterials}.
From a broader perspective, our results point to a generalized form of carrier multiplication, in which the absorption of a photon creates multiple electron-hole excitations~\cite{Nozik2009Nanophotonics:Photons}.
Moreover, due to the coherent near-field coupling to bulk modes demonstrated in this study, surface non-equilibrium Fermi liquids constitute a powerful platform for inducing collective interaction-driven phenomena that are otherwise inaccessible at thermal equilibrium, including phononic control of phase transitions~\cite{Rini2007ControlExcitation, Basov2017TowardsMaterials} and intrinsically driven Floquet phases~\cite{Hubener2018PhononMatter}.

\vspace{2mm}
\noindent\textbf
{Acknowledgments:}
We thank V.~Madhavan, M.~Knudtson, M.~Vergniory, and J.~Goldberger for insightful discussions.
R.A., N.B., H.E.A., Y.K., P.M., and F.M. acknowledge support from NSF Career Award No.~DMR-2144256 and from the EPiQS program of the Gordon and Betty Moore Foundation, Grant No.~GBMF11069.
C.B.-C., F.H.-M., and P.A. acknowledge support from the Center for Quantum Sensing and Quantum Materials, an Energy Frontier Research Center funded by the U.S. Department of Energy (DOE), Office of Science, Basic Energy Sciences (BES), under Award No.~DE-SC0021238, as well as additional support from the EPiQS program of the Gordon and Betty Moore Foundation, Grant No.~GBMF9452.
B.J.W. and E.G. acknowledge support from the European Union’s Horizon Europe research and innovation program (ERC-StG-101117835-TopoRosetta), ANR PIA funding (ANR-20-IDEES-0002), and CNRS IRP Project NP-Strong.
B.J.W. and E.G. further acknowledge the Laboratoire de Physique des Solides, Orsay for hosting during the preparation of this work.
Sample characterization was carried out in part in the Materials Research Laboratory Central Research Facilities, University of Illinois.
The first-principles calculations for this work were performed using the CCRT High-Performance Computing (HPC) facility under the Grant CCRT2025-gerbere awarded by the Fundamental Research Division (DRF) of the CEA.

\vspace{2mm}
\noindent\textbf{Data availability:} The data in this Article are available via the Illinois Data Bank~\cite{AcharyaDataLiquid}.

\bibliography{references, refs_extra}

\onecolumngrid
\newpage
\section*{End Matter for ``Plasmon-driven electron-hole pair formation in a non-equilibrium Fermi liquid"}
\subsection*{Time- and Angle-Resolved Photoemission Spectroscopy Measurements}

\begin{figure}[h]
\includegraphics{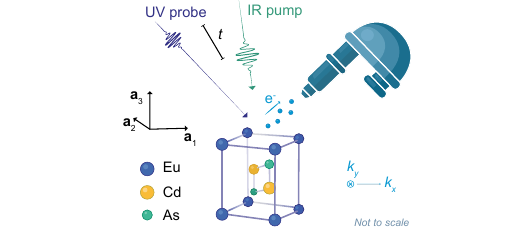}    \caption{\label{fig:experiment_setup}\textbf{Schematic of the Tr-ARPES experiment: } The UV probe and IR pump were incident along the $(x,z)$- and $(y,z)$-planes of the lab frame, both at \ang{45} to the sample normal (the $(001)$-surface) and were \textit{p}-polarized. 
The sample was aligned with $\overline{\Gamma}-\overline{\textrm{M}}$ along $k_x$. 
This measurement geometry probes both the $(001)$-surface and the bulk electronic structure at $k_{z}\approx 0$.}
\end{figure}

The Tr-ARPES experiments were performed in the home lab at the University of Illinois, Urbana-Champaign. The fifth harmonic (\qty{6}{eV}) of the \qty{1.2}{eV} (\qty{1030}{nm}) output of a Yb:KGW amplified laser (Pharos, Light Conversion) was used as the probe. The linearly polarized probe was transmitted through a quarter waveplate for CD-ARPES measurements. The IR excitation pulses of \qty{1.2}{eV} (\qty{1000}{nm}) and \qty{1.5}{eV} (\qty{825}{nm}) were provided by an OPA (Orpheus, Light Conversion) pumped by the same Yb:KGW laser. 
The samples were cleaved at \qty{9.5}{K} at a base pressure better than \qty{1E-10}{mB}. Measurements were performed using a hemispherical analyzer (DA-30L, Scienta Omicron) equipped with a deflection mode to observe dispersion along both in-plane axes without rotation of the sample. The geometry of the Tr-ARPES experiment is shown in Fig.~\ref{fig:experiment_setup}.
The sample crystal axes were aligned using the measured Fermi surface. The energy resolution (measured through Fermi edge broadening on \ce{Bi2Se3}) was set at \qty{10}{meV}. The momentum resolution was better than \qty{0.02}{A^{-1}}. 
The overall temporal resolution of the system (measured by cross-correlation on unoccupied topological surface states of \ce{Bi2Se3}) was set at \qty{200}{fs}.

\subsection*{First-Principles and Surface State Calculations}

We performed the \emph{ab initio} calculations of the electronic structure of the PM phase of EuCd$_{2}$As$_{2}$ within the density-functional theory (DFT) framework, as implemented in the \texttt{Quantum Espresso} software package~\cite{Giannozzi2009QUANTUMMaterials, Giannozzi2017AdvancedESPRESSO, Giannozzi2020QuantumExascale}.
The calculations included fully relativistic spin–orbit coupling (SOC) in the noncollinear scheme. 
In this setup, when no initial magnetic moments are specified, the self-consistent cycle converges to a nonmagnetic, time-reversal- ($\mathcal{T}$-) invariant solution. To properly localize the half-filled Eu $4f$ orbitals and improve the overall accuracy of the PM-state band structure, a Hubbard $U$ correction of $U=5$ eV on the Eu sites was implemented within the GGA $+\ U$ scheme~\cite{Dudarev1998Electron-energy-lossStudy}.
To analyze the spectral and topological properties of the PM state, we used maximally localized Wannier functions~\cite{Pizzi2020Wannier90Applications, Marzari1997MaximallyBands, Souza2001MaximallyBands} to construct a tight-binding model using the Eu 5\emph{d} and 4\emph{f}, Cd 5\emph{s}, and As 4\emph{p} orbitals with SOC, and symmetrized the model~\cite{Zhi2022WannSymm:Orbitals} using the crystal symmetries in space group 164 ($P$$\bar{3}m1$) and $\mathcal{T}$ symmetry. To study the $(001)$-surface states, we constructed a finite slab model using the \texttt{PythTB} software package~\cite{PythTB} with the Wannier model as input.
Within the slab model, we then computed the surface spectral function and spin and orbital angular momentum textures using surface Green's functions (see SM Sec.~\ref{theory_SI}~\cite{2026SeeURL} for complete DFT calculation details).

\subsection*{Electron Energy Loss Spectroscopy Measurements}

To increase the scattering intensity for momentum-resolved electron energy loss spectroscopy (M-EELS) measurements, five single crystals of \ce{FM-EuCd2As2} from the same batch used for the Tr-ARPES measurements were roughly vertically co-aligned on an oxygen-free, high-conductivity copper puck. 
The samples were secured with Ag-epoxy (EPOTEK H2OE) cured at 150°C for an hour. Custom posts were mounted on top of the samples using the same epoxy. After preparation, the samples were cleaved in a UHV chamber ($\sim$\qty{5E-10}{Torr}) at room temperature and transferred into the M-EELS measurement chamber held at $\sim$\qty{7E-11}{Torr}.

The M-EELS experiments were performed in a reflection geometry using a modified high-resolution EELS spectrometer (ELS5000, LK Technologies) with the sample placed in an eucentric cryogenic three-circle goniometer~\cite{Vig2017MeasurementM-EELS}. Samples were placed such that the out-of-plane momentum transfer was along the c-axis. An incident energy of \qty{50}{eV} with an energy resolution of \qty{6.6}{meV} was used. Due to the rough alignment of the individual crystals, the momentum in the scattering plane was considered integrated (\emph{i.e.}, a single value of $q$ contains information summed over the full Brillouin zone).

\subsection*{Sample Synthesis}

The \ce{EuCd2As2} samples were grown using a salt-flux method, with high-quality elemental \ce{Eu}, \ce{Cd}, and \ce{As} pieces, and an equimolar ratio of \ce{NaCl} and \ce{KCl}. The growth method -- as well as sample characterization through transport, scanning tunneling microscopy, and soft X-ray ARPES -- are described in Ref.~\cite{Roychowdhury2023AnomalousEuCd2As2}. We note that the salt flux growth method used for the samples in this work (as opposed to the \ce{Sn} flux method more commonly used for the growth of \ce{EuCd2As2}) leads to an approximately \qty{1}\% concentration of \ce{Eu} vacancies, which results in FM order appearing below $T_C\sim\qty{25}{K}$ (as opposed to antiferromagnetic ordering below $T_N\sim\qty{9}{K}$)~\cite{Jo2020ManipulatingEuCd2As2}.

%% file: arXiv_v3/SI_body.tex
\title{Supplementary Material for ``Plasmon-driven electron-hole pair formation in a non-equilibrium Fermi liquid''}

\author{Rishi Acharya} 
\affiliation{Department of Physics, The Grainger College of Engineering, University of Illinois at Urbana-Champaign, Urbana, 61801 IL, USA}
\affiliation{Materials Research Laboratory, The Grainger College of Engineering, University of Illinois at Urbana-Champaign, Urbana, 61801 IL, USA}

\author{Eli Gerber}
\affiliation{Institut de Physique Th\'eorique, Universit\'e Paris-Saclay, CEA, CNRS, F-91191 Gif-sur-Yvette, France}

\author{Nina Bielinski}
\affiliation{Department of Physics, The Grainger College of Engineering, University of Illinois at Urbana-Champaign, Urbana, 61801 IL, USA}
\affiliation{Materials Research Laboratory, The Grainger College of Engineering, University of Illinois at Urbana-Champaign, Urbana, 61801 IL, USA}

\author{Hannah E. Aguirre}
\affiliation{Department of Physics, The Grainger College of Engineering, University of Illinois at Urbana-Champaign, Urbana, 61801 IL, USA}
\affiliation{Materials Research Laboratory, The Grainger College of Engineering, University of Illinois at Urbana-Champaign, Urbana, 61801 IL, USA}

\author{Younsik Kim}
\affiliation{Department of Physics, The Grainger College of Engineering, University of Illinois at Urbana-Champaign, Urbana, 61801 IL, USA}
\affiliation{Materials Research Laboratory, The Grainger College of Engineering, University of Illinois at Urbana-Champaign, Urbana, 61801 IL, USA}

\author{Camille Bernal-Choban}
\affiliation{Department of Physics, The Grainger College of Engineering, University of Illinois at Urbana-Champaign, Urbana, 61801 IL, USA}
\affiliation{Materials Research Laboratory, The Grainger College of Engineering, University of Illinois at Urbana-Champaign, Urbana, 61801 IL, USA}

\author{Gaurav Tenkila}
\affiliation{Department of Physics, The Grainger College of Engineering, University of Illinois at Urbana-Champaign, Urbana, 61801 IL, USA}
\affiliation{Institute of Condensed Matter Theory,
University of Illinois at Urbana-Champaign, Urbana, 61801 IL, USA}

\author{Suhas Sheikh}
\affiliation{Department of Physics, The Grainger College of Engineering, University of Illinois at Urbana-Champaign, Urbana, 61801 IL, USA}
\affiliation{Institute of Condensed Matter Theory,
University of Illinois at Urbana-Champaign, Urbana, 61801 IL, USA}

\author{Pranav Mahaadev}
\affiliation{Department of Physics, The Grainger College of Engineering, University of Illinois at Urbana-Champaign, Urbana, 61801 IL, USA}
\affiliation{Materials Research Laboratory, The Grainger College of Engineering, University of Illinois at Urbana-Champaign, Urbana, 61801 IL, USA}

\author{Faren Hoveyda-Marashi}
\affiliation{Department of Physics, The Grainger College of Engineering, University of Illinois at Urbana-Champaign, Urbana, 61801 IL, USA}
\affiliation{Materials Research Laboratory, The Grainger College of Engineering, University of Illinois at Urbana-Champaign, Urbana, 61801 IL, USA}

\author{Subhajit Roychowdhury}
\affiliation{Max Planck Institute for Chemical Physics of Solids, Dresden 01187, Germany}
\affiliation{Department of Chemistry, Indian Institute of Science Education and Research Bhopal, Bhopal 462066, India}

\author{Chandra Shekhar}
\affiliation{Max Planck Institute for Chemical Physics of Solids, Dresden 01187, Germany}

\author{Claudia Felser}
\affiliation{Max Planck Institute for Chemical Physics of Solids, Dresden 01187, Germany}

\author{Peter Abbamonte}
\affiliation{Department of Physics, The Grainger College of Engineering, University of Illinois at Urbana-Champaign, Urbana, 61801 IL, USA}
\affiliation{Materials Research Laboratory, The Grainger College of Engineering, University of Illinois at Urbana-Champaign, Urbana, 61801 IL, USA}

\author{Benjamin J. Wieder}
\affiliation{Institut de Physique Th\'eorique, Universit\'e Paris-Saclay, CEA, CNRS, F-91191 Gif-sur-Yvette, France}

\author{Fahad Mahmood} 
\email{fahad@illinois.edu}
\affiliation{Department of Physics, The Grainger College of Engineering, University of Illinois at Urbana-Champaign, Urbana, 61801 IL, USA}
\affiliation{Materials Research Laboratory, The Grainger College of Engineering, University of Illinois at Urbana-Champaign, Urbana, 61801 IL, USA}

\maketitle
\onecolumngrid
\newpage
\tableofcontents
\newpage

\section{Evidence for a Weyl semimetal phase in ferromagnetic \ce{EuCd2As2}}
\label{Semimetal_SI}

\begin{figure}[H]
\includegraphics[width=0.96\linewidth]{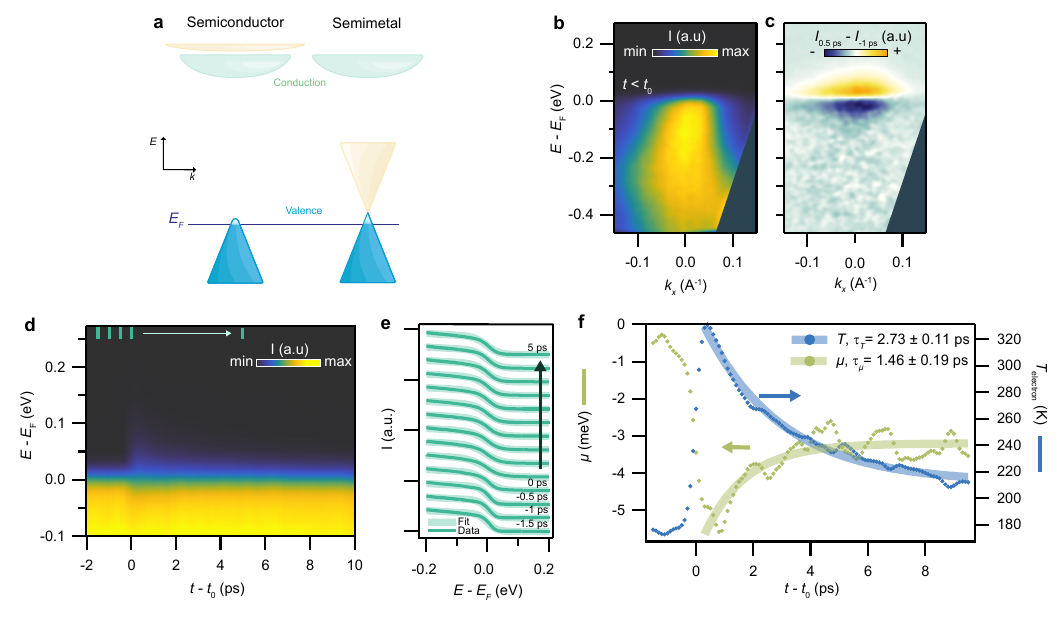}
\caption{\textbf{Evidence for a Weyl semimetal phase in ferromagnetic \ce{EuCd2As2}:} \textbf{(a)} Schematics of the competing descriptions of the bulk electronic band structure of \ce{EuCd2As2} near the Fermi energy ($E_F$) in the ferromagnetic (FM) state.
\textbf{(b)} $E-E_F$ vs $k_x$ spectra obtained from our ARPES measurements before photo-excitation, at a lattice temperature approximately at the FM transition $T_C \approx \qty{25}{K}$.
\textbf{(c)} Difference plot showing the equilibrium occupation of (b) subtracted from the time-integrated occupation between \qty{0} and \qty{1}{ps} after photo-excitation by \qty{1.2}{eV} photons at low fluence (\qty{15}{\micro J/cm^2}). 
\textbf{(d)} Time trace corresponding to the spectra shown in (b) and (c). 
\textbf{(e)} Waterfall plot showing Fermi-Dirac fits (each convolved with a \qty{10}{meV} Gaussian resolution function) for the electron distribution curves (EDCs) from the time trace shown in (d), taken at the times indicated by the vertical green bars at the top of (d).
For the Fermi-Dirac fits in (e), we take the density of states to vary quadratically around $E-E_F=\qty{50}{meV}$, and include a constant background offset to incorporate the effects of impurities and pump-induced scatter.
\textbf{(f)} Extracted temperature (blue circles, right axis) and chemical potential (green circles, left axis) for the electronic distribution as a function of time, from the fits shown in (e), as well as additional intermediate times. 
The increase (decrease) in chemical potential for decreasing (increasing) temperature indicates the presence of an electron-like conduction band near $E_F$, supporting the semimetal scenario in (a).
Shaded lines represent exponential decay fits. 
We note that due to charging effects, rigid energy offsets (relative to the Fermi edge found using a gold reference) were added in (d-e) to place the Fermi edge at $E_F$; for each time trace in this figure, the same offset was applied to all $E-E_F$ vs $k_x$ spectra.}
\label{fig:si_semimetal}
\end{figure}

There has been much controversy \cite{Ma2019SpinEuCd2As2, Jo2020ManipulatingEuCd2As2, Soh2019IdealExchange, Roychowdhury2023AnomalousEuCd2As2, Santos-Cottin2023EuCd2As2:Semiconductor, Shi2024AbsenceEffect, Cuono2023AbCompounds, Nelson2024RevealingLa-Doping} surrounding the subject of whether the ferromagnetic (FM) state of \ce{EuCd2As2} is a topological Weyl semimetal or a gapped semiconductor. Here, we present experimental evidence that the samples used in this work are semimetallic in the FM phase.

Initial work on \ce{EuCd2As2} described it as a Dirac semimetal above \qty{100}{K} (in the paramagnetic state), and argued that at intermediate temperatures, between \qty{100}{K} and the antiferromagnetic transition ($T_N=\qty{9}{K}$), ferromagnetic fluctuations lead to magnetic splitting of the Dirac point into a pair of Weyl points \cite{Ma2019SpinEuCd2As2}. 
In this scenario, the electronic band structure of the Weyl semimetal phase includes two pairs of band crossings with linear dispersion whose nodal degeneracies lie \qty{50}{meV} above the Fermi energy ($E_F$), as schematically depicted with blue and yellow bands in the right panel of Fig.~\ref{fig:si_semimetal}a.
We note that for clarity, only a single Weyl cone is schematically depicted in the right panel of Fig.~\ref{fig:si_semimetal}a; the second Weyl cone is predicted to lie at the same energy but offset along the $k_z$ direction.
The hole-like band (blue), as well as the splitting thereof under ferromagnetic fluctuations, was previously measured through synchrotron ARPES studies \cite{Ma2019SpinEuCd2As2}. 
However, because it lies above $E_F$, the electron-like band (yellow in Fig.~\ref{fig:si_semimetal}a, right panel) could not be clearly resolved through the same measurements. 
Nevertheless, the dispersion of an electron-like band was faintly resolved using quasiparticle interference in scanning tunneling microscopy measurements \cite{Roychowdhury2023AnomalousEuCd2As2}. 
Further experiments stabilized FM order below a transition temperature of $T_C=\qty{25}{K}$ through the introduction of \ce{Eu} vacancies \cite{Roychowdhury2023AnomalousEuCd2As2}; we use the same samples as in Ref. \cite{Roychowdhury2023AnomalousEuCd2As2} (with Eu vacancies) in this work.
Hence for the samples studied in this work, we expect the presence of bulk Weyl and (potentially massive) Dirac points respectively below and above $T_C=\qty{25}{K}$.

However, several recent studies instead characterize \ce{EuCd2As2} as a semiconductor \cite{Santos-Cottin2023EuCd2As2:Semiconductor, Cuono2023AbCompounds, Shi2024AbsenceEffect, Nelson2024RevealingLa-Doping}, questioning the robustness of the semimetallic characterization of \ce{EuCd2As2} in the earlier works. The semiconducting electronic band structure proposed in these studies is schematically depicted in the left panel of Fig.~\ref{fig:si_semimetal}a, and is based on the results of Tr-ARPES \cite{Santos-Cottin2023EuCd2As2:Semiconductor} and \ce{La}-doping-supported ARPES \cite{Nelson2024RevealingLa-Doping} experiments on samples grown with high-purity \ce{Eu}.
These measurements specifically identified a hole-like valence band with a maximum slightly above $E_F$ (in agreement with ARPES measurements that supported the Weyl semimetal characterization), as well as a conduction band minimum \qty{0.7}-\qty{0.8}{eV} above $E_F$, but did not show evidence of an electron-like band immediately above $E_F$ (the yellow band in the right panel of Fig.~\ref{fig:si_semimetal}a). The semiconducting gap was further observed to be temperature-independent; neither the Weyl nor the Dirac semimetal phases were observed \cite{Santos-Cottin2023EuCd2As2:Semiconductor}.
Nevertheless, we note that the existence of a large-mass conduction band \qty{0.7}{eV}-\qty{0.8}{eV} above $E_F$ (the green band in both panels of Fig.~\ref{fig:si_semimetal}a) has been confirmed by previous experimental observation of a strong peak at the corresponding infrared absorption spectrum, both in samples characterized as semiconducting \cite{Santos-Cottin2023EuCd2As2:Semiconductor} and in those characterized as semimetallic \cite{Wu2024TheSingularity}.

Our static \qty{6}{eV} laser ARPES measurements of \ce{EuCd2As2}, shown in Fig.~\ref{fig:si_semimetal}b, corroborate equilibrium ARPES studies taken using X-ray probe photons, which unanimously report a hole-like valence band with a maximum near $E_F$ (for both semimetallic \cite{Roychowdhury2023AnomalousEuCd2As2} and semiconducting \cite{Santos-Cottin2023EuCd2As2:Semiconductor} samples). 
However, our time-resolved (Tr-) ARPES experiments suggest that the samples analyzed in this study are semimetallic in the FM state via the observation of semimetallic thermodynamics.

Specifically, in Fig.~\ref{fig:si_semimetal}c, we show the change in the spectral weight near $E_F$ at the instant of photo-excitation with \qty{1.2}{eV} photons. 
We observe a transient increase in electronic occupation above $E_F$ (orange data points in Fig.~\ref{fig:si_semimetal}c), accompanied by a decrease below $E_F$ (dark blue data points in Fig.~\ref{fig:si_semimetal}c). Such broadening of the Fermi edge in response to photo-excitation is ubiquitous in Tr-ARPES experiments, and is generally interpreted as rapid heating of electrons by the pump pulse, followed by the slower cooling of electrons via coupling to phonons \cite{Gierz2013SnapshotsGraphene}. 
The time trace in Fig.~\ref{fig:si_semimetal}d shows electron distribution curves (EDCs) taken at various pump-probe delays for \ce{EuCd2As2} under \qty{1.2}{eV} photo-excitation at $T\approx T_C$. 
At $t_0$, as described earlier, there is a transient increase in population above $E_F$, along with a faint decrease in occupation below $E_F$; these two disturbances cool over the following few picoseconds. 
In Fig.~\ref{fig:si_semimetal}e, we demonstrate that the EDCs in Fig.~\ref{fig:si_semimetal}d of hot electrons at various time delays are well-fit by Fermi-Dirac distributions multiplied by a parabolic density of states centered \qty{50}{meV} above $E_F$. 
The transient electronic temperature $T_{e}$ and chemical potential $\mu$ as functions of time extracted from the fits in Fig.~\ref{fig:si_semimetal}e are shown in Fig.~\ref{fig:si_semimetal}f. 
We find that the increase in electronic temperature in response to the pump pulse at $t=t_0$ is accompanied by a decrease in the extracted chemical potential. 
Moreover, the recovery of the electronic temperature to a quasi-equilibrium value is exponential, with a timescale (\qty{2.73}{ps}) approximately double that of the exponential recovery of the chemical potential (\qty{1.46}{ps}).

To interpret this result, we recall that the relation between the chemical potential and temperature for a (semi)metal or doped semiconductor can be derived through a Sommerfeld expansion: $\Delta\mu\propto-T^2 g'(E_F)/g(E_F)$, where $g(E)$ is the density of states in energy \cite{Ashcroft1976SolidPhysics}. Crucially, this indicates that for an inherently p-doped semiconductor -- in which the density of states decreases with increasing energy -- an increase (decrease) in chemical potential with increasing (decreasing) temperature is expected. However, for a charge-neutral semimetal (\emph{i.e.}~a semimetal with $E_F$ at the band crossing), the density of states increases with increasing energy, leading to an increase (decrease) in chemical potential with decreasing (increasing) temperature. 
We therefore assert that when the temperature of the electrons in \ce{EuCd2As2} is increased by the pump pulse, much of the increase in spectral weight above $E_F$ is realized above the Weyl point in an electron-like band, which leads to the transient decrease in $\mu$ and its subsequent recovery at twice the decay rate of $T$.
For comparison, previous Tr-ARPES experiments on strongly p-type graphene ($E_0\approx E_F-0.2\,\mathrm{eV}$) show simultaneous increases in both $\mu$ and $T_e$ in response to photo-excitation \cite{Gierz2013SnapshotsGraphene}. 
We also note that similar observations have been made in the 3D Dirac semimetal \ce{Cd3As2} \cite{Bao2022PopulationCd3As2}, and in the 2D Dirac surface states of the topological insulator \ce{Bi2Se3} \cite{Ponzoni2023DiracMicroscopy}.

Finally, we note that the magnetism-induced splitting below $T_C$ does not affect the aforementioned arguments. 
Due to the large expected size of the splitting (on the order of \qty{0.1}{\angstrom^{-1}}), we expect that Fig.~\ref{fig:si_semimetal}b shows the dispersion of a single Weyl fermion. 
The theoretical description of the dynamics shown in Fig.~\ref{fig:si_semimetal}f similarly remains unchanged when we account for the presence of a second Weyl point. 
Moreover, across all of the experimental analysis in this study, the emergence of magnetic ordering across $T_C$ does not significantly affect our observations, aside from shifting the plasmon energy (see the main text for further details).

Overall, though our data is consistent with the presence of a Weyl semimetal phase in the FM state, the phase diagram of \ce{EuCd2As2} -- and the relative roles of magnetic impurities and \ce{Eu} vacancies -- are likely complex. 
\ce{EuCd2As2} may feature both semiconducting and semimetallic regimes, and its magnetic and (topological) electronic phase diagrams consequently merit further study.

\newpage
\section{Temperature-independence of the surface-state dispersion and circular dichroism ARPES measurements}
\label{CD_temp_SI}

\begin{figure}[H]
\centering
\includegraphics{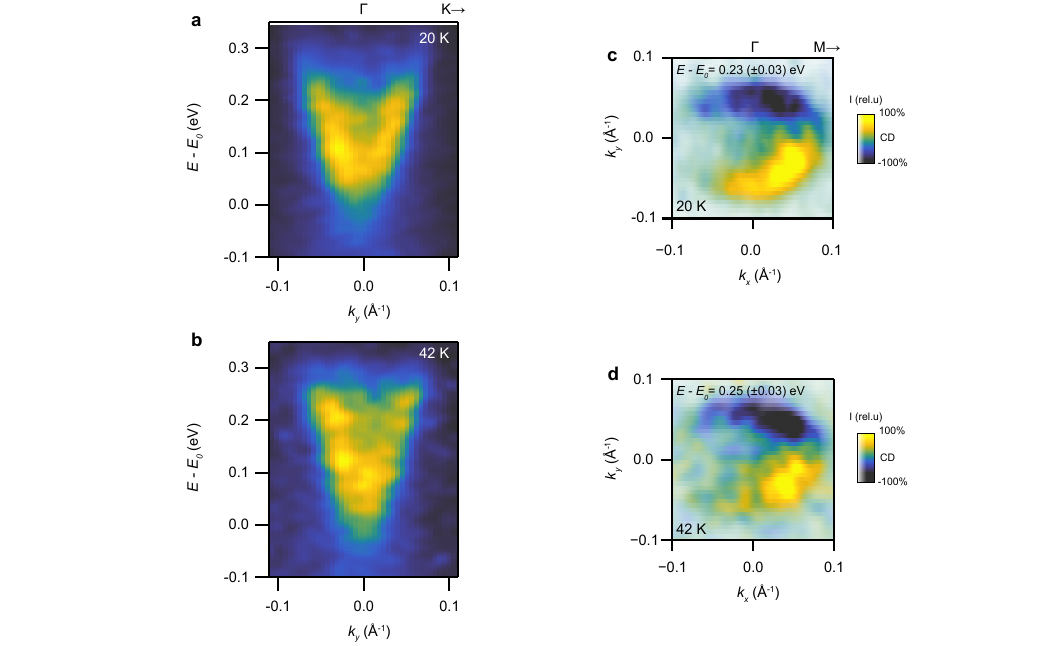}
\caption{\textbf{Temperature-invariance of the surface-state dispersion and circular dichroism ARPES spectra:} 
\textbf{(a, b)} ARPES spectra at $t_0$ in the ferromagnetic (FM) state ($T=\qty{20}{K}$) and paramagnetic (PM) state ($T=\qty{42}{K}$) of \ce{EuCd2As2}, respectively, where the PM to FM transition occurs at $T_{C}\approx\qty{25}{K}$ \cite{Roychowdhury2023AnomalousEuCd2As2}. 
In (a, b), it is important to note that the conduction band below $E_0$ is not visible (where $E_0\approx E_F+\qty{0.8}{eV}$ is the bottom of the unoccupied surface state, see Sec.~\ref{sislabgreens}), because the ARPES data were taken at the instant of photo-excitation, before electrons have scattered to energies significantly below those initially excited by the pump.
\textbf{(c, d)} Constant-energy cuts of the $t_0$ circular dichroism ARPES (CD-ARPES) measurements at \qty{20}{K} and \qty{42}{K}, respectively, integrated over $\pm$\qty{30}{meV} bins around the stated center energies.
Both the ARPES data in (a, b) and the CD-ARPES data in (c, d) are nearly identical above and below $T_{C}$, providing support for our focus on the PM state in our first-principles calculations (Sec.~\ref{theory_SI}).}
\label{fig:si_temperature}
\end{figure}

To theoretically characterize the bulk and surface spectrum of \ce{EuCd2As2}, we performed first-principles (DFT) calculations in the paramagnetic (PM) state in Sec.~\ref{theory_SI}.
To justify our focus on the PM state in our DFT calculations, we show in Fig.~\ref{fig:si_temperature} ARPES (panels (a, b)) and circular dichroism ARPES (CD-ARPES, panels (c, d)) measurements of \ce{EuCd2As2} both below and above the ferromagnetic (FM) transition temperature of $T_C\approx\qty{25}{K}$ (for further characterization and details of the samples used in this study, see Ref. \cite{Roychowdhury2023AnomalousEuCd2As2}).
The data taken above and below $T_{C}$ in Fig.~\ref{fig:si_temperature} are qualitatively identical, supporting the applicability of our PM-state DFT calculations in Sec.~\ref{theory_SI}.

We additionally note that the PM state of \ce{EuCd2As2} has both inversion ($\mathcal{I}$) and time-reversal ($\mathcal{T}$) symmetry in the 3D bulk. 
This provides further evidence that the CD-ARPES signal in Fig.~\ref{fig:si_temperature} corresponds to an (unoccupied) 2D surface state, because the CD-ARPES signal must vanish at all ${\bf k}$ in a system with $\mathcal{I}$ and $\mathcal{T}$ symmetries \cite{Sidilkover2025ReexaminingInsulator}.

\newpage

\section{First-Principles Calculations}
\label{theory_SI}

In this section we will perform and analyze \emph{ab initio} density-functional theory (DFT) calculations and construct Wannier tight-binding models to study the bulk electronic structure, surface states, and band topology of \ce{EuCd2As2}.
Because the ARPES measurements in this work produced qualitatively similar data both above and below the paramagnetic (PM) to FM transition temperature of $T_c \approx \qty{25}{K}$ (see Fig.~\ref{fig:si_DFT_ARPES_dispersions} and Sec.~\ref{CD_temp_SI}), we will focus our analysis below on the simpler case of \ce{EuCd2As2} in the PM state.
We will first in Sec.~\ref{siDFT} detail our DFT calculations and the construction of our Wannier model.
In Sec.~\ref{sislabgreens}, we will then use surface Green's functions in a finite slab Hamiltonian to obtain the $(001)$- ($\hat{z}$-normal) surface spectrum of \ce{EuCd2As2} in the PM state, and quantify the contribution of the Eu $4f$ states to the surface spectrum, as they exert a strong influence on the surface chemical potential via the Hubbard $U$ parameter in our DFT calculations (Sec.~\ref{siDFT}).
Our analysis in Sec.~\ref{sislabgreens} reveals the presence of unoccupied $(001)$-surface bands that coexist with the projected continuum of unoccupied bulk states (the projected bulk conduction manifold), consistent with our Tr-ARPES experiments (see Fig.~\ref{fig:overview}b of the main text).

To further our understanding of the $(001)$-surface states, we will next in Sec.~\ref{sispintext} compute the surface spin and orbital angular momentum (OAM) textures, which reveal the surface states to be a 2D Dirac cone with weak spin-orbit-coupling- (SOC-) driven splitting away from ${\bf k}={\bf 0}$.
We will also in Sec.~\ref{sispintext} introduce a simplified low-energy ${\bf k}\cdot{\bf p}$ Hamiltonian that qualitatively reproduces the dispersion and angular momentum textures of the $(001)$-surface states obtained from our DFT-based calculations.
Lastly, in Sec.~\ref{sirepeattopo}, we will apply the methods of Ref.~\cite{Vergniory2022AllMaterials} to our PM-state DFT calculation to analyze the band topology of \ce{EuCd2As2} at and away from $E_{F}$.
Through this analysis, we identify a large range of contiguous topological gaps in the bulk conduction manifold, from which we conclude that the unoccupied $(001)$-surface bands in \ce{EuCd2As2} can be viewed as the 2D topological surface states of a 3D topological band insulator phase at an electronic filling above $E_{F}$.

\subsection{First-principles calculations and Wannier tight-binding model of the paramagnetic state of \ce{EuCd2As2}}
\label{siDFT}

\begin{table}[H]
\centering
\begin{tabular}{c c c c c c}
\toprule
Site & Atom & Wyckoff Position &
Atomic Position (Direct) &
Wannier Center (Direct) &
Wannier Orbitals (Spinful) \\
\midrule

Eu &
Eu & 1a &
$(0.0000,\,0.0000,\,0.0000)$ &
$(0.0000,\,0.0000,\,0.0000)$ &
$5\times 5d \;+\; 7\times 4f$ (total $24$)
\\[4pt]

Cd$_1$ &
Cd & 2d &
$(0.6667,\,0.3333,\,0.3668)$ &
$(0.6667,\,0.3333,\,0.3668)$ &
$1\times 5s$ (total $2$)
\\[4pt]

Cd$_2$ &
Cd & 2d &
$(0.3333,\,0.6667,\,0.6332)$ &
$(0.3333,\,0.6667,\,0.6332)$ &
$1\times 5s$ (total $2$)
\\[4pt]

As$_1$ &
As & 2d &
$(0.6667,\,0.3333,\,0.7533)$ &
$(0.6667,\,0.3333,\,0.7533)$ &
$3\times 4p$ (total $6$)
\\[4pt]

As$_2$ &
As & 2d &
$(0.3333,\,0.6667,\,0.2467)$ &
$(0.3333,\,0.6667,\,0.2467)$ &
$3\times 4p$ (total $6$)
\\
\bottomrule
\end{tabular}
\caption{{\bf Atomic positions and Wannier centers of EuCd$_2$As$_2$ in the paramagnetic state.} 
In this table, we list for each atom in EuCd$_2$As$_2$ the site name, the atomic species, the site Wyckoff position, the position of the atom in direct coordinates (\emph{i.e.} in the units of $\mathbf{a}_{1,2,3}$ in Eq.~(\ref{primlattvecs})), the position in direct coordinates of the Wannier orbital associated to the atom in our Wannier model, and the number and indexing of the Wannier orbitals.
The atomic positions were obtained from Inorganic Crystal Structure Database (ICSD) entry number 422963~\cite{Schellenberg2011AYbCd2Sb2, Bergerhoff1983TheBase}, which we used in our DFT calculations without structural relaxation.
In the Wannier model, the up- and down-spin components of each orbital lie at exactly the same positions, and the orbital centers coincide with the atomic positions to within $10^{-4}$ in direct coordinates. 
This allows us to use the Wannier model to compute the orbital angular momentum texture in Sec.~\ref{sispintext}, as further detailed below in the text surrounding Eq.~(\ref{eq:LVector}).} 
\label{tab:wann_orbitals_direct}
\end{table}

To determine the bulk and surface spectrum of \ce{EuCd2As2} for comparison with our experimental data, we begin by performing \emph{ab initio} calculations within the DFT framework as implemented in the \texttt{Quantum Espresso} software package \cite{Giannozzi2009QUANTUMMaterials, Giannozzi2017AdvancedESPRESSO, Giannozzi2020QuantumExascale}. 
We use as input to our calculations the atomic positions and lattice parameters in the \texttt{CIF} structure file for \ce{EuCd2As2} from Inorganic Crystal Structure Database (ICSD) entry number 422963~\cite{Schellenberg2011AYbCd2Sb2, Bergerhoff1983TheBase}, and do not perform further structural relaxation.
Because the experiments in this study produced qualitatively similar data both above and below the FM transition temperature of $T_c \approx \qty{25}{K}$ (see Fig.~\ref{fig:si_DFT_ARPES_dispersions} and Sec.~\ref{CD_temp_SI}), and because the symmetry and topology of nonmagnetic 3D systems are simpler than those in magnetic phases~\cite{Elcoro2021MagneticChemistry, Xu2020High-throughputMaterials}, then we focus our first-principles calculations on the PM state.

In the PM phase, \ce{EuCd2As2} respects the symmetries of nonmagnetic space group (SG) 164 ($P$$\bar{3}m1$).
Each primitive (unit) cell of \ce{EuCd2As2} contains one Eu atom, two Cd atoms, and two As atoms, whose coordinates are provided in Table~\ref{tab:wann_orbitals_direct}. 
The primitive lattice vectors of \ce{EuCd2As2} are given by:
\begin{align}
\mathbf{a}_{1} &= (4.4499000 \,\textnormal{\AA})\ \hat{\mathbf{x}}, \nonumber \\
\mathbf{a}_{2} &= (-2.2249500\,\textnormal{\AA})\ \hat{\mathbf{x}} + (3.8537264\,\textnormal{\AA})\ \hat{\mathbf{y}}, \nonumber \\
\mathbf{a}_{3} &= (7.3499999\,\textnormal{\AA})\ \hat{\mathbf{z}}.\label{primlattvecs}
\end{align}
The primitive reciprocal lattice vectors of \ce{EuCd2As2} are then correspondingly given by:
\begin{align}
\mathbf{b}_{1} &= \frac{2\pi (\mathbf{a}_{2} \times \mathbf{a}_{3})}{\mathbf{a}_{1}\cdot (\mathbf{a}_{2}\times \mathbf{a}_{3})} = (1.41198348\,\textnormal{\AA}^{-1})\ \hat{\mathbf{x}} + (0.81520904\,\textnormal{\AA}^{-1})\ \hat{\mathbf{y}}, \nonumber \\
\mathbf{b}_{2} &= \frac{2\pi(\mathbf{a}_{3} \times \mathbf{a}_{1})}{\mathbf{a}_{1}\cdot (\mathbf{a}_{2}\times \mathbf{a}_{3})} = (1.63041809\,\textnormal{\AA}^{-1})\ \hat{\mathbf{y}}, \nonumber \\
\mathbf{b}_{3} &= \frac{2\pi(\mathbf{a}_{1} \times \mathbf{a}_{2})}{\mathbf{a}_{1}\cdot (\mathbf{a}_{2}\times \mathbf{a}_{3})} = (0.85485514\,\textnormal{\AA}^{-1})\ \hat{\mathbf{z}}.
\label{reciplattvecs}
\end{align}

In our DFT calculations, we employed the Perdew-Burke-Ernzerhof generalized gradient approximation for the exchange-correlation term \cite{Blochl1994ProjectorMethod, Kresse1999FromMethod, Perdew1996GeneralizedSimple} together with projected augmented-wave (PAW) pseudopotentials \cite{DalCorso2014PseudopotentialsPu}, including fully relativistic SOC in the noncollinear scheme.
In this setup, when no initial magnetic moments are specified, the self-consistent cycle converges to a nonmagnetic, time-reversal- ($\mathcal{T}$-) invariant solution (\emph{i.e.} a state in which the spin density vanishes at all positions in real space).
Core electrons were treated by norm-conserving, optimized nonlocal pseudopotentials. 
The self-consistent calculations were performed over a mesh of $10 \times 10 \times 5$ ${\bf k}$ points. 
The kinetic energy cutoff was set to 680 eV. 
To properly localize the half-filled Eu $4f$ orbitals in the PM state, correct self-interaction errors, and to improve the overall accuracy of the PM-state band structure, a Hubbard $U$ correction of $U=5$ eV on the Eu sites was implemented within the GGA $+$ $U$ scheme~\cite{Dudarev1998Electron-energy-lossStudy}.

\begin{figure}[t]
\centering
\includegraphics[width=0.85\textwidth]{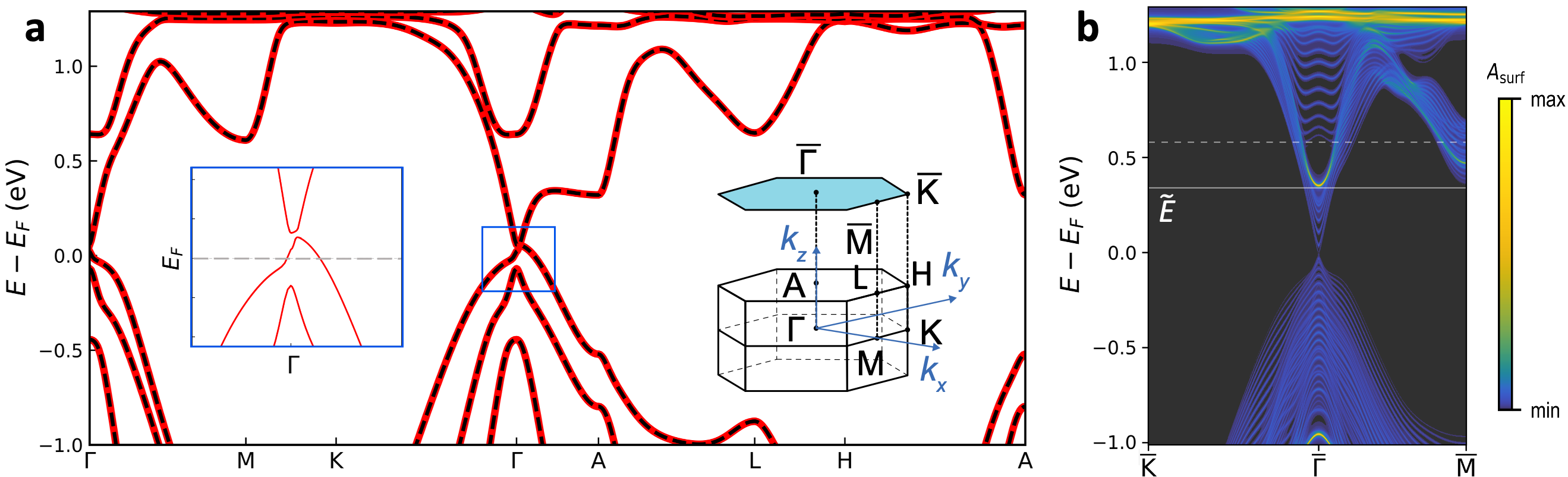}
\caption{\textbf{Calculated bulk band structure and surface spectral function of \ce{EuCd2As2}:} \textbf{(a)} Bulk band structure of the paramagnetic (PM) state of \ce{EuCd2As2} obtained from first-principles calculations (DFT, solid red lines) and the bands of a Wannier tight-binding model fit to the DFT calculation (dashed black lines, see Table~\ref{tab:wann_orbitals_direct} and the surrounding text).
At intrinsic filling ($\nu_{F}$ in Eq.~(\ref{eq:IntrinsicFilling})), the bulk is metallic, as evidenced by the Fermi pocket at $E_{F}$ in the left inset panel in (a).
In (a), we also depict the bulk and $(001)$- ($\hat{z}$-normal) surface-projected Brillouin zones (BZs) for the symmetry group (SG) of the PM phase of \ce{EuCd2As2}, nonmagnetic SG 164 ($P\bar{3}m1$).
The bands in (a) are doubly degenerate owing to the presence of bulk 3D spatial inversion ($\mathcal{I}$) and time-reversal ($\mathcal{T}$) symmetries~\cite{Vafek2014DiracSemimetals, Wieder2016Spin-orbitGroups}.
\textbf{(b)} The $(001)$-surface spectral function $A_{\mathrm{surf}}(\mathbf{k},E)$ (Eq.~(\ref{eq:Asurf_def})) of the Wannier tight binding model.
In (b), we observe a single bright surface spectral feature above $E_{F}$ that lies within the projected continuum of unoccupied bulk states and exhibits an energy minimum at $\widetilde{E} = E_F + \qty{0.34}{eV}$ (solid gray line).
Though the surface states in (b) are quadratic at low energies, we observe that the surface bands are generally well fit by a hyperbolic dispersion with a small mass (\emph{i.e.} a nearly linear dispersion relation).
The dashed line in (b) indicates the energy at which we will shortly compute constant-energy spectral functions and spin and orbital angular momentum textures (Fig.~\ref{fig:sxzconstenergy}).}
\label{fig:si_dft}
\end{figure}

The resulting band structure is plotted in Fig.~\ref{fig:si_dft}a with solid red lines.
Each of the bands in Fig.~\ref{fig:si_dft}a is doubly degenerate due to the presence of bulk 3D spatial inversion ($\mathcal{I}$) and $\mathcal{T}$ symmetries~\cite{Vafek2014DiracSemimetals, Wieder2016Spin-orbitGroups}.
At intrinsic filling, \ce{EuCd2As2} carries an odd number of valence electrons $\nu_{F}$ per unit cell:
\begin{equation}
\nu_{F} = 51.
\label{eq:IntrinsicFilling}
\end{equation}
Because $\nu_{F}\text{ mod }2=1$, then due to $\mathcal{T}$ symmetry in the PM state, \ce{EuCd2As2} is a filling-enforced metal at $E_{F}$ (indicated by the red dashed line in the left inset in Fig.~\ref{fig:si_dft}a)~\cite{Watanabe2015FillingCrystals,Wieder2016Spin-orbitGroups}.
Just above $E_{F}$, at a filling of $\nu_{F}+1$, there is instead an energy gap, and the system can be viewed overall as a weakly massive 3D Dirac semimetal.
We will shortly below in Sec. \ref{sirepeattopo} show that the gap at $\nu_{F}+1$ is topological, and more generally analyze the system topology at and away from $E_{F}$.

Our choice of $U=5$ eV and the accuracy of the DFT band structure in Fig.~\ref{fig:si_dft}a are supported by previous theoretical and experimental studies of \ce{EuCd2As2}.
First, previous DFT calculations performed with $U=5$ eV in the FM state were found to accurately reproduce the ARPES spectrum of the occupied bands, importantly including the energies of the Eu $4f$ orbitals, over a wide energy range~\cite{Roychowdhury2023AnomalousEuCd2As2, Soh2019IdealExchange}. 
DFT + $U$ calculations with $U=5$~eV in the PM state were also found to accurately capture the ARPES spectrum in the PM phase~\cite{Ma2019SpinEuCd2As2}, and closely match the band structure in Fig.~\ref{fig:si_dft}a over the energy range $E_{F} \pm 1$ eV.
Lastly, recent ARPES measurements of the \emph{unoccupied} band structure have determined the first large energy gap above $E_{F}$ near $\Gamma$ to be approximately $\sim 0.7$ eV in the PM state~\cite{Nelson2024RevealingLa-Doping,Santos-Cottin2023EuCd2As2:Semiconductor}, consistent with the DFT band structure in Fig.~\ref{fig:si_dft}a (specifically the gap between the doubly degenerate conduction bands in the left inset and the next lowest doubly degenerate conduction bands at $\Gamma$, which are nondispersing to leading order).

To analyze the spectrum and topology of the PM state, we next employ Wannier functions \cite{Pizzi2020Wannier90Applications, Marzari1997MaximallyBands, Souza2001MaximallyBands} to construct a tight-binding model of \ce{EuCd2As2}, constrained to closely reproduce the energy spectrum obtained from DFT within a few eV of $E_{F}$ (black dashed lines in Fig.~\ref{fig:si_dft}a). 
We specifically construct our Wannier model using the Eu 5\emph{d} and $4\emph{f}$, Cd 5\emph{s}, and As 4\emph{p} orbitals with SOC ($N_{a}=20$ spin-degenerate pairs of orbitals). 
We then symmetrize the Wannier model \cite{Zhi2022WannSymm:Orbitals} using the crystal symmetries in SG 164 ($P$$\bar{3}m1$) and $\mathcal{T}$.
The locations of the resulting Wannier centers are provided in Table~\ref{tab:wann_orbitals_direct}, and coincide with the atomic positions to within $10^{-4}$ in the units of $\mathbf{a}_{1,2,3}$ in Eq.~(\ref{primlattvecs}).

\subsection{Surface spectral function calculation details}
\label{sislabgreens}

We will next use the Wannier-based tight-binding Hamiltonian for the PM phase of \ce{EuCd2As2} previously obtained from DFT calculations in Sec.~\ref{siDFT} to compute the $(001)$-surface states.
We begin by cutting the Wannier model into a slab geometry that is finite in the $z$-direction while retaining periodicity in the $(x,y)$-plane, resulting in a slab Hamiltonian $H_{\mathrm{slab}}$.
We then partially Fourier transform $H_{\mathrm{slab}}$ to obtain the slab Bloch Hamiltonian $H_{\mathrm{slab}}(\mathbf{k})$, where $\mathbf{k}$ denotes a vector in the space of the in-plane crystal momenta:
\begin{equation}
\mathbf{k} \equiv (k_1,k_2),
\end{equation}
where $k_{1,2}$ respectively lie along the reciprocal lattice vectors $\mathbf{b}_{1,2}$ in Eq.~(\ref{reciplattvecs}).
The slab Bloch Hamiltonian $H_{\mathrm{slab}}(\mathbf{k})$ is consequently a square matrix of dimension:
\begin{equation}
N = N_s \times N_a \times N_L,
\label{eq:spinsOrbitals}
\end{equation}
where $N_s=2$ is the number of spin components, $N_a=20$ is the number of (Kramers pairs of) orbitals per layer (slab unit cell), and $N_L$ is the number of layers.
Throughout all of the slab-based calculations in this work, we fix $N_L=40$.

We next define the retarded Green's function at a given energy $E$ and in-plane momentum $\mathbf{k}$ to be:
\begin{equation}
G^R(\mathbf{k},E) = \bigl((E-E_{F})\mathbb{I}_{N\times N} - H_{\mathrm{slab}}(\mathbf{k}) + i\Omega\mathbb{I}_{N\times N}\bigr)^{-1},
\label{eq:GR_def}
\end{equation}
where $\mathbb{I}_{N\times N}$ is the $N\times N$ identity matrix and $\Omega$ is a phenomenological broadening parameter.
For all of the theoretical spectral function calculations in this work, we choose $\Omega=3.87$ meV such that $\Omega$ is equal to $k_{B}T$ for $T=45$ K, which lies just above the temperature range in which our PM-state ARPES measurements were performed (see for example Sec.~\ref{CD_temp_SI}).
To restrict consideration to spectral weight on the top slab surface, we then construct the top-surface projector:
\begin{equation}
P_t = \sum_{s=\uparrow,\downarrow}\sum_{a=1}^{N_{a}}|\psi_{s,a,N_L}\rangle\langle \psi_{s,a,N_L}|,
\label{eq:topProjDef}
\end{equation}
where $|\psi_{s,a,i}\rangle$ is a tight-binding (Wannier) basis state of the slab model with a spin index $s$, an orbital index $a$, and a layer (slab unit cell) index $i$ (see the text surrounding Eq.~(\ref{eq:spinsOrbitals})).
In Eq.~(\ref{eq:topProjDef}), $P_{t}$ is hence an $N\times N$ matrix.
In the basis of the underlying spins, orbitals, and layers, $P_{t}$ is a diagonal matrix whose elements are equal to $1$ for basis states belonging to the top layer, and are $0$ for all other basis states.
Lastly, we define the \emph{top-surface spectral function} $A_{\mathrm{surf}}(\mathbf{k},E)$ at each $\mathbf{k}$ and $E$ by taking the imaginary part of the trace of $G^R(\mathbf{k},E)$ in Eq.~(\ref{eq:GR_def}) projected to the top surface~\cite{Wu2018WannierTools:Materials, Saito2016Tight-bindingFilms}:
\begin{equation}
A_{\mathrm{surf}}(\mathbf{k},E)
    = -\frac{1}{\pi}\,\mathrm{Im}\,\mathrm{Tr}
    \bigl[P_t G^R(\mathbf{k},E) P_t\bigr] = -\frac{1}{\pi}\,\mathrm{Im}\,\mathrm{Tr}
    \bigl[P_t G^R(\mathbf{k},E) \bigr],
    \label{eq:Asurf_def}
\end{equation}
in which we have simplified using the projector idempotence $P_{t}^{2}=P_{t}$ and the cyclic property of the trace.

To make contact between $A_{\mathrm{surf}}(\mathbf{k},E)$  and the full density of states, we first examine the full spectral function $A(\mathbf{k},E)$ realized by neglecting (setting to identity) the factor of $P_{t}$ in Eq.~(\ref{eq:Asurf_def}).
We begin this analysis by considering the action of $G^R(\mathbf{k},E)$ on a Bloch eigenstate $|n\mathbf{k}\rangle$, for which:
\begin{equation}
H_{\mathrm{slab}}(\mathbf{k}) |n\mathbf{k}\rangle
    = E_n(\mathbf{k}) |n\mathbf{k}\rangle.
\label{eq:HslabEslab}
\end{equation}
Because $[H_{\mathrm{slab}}(\mathbf{k}), G^R(\mathbf{k},E)]=0$, then Eqs.~(\ref{eq:GR_def}) and~(\ref{eq:HslabEslab}) imply that:
\begin{equation}
G^R(\mathbf{k},E) |n\mathbf{k}\rangle
    = \left(\frac{1}
    {(E-E_{F}) - E_n(\mathbf{k}) + i\Omega} \right)|n\mathbf{k}\rangle,
\label{eq:GrEvals}
\end{equation}
such that $G^R(\mathbf{k},E)$ can be re-expressed through the spectral decomposition:
\begin{equation}
G^R(\mathbf{k},E) = \sum_{n=1}^{N}
    \frac{|n\mathbf{k}\rangle\langle n\mathbf{k}|}
    {(E-E_{F}) - E_n(\mathbf{k}) + i\Omega}.
\label{eq:GR_spectral}
\end{equation}
Inserting Eq.~(\ref{eq:GR_spectral}) into Eq.~(\ref{eq:Asurf_def}) with $P_{t}$ set to the identity then returns a spectral function $A(\mathbf{k},E)$ for the entire slab at each $\mathbf{k}$:
\begin{eqnarray}
A(\mathbf{k},E) &=& -\frac{1}{\pi}\,\mathrm{Im}\,\mathrm{Tr}
    \bigl[G^R(\mathbf{k},E) \bigr] \nonumber \\
    &=& - \sum_{n=1}^{N}\frac{1}{\pi}\,\mathrm{Im}\,
    \frac{1}{(E-E_{F}) - E_n(\mathbf{k}) + i\Omega} \nonumber \\
    &=& \sum_{n=1}^{N}\frac{1}{\pi}\frac{\Omega}{[E-E_{F} - E_n(\mathbf{k})]^2+\Omega^2} \nonumber \\
    &\equiv& \sum_{n=1}^{N}\delta_\Omega\bigl(E-E_{F} - E_n(\mathbf{k})\bigr),
\label{eq:FTDOS}
\end{eqnarray}
where in the final line, we recognize the preceding expression as the Lorentzian representation of the Dirac delta function $\delta_\Omega(E-E_{F} - E_n(\mathbf{k}))$.
The overall spectral function $A(\mathbf{k},E)$ in Eq.~(\ref{eq:FTDOS}) hence characterizes the Fourier-transformed density of states~\cite{Saito2016Tight-bindingFilms, Fang2013TheoryInsulators, Mahan1990Many-ParticlePhysics}.
Due to the strongly nonlinear (delta-function-like) profile of $A(\mathbf{k},E)$ in Eq.~(\ref{eq:FTDOS}), we therefore throughout this work plot all spectral functions on a logarithmic scale with arbitrary units (a.u.).

Next, to understand the role of $P_{t}$ in Eq.~(\ref{eq:Asurf_def}), we simplify the trace in Eq.~(\ref{eq:Asurf_def}) by combining Eqs.~(\ref{eq:topProjDef}) and~(\ref{eq:GR_spectral}):
\begin{equation}
\mathrm{Tr}\bigl[P_t G^R(\mathbf{k},E) \bigr]
    = \sum_n
    \frac{\langle n\mathbf{k}|P_t|n\mathbf{k}\rangle}
    {(E-E_{F}) - E_n(\mathbf{k}) + i\Omega} = \sum_{s=\uparrow,\downarrow}\sum_{a=1}^{N_{a}}\sum_{n=1}^{N}
    \frac{\left|\langle\psi_{s,a,N_L}|n\mathbf{k}\rangle\right|^{2}}
    {(E-E_{F}) - E_n(\mathbf{k}) + i\Omega}.
\label{eq:overlap}
\end{equation}
Because each overlap magnitude $|\langle\psi_{s,a,N_L}|n\mathbf{k}\rangle|^{2}$ in Eq.~(\ref{eq:overlap}) is real, then using Eq.~(\ref{eq:FTDOS}), the surface spectral function $A_{\mathrm{surf}}(\mathbf{k},E)$ in Eq.~(\ref{eq:Asurf_def}) becomes:
\begin{eqnarray}
A_{\mathrm{surf}}(\mathbf{k},E) &=& \sum_{s=\uparrow,\downarrow}\sum_{a=1}^{N_{a}}\sum_{n=1}^{N} \left|\langle\psi_{s,a,N_L}|n\mathbf{k}\rangle\right|^{2}
    \left[-\frac{1}{\pi}\,\mathrm{Im}\,
    \frac{1}{(E-E_{F}) - E_n(\mathbf{k}) + i\Omega}\right] \nonumber \\
&=& \sum_{s=\uparrow,\downarrow}\sum_{a=1}^{N_{a}}\sum_{n=1}^{N} \left|\langle\psi_{s,a,N_L}|n\mathbf{k}\rangle\right|^{2}\delta_\Omega\bigl(E-E_{F} - E_n(\mathbf{k})\bigr).
\label{eq:Asurf_eigen}
\end{eqnarray}
Overall, the form of Eq.~(\ref{eq:Asurf_eigen}) indicates that $A_{\mathrm{surf}}(\mathbf{k},E)$ is similar in structure to the total spectral function $A(\mathbf{k},E)$ in Eq.~(\ref{eq:FTDOS}), but instead measures the weight of each slab Bloch eigenstate $|n\mathbf{k}\rangle$ at $E-E_{F}$ on the top surface via the combination of $\delta_\Omega(E-E_{F} - E_n(\mathbf{k}))$ and the overlap magnitudes of $|n\mathbf{k}\rangle$ with the top-surface basis states.
In Fig.~\ref{fig:si_dft}b, we show the $(001)$-surface spectral function of the PM state of \ce{EuCd2As2} computed using Eq.~(\ref{eq:Asurf_def}). 
The surface spectrum consists of a single hyperbolic feature that lies within the continuum of unoccupied bulk states (the projected bulk conduction manifold).

Before analyzing the angular momentum textures and topological origin of the unoccupied $(001)$-surface states in Fig.~\ref{fig:si_dft}b (which will shortly be done in Secs.~\ref{sispintext} and~\ref{sirepeattopo}), we will establish the degree to which the chemical potential of the surface states $\widetilde{E} = E_F + \qty{0.34}{eV}$ depends on our bulk calculation details.
This information is especially important because the unoccupied surface states in our ARPES measurements appear at relatively higher energies, specifically instead exhibiting a band minimum at $E_{0} \approx E_F+\qty{0.8}{eV}$ as estimated from the kinetic energy of photoelectrons.
First, $\widetilde{E}$ depends on the choice of surface termination atoms in our slab model, an effect that itself is typically significant in models of the surface states of real solid-state materials~\cite{Shockley1939OnPotential, Lee2023Layer-by-layerStates, Wu2020DistinctMnBi4Te7}.
Furthermore, in our \ce{EuCd2As2} samples, we do not have atomistic knowledge of the $(001)$-surface probed in our ARPES experiments with which to calibrate the slab surface termination in our DFT- and Wannier-based surface-state calculations.
Next, $\widetilde{E}$ in Fig.~\ref{fig:si_dft}b is also dependent on the value of $U$ employed in our bulk DFT $+$ SOC $+$ $U$ calculations (Sec.~\ref{siDFT}).
Specifically, in modeling the PM state of \ce{EuCd2As2}, we have incorporated a large Coulomb term $U=5$ eV in our DFT $+$ SOC $+$ $U$ calculations to shift the half-filled, nearly-flat Eu~4$f$ states away from $E_{F}$. 
The choice of $U$ in turn influences the surface-state chemical potential $\widetilde{E}$ via the spectral weight of the Eu~4$f$ orbitals in the surface bands.

We therefore next quantify the contribution of Eu~4$f$ states to the surface spectrum.
To accomplish this, we first construct a projector $P_{4f}$ onto the Eu $4f$ states in the slab model:
\begin{equation}
P_{4f} = \sum_{s=\uparrow,\downarrow}\sum_{a \in \{\text{Eu}_{4f}\}}\sum_{i=1}^{N_L} |\psi_{s,a,i}\rangle\langle \psi_{s,a,i}|,
\label{eq:EuProjDef}
\end{equation}
where $|\psi_{s,a,i}\rangle$ and $N_L$ are defined in the text surrounding Eq.~(\ref{eq:topProjDef}).
Similar to $P_{t}$ in Eq.~(\ref{eq:topProjDef}), when $P_{4f}$ is placed in the basis of the spins, orbitals, and layers of the slab model, $P_{4f}$ is a diagonal matrix whose elements are equal to $1$ for basis states belonging to the Eu $4f$ orbital subspace, and are $0$ for all other basis states.
Using Eqs.~(\ref{eq:GR_def}),~(\ref{eq:topProjDef}), and~(\ref{eq:EuProjDef}), we then define a spectral function $\langle\text{Eu}_{4f}(\mathbf{k},E)\rangle$ that is only large for the subset of the surface spectral weight that derives from the Eu $4f$ orbitals:
\begin{equation}
\langle\text{Eu}_{4f}(\mathbf{k},E)\rangle = -\frac{1}{\pi}\,\mathrm{Im}\,\mathrm{Tr}
    \bigl[P_{4f}P_t G^R(\mathbf{k},E) P_t\bigr].
\label{eq:Atauf_def}
\end{equation}

\begin{figure*}[t]
\centering
\includegraphics[width=\textwidth]{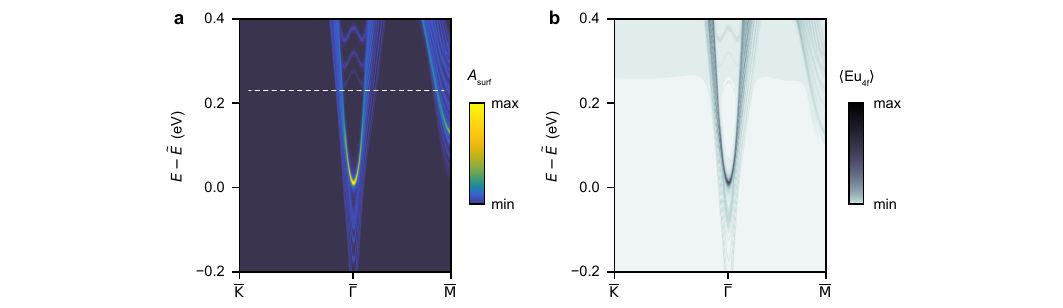}
\caption{{\bf Weight of the Eu $4f$ states in the $(001)$-surface spectral function:} \textbf{(a)} The $(001)$-surface spectral function $A_{\mathrm{surf}}(\mathbf{k},E)$ (Eq.~(\ref{eq:Asurf_def})) reproduced from Fig.~\ref{fig:si_dft}b and plotted relative to the minimum energy of the unoccupied surface states $\widetilde{E}$.
\textbf{(b)} The projected weight of the Eu $4f$ orbitals $\langle\text{Eu}_{4f}(\mathbf{k},E)\rangle$ in the $(001)$-surface states (Eq.~(\ref{eq:Atauf_def})).
Taken together, the data in (a, b) indicate that the unoccupied surface states carry a very large contribution from the Eu $4f$ orbitals, whose bulk chemical potential is set by the value of the Hubbard correction $U=5$ eV in our DFT calculations (see the text surrounding Eq.~(\ref{eq:IntrinsicFilling})).}
\label{fig:si_Eu_projections}
\end{figure*}

In Fig.~\ref{fig:si_Eu_projections}b, we plot $\langle\text{Eu}_{4f}(\mathbf{k},E)\rangle$ for the $(001)$-surface of \ce{EuCd2As2}.  
Comparing $\langle\text{Eu}_{4f}(\mathbf{k},E)\rangle$ in Fig.~\ref{fig:si_Eu_projections}b to the overall surface spectral function $A_{\mathrm{surf}}(\mathbf{k},E)$ in Fig.~\ref{fig:si_Eu_projections}a, we find that the hyperbolic surface state at $\widetilde{E}$ carries a very large contribution from the Eu $4f$ states.
This implies that beyond surface termination effects, $\widetilde{E}$ is highly dependent on the value of $U$ employed in our DFT $+$ SOC $+$ $U$ calculations (which was chosen to match previous bulk DFT calculations and ARPES measurements, see the text following Eq.~(\ref{eq:IntrinsicFilling})).

Hence, $\widetilde{E}$ in our theoretical calculations can be directly tuned by varying the slab surface chemical potential and the bulk value of $U$.
However, the dispersion relation and matrix elements of the $(001)$-surface states derive in a more complicated manner from the bulk band structure (and topology, see Sec.~\ref{sirepeattopo}). 
Furthermore, as discussed in the text following Eq.~(\ref{eq:IntrinsicFilling}), the \emph{bulk} band structure of the Wannier model (dashed black lines in Fig.~\ref{fig:si_dft}a) shows close agreement with previous studies of \ce{EuCd2As2}~\cite{Roychowdhury2023AnomalousEuCd2As2, Soh2019IdealExchange, Ma2019SpinEuCd2As2,Nelson2024RevealingLa-Doping,Santos-Cottin2023EuCd2As2:Semiconductor}.
Together, this suggests the possibility that we may interpret the shift in $\widetilde{E}$ relative to its experimental value $E_{0}$ as a leading order correction, and more simply treat the DFT-obtained $(001)$-surface states at $\widetilde{E}$ in Figs.~\ref{fig:si_dft}b and~\ref{fig:si_Eu_projections}a as representative of the physical unoccupied surface states in our ARPES experiments shifted by a \emph{surface} chemical potential.

\begin{figure*}[t]
\centering
\includegraphics{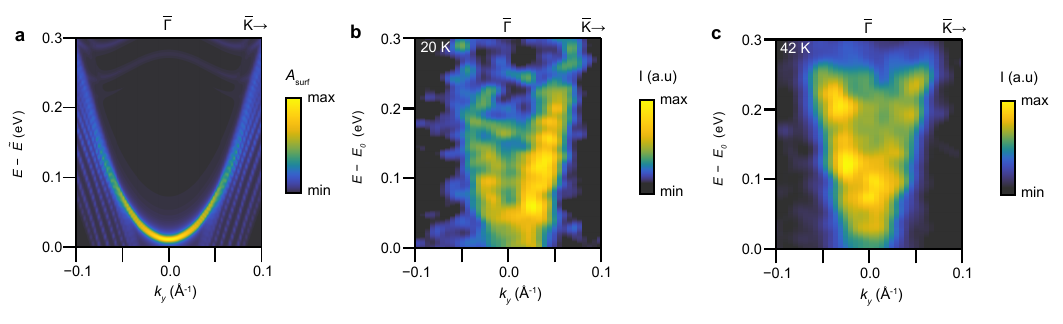}
\caption{{\bf Comparison between the DFT-obtained surface spectral function and the unoccupied surface-state dispersion in Tr-ARPES measurements:} \textbf{(a)} The $(001)$-surface spectral function $A_{\mathrm{surf}}(\mathbf{k},E)$ (Eq.~(\ref{eq:Asurf_def})) of the Wannier-based slab model obtained from our DFT calculations (Fig.~\ref{fig:si_dft}b), plotted relative to the minimum energy of the surface states $\widetilde{E}$ along $k_{y}$ ($\overline{\Gamma}-\overline{\mathrm{K}}$ in the surface BZ in Fig.~\ref{fig:si_dft}a), and zoomed into the momentum range accessible in our Tr-ARPES experiments.
\textbf{(b, c)} ARPES spectra of the unoccupied band structure plotted relative to the minimum energy of the surface states $E_{0}$ along $\overline{\Gamma}-\overline{\mathrm{K}}$ in the surface BZ, respectively taken below ($T=\qty{20}{K}$) and above ($T=\qty{42}{K}$) the PM to FM transition temperature $T_{C}\approx\qty{25}{K}$.
The ARPES data in (b) were obtained by integrating over a small window after photo-excitation ($t-t_0\in[\qty{0}{ps}, \qty{2}{ps}]$), whereas the spectrum in (c) was measured at $t=t_0$ and generated by summing spectra acquired with left- and right-circularly polarized probes (further circular dichroism measurements are provided in Sec.~\ref{CD_temp_SI}).
The experimental dispersions relative to $E_{0}$ in (b, c) show strong qualitative agreement with the DFT-obtained $(001)$-surface states viewed relative to $\widetilde{E}$ in (a).}
\label{fig:si_DFT_ARPES_dispersions}
\end{figure*}

To support this interpretation, in Fig.~\ref{fig:si_DFT_ARPES_dispersions}, we plot the DFT-obtained surface spectrum as a function of $k_{y}$ and $E-\widetilde{E}$ compared to ARPES measurements of the $(001)$-surface states as functions of $k_{y}$ and $E-E_{0}$ both above and below the PM to FM transition temperature $T_{C}\approx\qty{25}{K}$.
Though the ARPES measurements show a hyperbolic surface state with a slightly higher velocity, the theoretical dispersion relation in Fig.~\ref{fig:si_DFT_ARPES_dispersions}a and the experimental $(001)$-surface-state dispersions both above (Fig.~\ref{fig:si_DFT_ARPES_dispersions}c) and below (Fig.~\ref{fig:si_DFT_ARPES_dispersions}b) $T_{C}$ show close qualitative agreement.
Even stronger support for our theoretical treatment of the $(001)$-surface states can be obtained by comparing our circular dichroism (CD-) ARPES data to the OAM texture at the same energy $0.23$ eV relative to the minimum of the surface band.
Specifically, as highlighted in Fig.~\ref{fig:overview}(c, d) of the main text, the $\langle L_{xz}\rangle = \langle \frac{L_{x} + L_{z}}{\sqrt{2}} \rangle$ component of the OAM-dependent spectral function vector in the Wannier-based model (Eq.~(\ref{eq:LVector}) and Fig.~\ref{fig:sxzconstenergy}c) at $\widetilde{E} + 0.23$ eV (the dashed lines in Figs.~\ref{fig:si_dft}b and~\ref{fig:si_Eu_projections}a) shows excellent agreement with CD-ARPES measurements taken both above and below $T_{C}$ at approximately $0.23$ eV above $E_{0}$ (Fig.~\ref{fig:si_temperature}(c, d)).

\subsection{Spin and orbital textures of the $(001)$-surface states}
\label{sispintext}

\begin{figure*}[t]
\centering
\includegraphics[width=\textwidth]{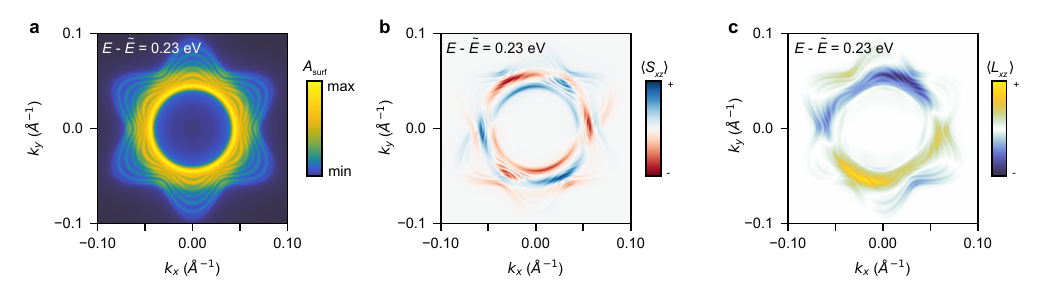}
\caption{{\bf Constant-energy surface spectrum and spin and orbital angular momentum textures in the Wannier-based model:}
\textbf{(a)} The $(001)$-surface spectral function $A_{\mathrm{surf}}(\mathbf{k},E)$ (Eq.~(\ref{eq:Asurf_def})) of the Wannier-based slab model obtained from our DFT calculations of the electronic structure of \ce{EuCd2As2} in the PM state (Fig.~\ref{fig:si_dft}a), plotted at a constant energy $0.23$ eV (the dashed line in Figs.~\ref{fig:si_dft}b and~\ref{fig:si_Eu_projections}a) above the minimum energy of the surface band $\widetilde{E}$ (the solid gray line in Fig.~\ref{fig:si_dft}b).
The value of $E-\widetilde{E}=0.23$ eV was chosen to match our constant-energy (CD-) ARPES measurements, which were taken approximately 0.23 eV above the experimentally observed surface band minimum $E_{0}$ (see Sec.~\ref{CD_temp_SI} and the text following Eq.~(\ref{eq:Atauf_def})).
The total spectral function $A_{\mathrm{surf}}(\mathbf{k},E)$ in (a) is dominated by a nearly circular feature surrounding ${\bf k}={\bf 0}$ with weaker hexagonal corrections.
\textbf{(b)} The $\langle S_{xz}\rangle = \langle \frac{S_{x} + S_{z}}{\sqrt{2}} \rangle$ component of the surface spin texture vector $\langle {\bf S}(\mathbf{k},E)\rangle$ (Eq.~(\ref{eq:SpinVector})), plotted at the same constant energy as $A_{\mathrm{surf}}(\mathbf{k},E)$ in (a).
$\langle S_{xz}({\bf k},E)\rangle$ in (b) exhibits two concentric rings with alternatingly signed $\langle S_{xz}\rangle$ spin polarizations, revealing that the circular surface spectral feature in (a) consists of two narrowly split surface Fermi pockets that together originate from a 2D surface Dirac cone at ${\bf k}={\bf 0}$ with extremely weak spin-orbit-coupling- (SOC-) driven splitting.
\textbf{(c)} The $\langle L_{xz}\rangle = \langle \frac{L_{x} + L_{z}}{\sqrt{2}} \rangle$ component of the surface orbital angular momentum (OAM) texture vector $\langle {\bf L}(\mathbf{k},E)\rangle$ (Eq.~(\ref{eq:LVector})), plotted at the same constant energy as the spectra in (a, b).
We selected $\langle L_{xz}\rangle$ in (c) to align with our experimental CD-ARPES measurement geometry (see the text following Eq.~(\ref{eq:oam_L2}), Fig.~\ref{fig:overview}c of the main text, and the End Matter).
Unlike $\langle S_{xz}({\bf k},E)\rangle$ in (b), $\langle L_{xz}({\bf k},E)\rangle$ in (c) exhibits the same sign on each ring-like Fermi pocket of the surface Dirac cone.
Overall, the OAM texture in (c) shows excellent agreement with our CD-ARPES measurements, both above and below the PM to FM transition temperature $T_{C}\approx\qty{25}{K}$ (see Fig.~\ref{fig:si_temperature}(c, d) and the text following Eq.~(\ref{eq:Atauf_def})).}
\label{fig:sxzconstenergy} 
\end{figure*}

To better understand the properties of the unoccupied surface bands in Fig.~\ref{fig:si_dft}b, and to draw connection with our circular dichroism (CD-) ARPES experiments (Sec.~\ref{CD_temp_SI}), we will next extract the spin and OAM textures of the $(001)$-surface states.
We will then introduce and analyze a simple ${\bf k}\cdot {\bf p}$ model that reproduces the dispersion and angular momentum textures of the surface conduction bands in \ce{EuCd2As2}.

First, to compute the spin texture, we introduce the surface-projected, spin-dependent spectral function vector $\langle {\bf S}(\mathbf{k},E)\rangle$, whose components are given by~\cite{Saito2016Tight-bindingFilms, Kohsaka2017Spin-orbitInterference, Chang2017UnconventionalRhSi, Schirmann2025Geometry-EnforcedMetals}:
\begin{equation}
\langle S_{\alpha}(\mathbf{k},E)\rangle
    = -\frac{1}{\pi}\,\mathrm{Im}\,\mathrm{Tr}
    \bigl[\hat{S}_{\alpha}\,P_t G^R(\mathbf{k},E) P_t\bigr],
    \label{eq:SpinVector}
\end{equation}
where $G^R(\mathbf{k},E)$ and $P_{t}$ are respectively defined in Eqs.~(\ref{eq:GR_def}) and~(\ref{eq:topProjDef}), and where $\hat{S}_{\alpha}$ is the $\alpha$-th component of the spin operator vector $\hat{\bf S} = (\hat{S}_{x},\hat{S}_{y},\hat{S}_{z})$.
In the basis of the spins and orbitals in the slab Wannier model of \ce{EuCd2As2} constructed in Secs.~\ref{siDFT} and~\ref{sislabgreens}, and in reduced units in which the spin eigenvalues are $\pm 1$, the spin operator components $\hat{S}_{\alpha}$ are represented as $N_{a}N_{L}N_{s}\times N_{a}N_{L}N_{s} = N\times N$ matrices $S_{\alpha}$:
\begin{equation}
    S_\alpha = \mathbb{I}_{N_{a}N_{L}\times N_{a}N_{L}}\,\sigma_\alpha,
    \qquad \alpha = x,y,z,
\label{eq:spinMats}
\end{equation}
where $\sigma_{x,y,z}$ are $2\times 2$ Pauli matrices that act in the spin subspace, $\mathbb{I}_{N_{a}N_{L}\times N_{a}N_{L}}$ is the $N_{a}N_{L}\times N_{a}N_{L}$ identity matrix, and where $N_{a}=20$ and $N_{L}=40$ are respectively the number of Kramers pairs of orbitals in each unit cell of the bulk Wannier model and the number of layers in the slab Hamiltonian (Eq.~(\ref{eq:spinsOrbitals})).
In order to ensure that the representation $S_{\alpha}$ of $\hat{S}_{\alpha}$ is ${\bf k}$-independent in the tight-binding (periodic) gauge of $H_{\mathrm{slab}}(\mathbf{k})$ (Eq.~(\ref{eq:HslabEslab})) in which physical and topological quantities are correctly embedded~\cite{Yu2011EquivalentConnection,Benalcazar2017ElectricInsulators,Lin2024Spin-resolvedInsulators}, we have constructed the Wannier model in a manner in which the up- and down-spin components of each Wannier orbital in Table~\ref{tab:wann_orbitals_direct} occupy the same positions in real space.
Because $\langle {\bf S}(\mathbf{k},E)\rangle$ is odd under $\mathcal{T}$ and even under $\mathcal{I}$ (\emph{i.e.} transforms as a $\mathcal{T}$-odd axial vector), $\langle {\bf S}(\mathbf{k},E)\rangle$ must vanish in bulk systems with $\mathcal{I}$ and $\mathcal{T}$ symmetry.
However, on the 2D surfaces of 3D systems, which cannot carry $\mathcal{I}$ as a symmetry~\cite{Wieder2016Spin-orbitGroups}, $\langle {\bf S}(\mathbf{k},E)\rangle$ can be nonvanishing.
Like the surface spectral function $A_{\mathrm{surf}}(\mathbf{k},E)$ in Eq.~(\ref{eq:Asurf_def}), we throughout this work plot $\langle S_{\alpha}(\mathbf{k},E)\rangle$ on a log scale with arbitrary units (a.u.), but now allow for signed data represented using a color map in which blue is positive and red is negative.

To provide context for the spin texture, we first in Fig.~\ref{fig:sxzconstenergy}a plot a constant-energy cut of the total $(001)$-surface spectral function $A_{\mathrm{surf}}(\mathbf{k},E)$ (Eq.~(\ref{eq:Asurf_def})) taken $0.23$ eV above the surface-state minimum $\widetilde{E}$ (the dashed line in Figs.~\ref{fig:si_dft}b and~\ref{fig:si_Eu_projections}a), which approximately corresponds to the same energy offset from the experimental surface-state band minimum $E_{0}$ as our constant-energy ARPES measurements (see Sec.~\ref{CD_temp_SI} and the text following Eq.~(\ref{eq:Atauf_def})). 
The total surface spectral function in Fig.~\ref{fig:sxzconstenergy}a is dominated by a nearly circular feature surrounding ${\bf k}={\bf 0}$ with weaker hexagonal corrections.
We then in Fig.~\ref{fig:sxzconstenergy}b plot the $\langle S_{xz}\rangle = \langle \frac{S_{x} + S_{z}}{\sqrt{2}} \rangle$ component of the spin-dependent spectral function $\langle {\bf S}(\mathbf{k},E)\rangle$ (Eq.~(\ref{eq:SpinVector})) at the same constant energy as $A_{\mathrm{surf}}(\mathbf{k},E)$ in Fig.~\ref{fig:sxzconstenergy}a, where we have selected the $\langle S_{xz}\rangle$ component to match the alignment of our experimental CD-ARPES probe of the OAM texture (further detailed below in the text following Eq.~(\ref{eq:oam_L2})).
Surprisingly, $\langle S_{xz}({\bf k},E)\rangle$ in Fig.~\ref{fig:sxzconstenergy}b consists of two concentric rings with alternatingly signed $\langle S_{xz}\rangle$ spin polarizations.
This indicates that the circular surface spectral feature in Fig.~\ref{fig:sxzconstenergy}a is in fact not one surface Fermi pocket, but is rather \emph{two} narrowly split Fermi pockets that together originate from a 2D surface Dirac cone with extremely weak SOC splitting.
As we will discuss below in Sec.~\ref{sirepeattopo}, the $(001)$-surface Dirac cone at ${\bf k}={\bf 0}$ in our Wannier-based calculations of \ce{EuCd2As2} can more specifically be viewed as the 2D topological surface state of a 3D topological band insulator phase~\cite{Hasan2010Colloquium:Insulators,Wieder2022TopologicalSymmetry} at an electronic filling above the Fermi level $\nu > \nu_{F}$ (Eq.~(\ref{eq:IntrinsicFilling})), which can be accessed by photo-doping experiments like those in the present study.

We next turn our attention to the orbital texture of the $(001)$-surface states, which is closely linked to our CD-ARPES experiments.
Specifically, of the different ${\bf k}$-resolved contributions to the orbital magnetization, CD-ARPES most directly couples to the local (in real space) contribution to the OAM from the underlying atomic orbitals~\cite{Park2011Orbital-angular-momentumSplitting, Park2012ChiralBi2Se3,Wang2013CircularInsulators,Yen2024ControllableSemimetals}.
To probe the atomic-orbital contribution to the OAM, we construct a surface-projected, OAM-dependent spectral function vector $\langle {\bf L}(\mathbf{k},E)\rangle$, analogous to the surface-projected, spin-dependent spectral function vector $\langle {\bf S}(\mathbf{k},E)\rangle$ in Eq.~(\ref{eq:SpinVector}), whose components are given by~\cite{Sidilkover2025ReexaminingInsulator,Schirmann2025Geometry-EnforcedMetals}:
\begin{equation}
\langle L_{\alpha}(\mathbf{k},E)\rangle
    = -\frac{1}{\pi}\,\mathrm{Im}\,\mathrm{Tr}
    \bigl[\hat{L}_{\alpha}\,P_t G^R(\mathbf{k},E) P_t\bigr],
\label{eq:LVector}
\end{equation}
where $G^R(\mathbf{k},E)$ and $P_{t}$ are respectively defined in Eqs.~(\ref{eq:GR_def}) and~(\ref{eq:topProjDef}), and where $\hat{L}_{\alpha}$ is the $\alpha$-th component of the OAM operator vector $\hat{\bf L} = (\hat{L}_{x},\hat{L}_{y},\hat{L}_{z})$.
Like $\langle {\bf S}(\mathbf{k},E)\rangle$ in Eq.~(\ref{eq:SpinVector}), $\langle {\bf L}(\mathbf{k},E)\rangle$ transforms as a $\mathcal{T}$-odd axial vector, and must hence vanish at each ${\bf k}$ in bulk systems with $\mathcal{I}$ and $\mathcal{T}$ symmetry.
However, on the 2D surfaces of 3D systems, $\langle {\bf L}(\mathbf{k},E)\rangle$ can similarly be nonvanishing due to the absence of 3D $\mathcal{I}$ (specifically $\mathcal{I}\times\mathcal{T}$) symmetry.

Unlike the spin operator components $\hat{S}_{\alpha}$, the OAM operator components $\hat{L}_{\alpha}$ have more complicated representations in the Wannier basis, and are also more sensitive to the numerical Wannierization procedure.
Specifically, in order to interpret $\langle L_{\alpha}(\mathbf{k},E)\rangle$ as a physical observable, it is crucial that the value of $\langle L_{\alpha}(\mathbf{k},E)\rangle$ at each ${\bf k}$ does not depend on the arbitrary piece of the gauge of the wavefunctions $|n\mathbf{k}\rangle$ (Eq.~(\ref{eq:HslabEslab})) of the slab Bloch Hamiltonian $H_{\mathrm{slab}}(\mathbf{k})$ (\emph{i.e.} the free gauge factors that do not affect the real-space model embedding~\cite{Yu2011EquivalentConnection,Benalcazar2017ElectricInsulators,Lin2024Spin-resolvedInsulators}).
This can most straightforwardly be accomplished by requiring that, like $S_{\alpha}$ in Eq.~(\ref{eq:spinMats}), each OAM operator component $\hat{L}_{\alpha}$ is represented as a ${\bf k}$-independent matrix $L_{\alpha}$ in the tight-binding (periodic) gauge of $H_{\mathrm{slab}}(\mathbf{k})$.
In turn, in order for the matrix representatives $L_{\alpha}$ of $\hat{L}_{\alpha}$ to be ${\bf k}$-independent, each of the Wannier orbitals associated to the same atomic shell of each atom -- specifically the Wannier orbitals for each atom that are assigned the same principal ($n$) and azimuthal ($\ell$, \emph{i.e.} OAM) quantum numbers -- must occupy the same positions in real space.
This ensures that each OAM operator component $\hat{L}_{\alpha}$ acts locally on the real-space tight-binding basis states, such that the basis-state positions remain invariant under the action of $\hat{L}_{\alpha}$ for all $\alpha$. 
In the Wannier tight-binding model that we constructed for the PM state of \ce{EuCd2As2} in Sec.~\ref{siDFT}, the basis-state positions (Table~\ref{tab:wann_orbitals_direct}) not only satisfy the above requirement, but are even more tightly grouped.
Specifically, as shown in Table~\ref{tab:wann_orbitals_direct}, each of the Wannier orbitals in our model coincides with the position of its associated atom to within $10^{-4}$ in the units of the lattice vectors $\mathbf{a}_{1,2,3}$ in Eq.~(\ref{primlattvecs}).
Our Wannier model is hence suitably constructed for computing the surface-state OAM texture $\langle {\bf L}(\mathbf{k},E)\rangle$ in Eq.~(\ref{eq:LVector}).

Having established that the surface OAM texture $\langle {\bf L}(\mathbf{k},E)\rangle$ can be computed by applying Eq.~(\ref{eq:LVector}) to our Wannier tight-binding model, we next turn our attention to constructing the OAM matrix representatives $L_{\alpha}$ in the Wannier basis.
In reduced units in which $\hbar=1$, the OAM operator components $\hat{L}_{\alpha}$ are specifically represented in the slab Wannier model constructed in Sec.~\ref{sislabgreens} as $N_{a}N_{L}N_{s}\times N_{a}N_{L}N_{s} = N\times N$ matrices $L_{\alpha}$ that each take a block form:
\begin{equation}
L_{\alpha} = \left\{\bigoplus_{\xi, \ell_\xi \neq 0} L^{\xi,\ell_{\xi}}_{\alpha}\right\} \oplus \left\{\bigoplus_{\xi, \ell_\xi = 0} \mathbb{0}_{N_{s}N_{L}\times N_{s}N_{L}}\right\},
\label{eq:Lbreakdown}
\end{equation}
where each $L^{\xi,\ell_{\xi}}_{\alpha}$ is proportional to the identity in the $N_{s}N_{L}\times N_{s}N_{L} = 2N_{L}\times 2N_{L}$ spin and layer subspace (Eq.~(\ref{eq:spinsOrbitals})), and where $\mathbb{0}_{N_{s}N_{L}\times N_{s}N_{L}}$ is the $N_{s}N_{L}\times N_{s}N_{L}$ matrix of zeroes.
In Eq.~(\ref{eq:Lbreakdown}), $\xi$ and $\ell_\xi$ respectively run in Table~\ref{tab:wann_orbitals_direct} over the ``Site'' column and the OAM quantum numbers $\ell_{\xi}$ per site in the ``Wannier Orbitals'' column, such that for example the $\ell=2$ $d$ and $\ell=3$ $f$ orbitals of the $1a$ Eu atom are included in the first direct sum ($\ell_{\xi}\neq 0$) and excluded in the second ($\ell_{\xi} = 0$), whereas for example the $\ell=0$ $s$ orbitals of the $2d$ Cd atoms are excluded from the first direct sum and included in the second.
Each $L^{\xi,\ell_{\xi}}_{\alpha}$ in Eq.~(\ref{eq:Lbreakdown}) is hence an $N_{L}N_{s}(2\ell_{\xi}+1) \times N_{L}N_{s}(2\ell_{\xi}+1)$ matrix~\cite{Griffiths2016IntroductionMechanics}.

To obtain the matrices $L^{\xi,\ell_{\xi}}_{\alpha}$ in Eq.~(\ref{eq:Lbreakdown}) in the Wannier tight-binding basis, we first recognize that each atomic orbital in our model is expressed in the real cubic–harmonic basis~\cite{Muggli1972CubicHarmonics}:
\begin{equation}
\bigl\{s,p_x,p_y,p_z,
      d_{z^2}, d_{xz}, d_{yz}, d_{x^2-y^2}, d_{xy}, \ldots \bigr\},
\label{eq:cubicListing}
\end{equation}
rather than in the complex spherical–harmonic basis $\{\,Y_\ell^m\,\}$ in which $L^{\xi,\ell_{\xi}}_{z}$ is a diagonal matrix whose only nonzero elements take values given by the orbital magnetic quantum numbers $m$~\cite{Griffiths2016IntroductionMechanics,McQuarrie1997PhysicalApproach}.
Our strategy for constructing each $L^{\xi,\ell_{\xi}}_{x,y,z}$ matrix therefore consists of two steps: we first construct $L^{\xi,\ell_{\xi}}_{x,y,z}$ in the familiar basis of spherical harmonics, and we then apply a unitary transformation to rotate $L^{\xi,\ell_{\xi}}_{x,y,z}$ into the cubic-harmonic Wannier basis.

We next recognize that given an OAM matrix representative $(L^{\xi,\ell_{\xi}}_{z})^{(\ell)}$ in the spherical-harmonic basis (denoted by the superscript $(\ell)$ here and below) with the eigenstates $|\ell,m\rangle$ such that:
\begin{equation}
(L^{\xi,\ell_{\xi}}_{z})^{(\ell)}|\ell,m\rangle = m|\ell,m\rangle,
\end{equation}
there exist ladder operators $\hat{L}_{\pm}^{\xi,\ell_\xi}$ that change $m$ by one unit within the $\ell_\xi$ manifold~\cite{Sakurai2017ModernMechanics}:
\begin{equation}
\hat L^{\xi,\ell_\xi}_\pm\,|\ell,m\rangle = \sqrt{\ell_\xi(\ell_\xi+1)-m(m\pm 1)}\;|\ell,m\pm 1\rangle.
\label{eq:Lpm_action}
\end{equation}
Using Eq.~(\ref{eq:Lpm_action}), we then construct the matrix representatives $(L^{\xi,\ell_\xi}_\pm)^{(\ell)}$ of $\hat L^{\xi,\ell_\xi}_\pm$, whose matrix elements are given by:
\begin{equation}
\left[(L^{\xi,\ell_\xi}_\pm)^{(\ell)}\right]_{m,m'}
  = \sqrt{\ell_\xi(\ell_\xi+1)-m'(m'\pm 1)}\;\delta_{m,m'\pm 1}.
  \label{eq:Lpm_matrix_mbasis}
\end{equation}
The $\alpha=x,y,z$ OAM matrix representatives in the spherical-harmonic basis are then by definition given by~\cite{Sakurai2017ModernMechanics,Griffiths2016IntroductionMechanics}:
\begin{equation}
(L^{\xi,\ell_\xi}_x)^{(\ell)}=\frac{1}{2}\left\{(L^{\xi,\ell_\xi}_+)^{(\ell)} + (L^{\xi,\ell_\xi}_-)^{(\ell)}\right\},\qquad (L^{\xi,\ell_\xi}_y)^{(\ell)}=\frac{1}{2i}\left\{(L^{\xi,\ell_\xi}_+)^{(\ell)} - (L^{\xi,\ell_\xi}_-)^{(\ell)}\right\}.
\label{eq:LxLy_from_ladders}
\end{equation}
Lastly, each cubic-harmonic (Wannier-basis) eigenstate $|\ell,\mu\rangle_{\rm cub}$ represents a linear combination of the spherical-harmonic eigenstates $|\ell,m\rangle$~\cite{Muggli1972CubicHarmonics}:
\begin{equation}
|\ell,\mu\rangle_{\rm cub}
  = \sum_{m=-\ell}^{+\ell} \left[U^{\ell}\right]_{\mu m}\,|\ell,m\rangle.
  \label{eq:U_def}
\end{equation}
Using Eq.~(\ref{eq:U_def}), we hence obtain the unitary matrix $U^{\ell}$ that transforms the OAM matrix representatives $(L^{\xi,\ell_\xi}_\alpha)^{(\ell)}$ in the spherical-harmonic basis into the OAM matrix representatives $L^{\xi,\ell_{\xi}}_{\alpha}$ in the cubic-harmonic basis of our Wannier tight-binding model (Eq.~(\ref{eq:Lbreakdown})):
\begin{equation}
L^{\xi,\ell_\xi}_\alpha = U^{\ell}\, (L^{\xi,\ell_\xi}_\alpha)^{(\ell)}\, \big(U^{\ell}\big)^\dagger,
  \qquad \alpha\in\{x,y,z\}.
\label{eq:similarity_transform}
\end{equation}

As shown in Table~\ref{tab:wann_orbitals_direct}, our Wannier model contains $s$ ($\ell=0$), $p$ ($\ell=1$), $d$ ($\ell=2$), and $f$ ($\ell=3$) orbitals.
Using the procedure detailed in Eqs.~(\ref{eq:Lbreakdown}) through~(\ref{eq:similarity_transform}), we may now construct the OAM matrix representatives $L_{\alpha}$ required to compute $\langle {\bf L}(\mathbf{k},E)\rangle$ in Eq.~(\ref{eq:LVector}).
For each of the orbital shells with $\ell\neq 0$ enumerated below, we will specifically provide explicit matrix representatives in the form of $(2\ell+1)\times (2\ell+1)$ square matrices, each of which is taken to be shorthand notation for a larger $N_{L}N_{s}(2\ell+1) \times N_{L}N_{s}(2\ell +1)$ matrix that is proportional to the identity in the $N_{s}N_{L}\times N_{s}N_{L}$ spin and layer subspace (Eq.~(\ref{eq:spinsOrbitals})).

First, for the $p$ shell of the $2d$ As atoms (denoted with $\xi,\ell_\xi \equiv (p)$), the real cubic harmonics are expressed in terms of the spherical harmonics $Y_{1}^{m}$ as:
\begin{equation}
p_z = Y_1^0, \qquad
p_x = \frac{1}{\sqrt{2}}\bigl(Y_1^{-1}-Y_1^{1}\bigr), \qquad
p_y = \frac{i}{\sqrt{2}}\bigl(Y_1^{-1}+Y_1^{1}\bigr).
\label{eq:pcub}
\end{equation}
In the cubic-harmonic ordering $\{p_z,p_x,p_y\}$, the ladder operators $L^{(p)}_{\pm}$ in the Wannier basis are then represented by:
\begin{equation}
L^{(p)}_{+}=
\begin{pmatrix}
0 & 1 & - i \\
-1 & 0 & 0 \\
i & 0 & 0
\end{pmatrix},
\qquad
L^{(p)}_{-}=
\begin{pmatrix}
0 & -1 & - i \\
1 & 0 & 0 \\
i & 0 & 0
\end{pmatrix},
\label{eq:Lpm_p}
\end{equation}
and the block $3\times3$ OAM representatives $L_{\alpha}^{(p)}$, $\alpha \in \{x,y,z\}$ in the Wannier basis are given by:
\begin{equation}
L_x^{(p)}=
\begin{pmatrix}
0&0&-i\\
0&0&0\\
i&0&0
\end{pmatrix}, \qquad
L_y^{(p)}=
\begin{pmatrix}
0&-i&0\\
i&0&0\\
0&0&0
\end{pmatrix}, \qquad
L_z^{(p)}=
\begin{pmatrix}
0&0&0\\
0&0&i\\
0&-i&0
\end{pmatrix}. 
\label{eq:lp}
\end{equation}

Next, for the $d$ shell of the $1a$ Eu atom (denoted with $\xi,\ell_\xi \equiv (d)$), the real cubic harmonics are expressed in terms of the spherical harmonics $Y_2^{m}$ as:
\begin{equation}
\begin{aligned}
d_{z^2} &= Y_2^{0},\\[4pt]
d_{xz} &= \frac{1}{\sqrt{2}}\!\left( Y_2^{-1} - Y_2^{1} \right),\\[4pt]
d_{yz} &= \frac{i}{\sqrt{2}}\!\left( Y_2^{-1} + Y_2^{1} \right),\\[4pt]
d_{x^2-y^2} &= \frac{1}{\sqrt{2}}\!\left( Y_2^{2} + Y_2^{-2} \right),\\[4pt]
d_{xy} &= \frac{i}{\sqrt{2}}\!\left( Y_2^{2} - Y_2^{-2} \right).
\end{aligned} \label{eq:dcub}
\end{equation}
In the cubic-harmonic ordering $\{d_{z^2},d_{xz},d_{yz},d_{x^2-y^2},d_{xy}\}$, the ladder operators $L^{(d)}_{\pm}$ in the Wannier basis are then represented by:
\begin{equation}
L^{(d)}_{+}=
\begin{pmatrix}
0 & 0 & -1 & - i & 0 \\
0 & 0 & i & -1 & 0 \\
1 & - i & 0 & 0 & - \sqrt{3} \\
i & 1 & 0 & 0 & \sqrt{3} i \\
0 & 0 & \sqrt{3} & - \sqrt{3} i & 0
\end{pmatrix},
\qquad
L^{(d)}_{-}=
\begin{pmatrix}
0 & 0 & 1 & - i & 0 \\
0 & 0 & i & 1 & 0 \\
-1 & - i & 0 & 0 & \sqrt{3} \\
i & -1 & 0 & 0 & \sqrt{3} i \\
0 & 0 & -\sqrt{3} & - \sqrt{3} i & 0
\end{pmatrix},
\label{eq:Lpm_d}
\end{equation}
and the block $5\times 5$ OAM representatives $L_{\alpha}^{(d)}$, $\alpha \in \{x,y,z\}$ in the Wannier basis are given by:
\begin{equation}
L_x^{(d)}=
\begin{pmatrix}
0&0&0&-i&0\\
0&0&i&0&0\\
0&-i&0&0&0\\
i&0&0&0&\sqrt{3}\,i\\
0&0&0&-\sqrt{3}\,i&0
\end{pmatrix}, \qquad
L_y^{(d)}=
\begin{pmatrix}
0&0&i&0&0\\
0&0&0&i&0\\
-i&0&0&0&\sqrt{3}\,i\\
0&-i&0&0&0\\
0&0&-\sqrt{3}\,i&0&0
\end{pmatrix}, \qquad
L_z^{(d)}=
\begin{pmatrix}
0&2i&0&0&0\\
-2i&0&0&0&0\\
0&0&0&i&0\\
0&0&-i&0&0\\
0&0&0&0&0
\end{pmatrix}. \label{eq:ld}
\end{equation}

Lastly, for the $f$ shell of the $1a$ Eu atom (denoted with $\xi,\ell_\xi \equiv (f)$), the real cubic harmonics are expressed in terms of the spherical harmonics $Y_3^{m}$ as:
\begin{equation}
\begin{aligned}
f_{z^3} &= Y_3^{0},\\
f_{xz^2} &= \frac{1}{\sqrt{2}}\bigl(Y_3^{-1}-Y_3^{1}\bigr),\\
f_{yz^2} &= \frac{i}{\sqrt{2}}\bigl(Y_3^{-1}+Y_3^{1}\bigr),\\
f_{z(x^2-y^2)} &= \frac{1}{\sqrt{2}}\bigl(Y_3^{2}+Y_3^{-2}\bigr),\\
f_{xyz} &= \frac{i}{\sqrt{2}}\bigl(Y_3^{2}-Y_3^{-2}\bigr),\\
f_{x(x^2-3y^2)} &= \frac{1}{\sqrt{2}}\bigl(Y_3^{3}-Y_3^{-3}\bigr),\\
f_{y(3x^2-y^2)} &= \frac{i}{\sqrt{2}}\bigl(Y_3^{3}+Y_3^{-3}\bigr).
\end{aligned}\label{eq:fcub}
\end{equation}
In the cubic-harmonic ordering $\{f_{z^3},f_{xz^2},f_{yz^2},f_{z(x^2-y^2)},f_{xyz},f_{x(x^2-3y^2)},f_{y(3x^2-y^2)}\}$, the ladder operators $L^{(f)}_{\pm}$ in the Wannier basis are then represented by:
\begin{equation}
L^{(f)}_{+}=
\begin{pmatrix}
0 & \sqrt{6} & - \sqrt{6} i & 0 & 0 & 0 & 0 \\
- \sqrt{6} & 0 & 0 & \frac{\sqrt{10}}{2} & \frac{\sqrt{10} i}{2} & 0 & 0 \\
\sqrt{6} i & 0 & 0 & \frac{\sqrt{10} i}{2} & -\frac{\sqrt{10}}{2} & 0 & 0 \\
0 & -\frac{\sqrt{10}}{2} & -\frac{\sqrt{10} i}{2} & 0 & 0 & \frac{\sqrt{6}}{2} & -\frac{\sqrt{6} i}{2} \\
0 & -\frac{\sqrt{10} i}{2} & \frac{\sqrt{10}}{2} & 0 & 0 & -\frac{\sqrt{6} i}{2} & -\frac{\sqrt{6}}{2} \\
0 & 0 & 0 & -\frac{\sqrt{6}}{2} & \frac{\sqrt{6} i}{2} & 0 & 0 \\
0 & 0 & 0 & \frac{\sqrt{6} i}{2} & \frac{\sqrt{6}}{2} & 0 & 0
\end{pmatrix}, \nonumber
\label{eq:Lplus_f}
\end{equation}
\begin{equation}
L^{(f)}_{-}=
\begin{pmatrix}
0 & - \sqrt{6} & - \sqrt{6} i & 0 & 0 & 0 & 0 \\
\sqrt{6} & 0 & 0 & -\frac{\sqrt{10}}{2} & \frac{\sqrt{10} i}{2} & 0 & 0 \\
\sqrt{6} i & 0 & 0 & \frac{\sqrt{10} i}{2} & \frac{\sqrt{10}}{2} & 0 & 0 \\
0 & \frac{\sqrt{10}}{2} & -\frac{\sqrt{10} i}{2} & 0 & 0 & -\frac{\sqrt{6}}{2} & -\frac{\sqrt{6} i}{2} \\
0 & -\frac{\sqrt{10} i}{2} & -\frac{\sqrt{10}}{2} & 0 & 0 & -\frac{\sqrt{6} i}{2} & \frac{\sqrt{6}}{2} \\
0 & 0 & 0 & \frac{\sqrt{6}}{2} & \frac{\sqrt{6} i}{2} & 0 & 0 \\
0 & 0 & 0 & \frac{\sqrt{6} i}{2} & -\frac{\sqrt{6}}{2} & 0 & 0
\end{pmatrix},
\label{eq:Lminus_f}
\end{equation}
and the block $7\times 7$ OAM representatives $L_{\alpha}^{(f)}$, $\alpha \in \{x,y,z\}$ in the Wannier basis are given by:
\begin{equation}
L_x^{(f)}=
\begin{pmatrix}
0&0&-\sqrt{6}\,i&0&0&0&0\\
0&0&0&0&\frac{\sqrt{10}}{2}i&0&0\\
\sqrt{6}\,i&0&0&\frac{\sqrt{10}}{2}i&0&0&0\\
0&0&-\frac{\sqrt{10}}{2}i&0&0&0&-\frac{\sqrt{6}}{2}i\\
0&-\frac{\sqrt{10}}{2}i&0&0&0&-\frac{\sqrt{6}}{2}i&0\\
0&0&0&0&\frac{\sqrt{6}}{2}i&0&0\\
0&0&0&\frac{\sqrt{6}}{2}i&0&0&0
\end{pmatrix}, \nonumber 
\label{eq:lxf}
\end{equation}
\begin{equation}
L_y^{(f)}=
\begin{pmatrix}
0&-\sqrt{6}\,i&0&0&0&0&0\\
\sqrt{6}\,i&0&0&-\frac{\sqrt{10}}{2}i&0&0&0\\
0&0&0&0&\frac{\sqrt{10}}{2}i&0&0\\
0&\frac{\sqrt{10}}{2}i&0&0&0&-\frac{\sqrt{6}}{2}i&0\\
0&0&-\frac{\sqrt{10}}{2}i&0&0&0&\frac{\sqrt{6}}{2}i\\
0&0&0&\frac{\sqrt{6}}{2}i&0&0&0\\
0&0&0&0&-\frac{\sqrt{6}}{2}i&0&0
\end{pmatrix}, \qquad
L_z^{(f)}=
\begin{pmatrix}
0&0&0&0&0&0&0\\
0&0&i&0&0&0&0\\
0&-i&0&0&0&0&0\\
0&0&0&0&-2i&0&0\\
0&0&0&2i&0&0&0\\
0&0&0&0&0&0&3i\\
0&0&0&0&0&-3i&0
\end{pmatrix}. 
\label{eq:lyzf}
\end{equation}

We have importantly confirmed by direct computation that the OAM matrix representatives $L_{\alpha}^{\xi,\ell_\xi}$ in Eqs.~(\ref{eq:lp}) through~(\ref{eq:lyzf}) satisfy the canonical angular momentum commutation relations~\cite{Griffiths2016IntroductionMechanics}:
\begin{equation}
[L_{\alpha}^{\xi,\ell_\xi},L_{\beta}^{\xi,\ell_\xi}] = i\,\varepsilon^{\alpha\beta\gamma} L_{\gamma}^{\xi,\ell_\xi},
\label{eq:oam_comm}
\end{equation}
where $\varepsilon^{\alpha\beta\gamma}$ is the Levi–Civita tensor and
$\alpha,\beta,\gamma\in\{x,y,z\}$.
We have also numerically confirmed that the $L_{\alpha}^{\xi,\ell_\xi}$ matrices for each angular momentum shell satisfy the total angular momentum summation relation:
\begin{equation}
  (L^{\xi,\ell_\xi})^2 = \sum_{\alpha=x,y,z} (L_{\alpha}^{\xi,\ell_\xi})^2 = \ell_\xi(\ell_\xi+1)\,\mathbb{I}_{N_{L}N_{s}(2\ell_{\xi}+1) \times N_{L}N_{s}(2\ell_{\xi}+1)},
  \label{eq:oam_L2}
\end{equation}
where $\mathbb{I}_{N_{L}N_{s}(2\ell_{\xi}+1) \times N_{L}N_{s}(2\ell_{\xi}+1)}$ is the $N_{L}N_{s}(2\ell_{\xi}+1) \times N_{L}N_{s}(2\ell_{\xi}+1)$ identity matrix (see the text surrounding Eq.~(\ref{eq:Lbreakdown})).

Having established our OAM formalism, we may now finally compute the $(001)$-surface OAM texture $\langle {\bf L}(\mathbf{k},E)\rangle$ (Eq.~(\ref{eq:LVector})) in our Wannier-based model of \ce{EuCd2As2}.
In our CD-ARPES experiments, measurements were performed with the photon scattering plane aligned along the $(x,z)$-plane of the sample (see Fig.~\ref{fig:overview}c of the main text and the End Matter). 
In momentum space, the photon scattering plane coincides with a crystallographic mirror plane whose normal vector lies along $k_{y}$ and is related to the BZ line $\overline{\Gamma}-\overline{\mathrm{K}}$ in Fig.~\ref{fig:si_dft}a by crystal symmetry.
We specifically employed in our CD-ARPES measurements \qty{6}{eV} circularly polarized photons with bulk and surface sensitivity that were incident along an axis aligned \ang{45} between the $x$ and $z$ axes.
We then extracted the CD-ARPES signature from the respective ARPES intensities $I_{CD}=I_{RCP}-I_{LCP}$.
We hence interpret our CD-ARPES measurements as probing the $\langle L_{xz}\rangle = \langle \frac{L_{x} + L_{z}}{\sqrt{2}} \rangle$ component of the $(001)$-surface OAM texture.
In Fig.~\ref{fig:sxzconstenergy}c, we therefore plot $\langle L_{xz}({\bf k},E)\rangle$ for the Wannier-based model at the same constant energy as the overall surface spectral function $A_{\mathrm{surf}}(\mathbf{k},E)$ in Fig.~\ref{fig:sxzconstenergy}a, using a signed log scale with a color map in which yellow is positive and dark blue is negative.
The $\langle L_{xz}({\bf k},E)\rangle$ OAM texture component in Fig.~\ref{fig:sxzconstenergy}c shows excellent agreement with our experimental CD-ARPES data measured at the same energy relative to the surface band minimum both above and below the PM to FM transition temperature $T_{C}\approx\qty{25}{K}$ (see Fig.~\ref{fig:si_temperature}(c, d) and the text following Eq.~(\ref{eq:Atauf_def})).

Interestingly, unlike the $\langle S_{xz}({\bf k},E)\rangle$ spin texture component in Fig.~\ref{fig:sxzconstenergy}b, $\langle L_{xz}({\bf k},E)\rangle$ in Fig.~\ref{fig:sxzconstenergy}c exhibits the \emph{same} sign for both concentric ring-like Fermi pockets of the surface Dirac cone.
To demonstrate a scenario that qualitatively reproduces the spin and OAM textures in Fig.~\ref{fig:sxzconstenergy}(b, c), we introduce a simple ${\bf k}\cdot {\bf p}$ model $\mathcal{H}_{kp}(k)$ that represents a 1D line cut along $k$ through a $\mathcal{T}$-invariant point $k=0$ in a 2D surface BZ:
\begin{equation}
\mathcal{H}_{kp}(k) = k \tau_{z} + M \tau_{x} + v \tau_{z} \sigma_{z}. 
\label{eq:kdotp}
\end{equation}
In Eq.~(\ref{eq:kdotp}), there are four (surface-localized) basis states that arise from spin-degenerate $p_{x}\pm ip_{y}$ orbitals with the orbital magnetic quantum numbers ($L_{z}$ eigenvalues) $m=\pm 1$ (see the text surrounding Eq.~(\ref{eq:cubicListing})), and hence overall form a total angular momentum $j=3/2$ representation~\cite{Griffiths2016IntroductionMechanics}.
We employ a notation in Eq.~(\ref{eq:kdotp}) in which:
\begin{equation}
\tau_{i}\sigma_{j} = s_j \otimes s_i,
\end{equation}
where $s_{i},\ i\in\{x,y,z\}$ are the $2\times 2$ Pauli matrices, $s_{0}$ is the $2\times 2$ identity matrix, and where factors of $\tau_{0}$ and $\sigma_{0}$ have been suppressed for notation simplicity.
The matrices $\tau_{z}$ and $\sigma_{z}$ respectively index the $p_{x}\pm ip_{y}$ orbital and spin-1/2 system subspaces, such that the matrix representatives of $\hat{S}_{z}$ (Eq.~(\ref{eq:SpinVector})) and $\hat{L}_{z}$ (Eq.~(\ref{eq:LVector})) are respectively given by:
\begin{equation}
S_{z} = \sigma_z, \qquad L_{z} = \tau_{z},
\label{eq:KPspinOAMdef}
\end{equation}
and $\mathcal{T}$ symmetry is represented through the action:
\begin{equation}
\mathcal{T}\mathcal{H}_{kp}(k)\mathcal{T}^{-1} = \tau_{x}\sigma_{y}\mathcal{H}^{*}_{kp}(-k)\tau_{x}\sigma_{y}.
\label{eq:TkpRep}
\end{equation}
In $\mathcal{H}_{kp}(k)$ in Eq.~(\ref{eq:kdotp}), the velocity term $k\tau_z$ hence induces fully OAM-polarized dispersion $kL_{z}$, $M \tau_{x}$ is a ${\bf k}$-independent OAM coupling term that breaks $L_{z}$ symmetry, and $v \tau_{z} \sigma_{z}$ is an $S_{z}$-symmetric SOC splitting term.

\begin{figure*}[t]
\centering
\includegraphics{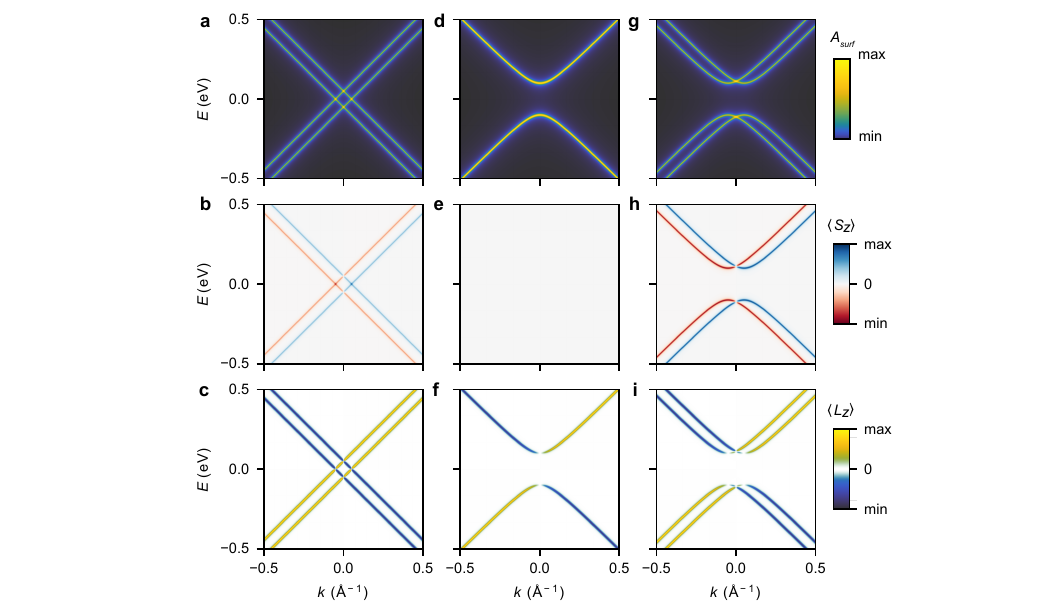}
\caption{{\bf Low-energy toy model of 2D Dirac surface states with contrasting spin and OAM polarizations:} 
In this figure, we plot the spectral function $A_{\mathrm{surf}}(k,E)$ (Eq.~(\ref{eq:Asurf_def}), $\langle S_{z}(k,E)\rangle$ spin texture component (Eq.~(\ref{eq:SpinVector})), and $\langle L_{z}(k,E)\rangle$ OAM texture component (Eq.~(\ref{eq:LVector})) of a simplified, $\mathcal{T}$-invariant ${\bf k}\cdot {\bf p}$ model Hamiltonian $\mathcal{H}_{kp}(k)$ along a 1D line parameterized by $k$ in a 2D surface BZ (Eqs.~(\ref{eq:kdotp}),~(\ref{eq:KPspinOAMdef}),~(\ref{eq:TkpRep}), and~(\ref{eq:identitySmallProj})).
\textbf{(a, b, c)} $A_{\mathrm{surf}}(k,E)$, $\langle S_{z}(k,E)\rangle$, and $\langle L_{z}(k,E)\rangle$, respectively, for $\mathcal{H}_{kp}(k)$ with $M=0$ and $v=-0.05$.
The spectrum in (a) consists of four linear bands with (b) narrow spin splitting, where the bands within each spin-split pair exhibit (c) $\langle L_{z} \rangle$ OAM polarizations with the same sign originating from the OAM dispersion term $k_{z}\tau_{z}=kL_{z}$ in Eqs.~(\ref{eq:kdotp}) and~(\ref {eq:KPspinOAMdef}).
\textbf{(d, e, f)} $A_{\mathrm{surf}}(k,E)$, $\langle S_{z}(k,E)\rangle$, and $\langle L_{z}(k,E)\rangle$, respectively, for $\mathcal{H}_{kp}(k)$ with vanishing SOC and significant OAM coupling ($M=-0.1$, $v=0$).
The spectrum in (d) consists of two pairs of spin-degenerate bands with hyperbolic dispersion that (e) exhibit vanishing $\langle S_{z} \rangle$ at all $k$, due to the spin degeneracy, and (f) carry the same sign of $\langle L_{z} \rangle$ within each spin-degenerate pair.
\textbf{(g, h, i)} $A_{\mathrm{surf}}(k,E)$, $\langle S_{z}(k,E)\rangle$, and $\langle L_{z}(k,E)\rangle$, respectively, for the previous case in (d, e, f) subject to additional weak SOC ($M=-0.1$, $v=-0.05$).
The spectrum in (g) consists of two $\mathcal{T}$-enforced twofold Dirac degeneracies at $k=0$, where the bands from each Dirac point exhibit linear dispersion at small $k$, but otherwise largely follow the hyperbolic energy dispersion in (d).
As in our Wannier-based calculations of the unoccupied $(001)$-surface states of \ce{EuCd2As2} (Fig.~\ref{fig:sxzconstenergy}), the two singly degenerate branches of each Dirac cone in (g) exhibit (h) oppositely signed $\langle S_{z}\rangle$ spin polarizations, but (i) carry $\langle L_{z} \rangle$ OAM polarizations with the same sign.}
\label{fig:kdotp}
\end{figure*}

In Fig.~\ref{fig:kdotp}, we respectively plot the total spectral function $A_{\mathrm{surf}}(k,E)$ (Eq.~(\ref{eq:Asurf_def}), panels a, d, and g in the top row), the $\langle S_{z} \rangle$ component of the spin texture $\langle {\bf S}(k,E)\rangle$ (Eq.~(\ref{eq:SpinVector}), panels b, e, and h in the middle row), and the $\langle L_{z}\rangle$ component of the OAM texture $\langle {\bf L}(k,E)\rangle$ (Eq.~(\ref{eq:LVector}), panels c, f, and i in the bottom row) as functions of $E$ and $k$ for $\mathcal{H}_{kp}(k)$ with varying values of $M$ and $v$ in Eq.~(\ref{eq:kdotp}).
For all panels in Fig.~\ref{fig:kdotp}, we have taken the top-surface projector $P_{t}$ (Eq.~(\ref{eq:topProjDef})) to be the $4\times 4$ identity matrix:
\begin{equation}
P_{t} = \mathbb{I}_{4\times 4},
\label{eq:identitySmallProj}
\end{equation}
because the low-energy Hamiltonian $\mathcal{H}_{kp}(k)$ is already projected into a (sub)space of surface-localized basis states.

We first in Fig.~\ref{fig:kdotp}(a, b, c) plot $A_{\mathrm{surf}}(k,E)$, $\langle S_{z}(k,E)\rangle$, and $\langle L_{z}(k,E)\rangle$ for $\mathcal{H}_{kp}(k)$ with $M=0$ and $v=-0.05$.
The spectral function $A_{\mathrm{surf}}(k,E)$ in Fig.~\ref{fig:kdotp}a consists of four linear bands with narrow spin splitting (Fig.~\ref{fig:kdotp}b), where the bands within each spin-split pair exhibit $\langle L_{z}\rangle$ OAM polarizations with the same sign (Fig.~\ref{fig:kdotp}c).
Next, in Fig.~\ref{fig:kdotp}(d, e, f), we plot $A_{\mathrm{surf}}(k,E)$, $\langle S_{z}(k,E)\rangle$, and $\langle L_{z}(k,E)\rangle$ for a case with vanishing SOC and significant OAM coupling in which $M=-0.1$ and $v=0$.
In this case, $A_{\mathrm{surf}}(k,E)$ in Fig.~\ref{fig:kdotp}d consists of two pairs of spin-degenerate bands with hyperbolic dispersion, for which $\langle S_{z}(k,E)\rangle$ is vanishing at all k due to the spin degeneracy (Fig.~\ref{fig:kdotp}e).
Importantly, however, the OAM texture $\langle L_{z}(k,E)\rangle$ in Fig.~\ref{fig:kdotp}f remains nonvanishing, and the two doubly degenerate bands within each pair continue to exhibit the \emph{same} sign of $\langle L_{z} \rangle$ at each k.
We lastly consider the case of weakly introducing SOC into the previous scenario, and therefore in Fig.~\ref{fig:kdotp}(g, h, i) plot $A_{\mathrm{surf}}(k,E)$, $\langle S_{z}(k,E)\rangle$, and $\langle L_{z}(k,E)\rangle$ for $\mathcal{H}_{kp}(k)$ with $M=-0.1$ and $v=-0.05$.
The energy spectrum in Fig.~\ref{fig:kdotp}g now consists of two $\mathcal{T}$-enforced Dirac cones whose two branches are only weakly spin-split (Fig.~\ref{fig:kdotp}h) and pairwise disperse in the same directions except at very small values of $k$.
Specifically, focusing on the upper Dirac cone in energy in Fig.~\ref{fig:kdotp}g, the two Dirac bands together form an essentially hyperbolic dispersion with positive velocities for all $k>0$ outside of a very small momentum range in the vicinity of $k=0$.
This closely resembles the dispersion and spin texture distribution of the unoccupied $(001)$-surface states in our Wannier-based model of \ce{EuCd2As2} (Figs.~\ref{fig:si_dft}b,~\ref{fig:si_DFT_ARPES_dispersions}a, and~\ref{fig:sxzconstenergy}a, b).
Also like the $(001)$-surface Dirac cone in our Wannier-based calculations (Fig.~\ref{fig:sxzconstenergy}c), the two singly degenerate branches of each Dirac cone in Fig.~\ref{fig:kdotp}i exhibit $\langle L_{z} \rangle$ OAM polarizations with the same sign, despite exhibiting oppositely signed $\langle S_{z} \rangle$ spin polarizations (Fig.~\ref{fig:kdotp}h), owing to the dominant OAM dispersion term $k\tau_z=kL_z$ in Eqs.~(\ref{eq:kdotp}) and~(\ref {eq:KPspinOAMdef}).

To conclude, we have introduced a simple ${\bf k}\cdot{\bf p}$ model that exhibits a 2D surface Dirac fermion with hyperbolic dispersion, weak spin splitting, and an OAM polarization distribution that mirrors our first-principles calculations of the $(001)$-surface spectrum of \ce{EuCd2As2}.
It is important to note, however, that the simple model in Eq.~(\ref{eq:kdotp}) only represents a demonstration that it is feasible to realize a surface Dirac cone with these characteristics, and we leave for future works the more detailed question of the specific physical mechanism that gives rise to the $(001)$-surface energy dispersion and angular momentum textures in \ce{EuCd2As2}.

\subsection{Topological classification and band topology above the Fermi level in \ce{EuCd2As2}}
\label{sirepeattopo}

In our calculations of the $(001)$-surface spectrum of the PM phase of \ce{EuCd2As2} in Secs.~\ref{sislabgreens} and~\ref{sispintext}, we observed (weakly SOC-split) 2D surface Dirac bands around ${\bf k}={\bf 0}$ with an energy minimum $\widetilde{E}$ above the Fermi level $E_{F}$ (Figs.~\ref{fig:si_dft}b,~\ref{fig:si_DFT_ARPES_dispersions}a, and~\ref{fig:sxzconstenergy}).
We will next investigate whether it is possible to associate the unoccupied Dirac surface states in \ce{EuCd2As2} to bulk topological bands.

To address this question, we employ the approach of Ref.~\cite{Vergniory2022AllMaterials}, in which the methods of Topological Quantum Chemistry~\cite{Bradlyn2017TopologicalChemistry,Elcoro2021MagneticChemistry,Wieder2022TopologicalSymmetry} and symmetry-based indicators (SIs)~\cite{Kruthoff2017TopologicalCombinatorics,PoHoiChun2017Symmetry-basedGroups, SongZhida2018QuantitativeSolids,Khalaf2018SymmetryInsulators} were applied to efficiently screen band structures in nonmagnetic materials (or material phases) at and away from $E_{F}$ for nontrivial topology that is detectable from crystal symmetry eigenvalues.
The authors of Ref.~\cite{Vergniory2022AllMaterials} specifically first performed DFT calculations using the Vienna Ab-initio Simulation Package (\texttt{VASP})~\cite{Kresse1996EfficiencySet,Kresse1996EfficientSet}, then applied the freely accessible Fortran program~\href{https://github.com/zjwang11/irvsp}{\texttt{VASP2Trace}} (\url{https://github.com/zjwang11/irvsp})~\cite{Gao2021Irvsp:VASP} to extract the eigenvalues of the unitary crystal symmetries, and lastly uploaded the extracted symmetry eigenvalues to the~\href{https://cryst.ehu.es/cgi-bin/cryst/programs/magnetictopo.pl?tipog=gesp}{\texttt{Check Topological Mat}} tool on the Bilbao Crystallographic Server (BCS) to determine the band topology (accessible by navigating from~\url{http://webbdcrista2.ehu.es/} to~\url{https://cryst.ehu.es/cgi-bin/cryst/programs/magnetictopo.pl?tipog=gesp})~\cite{Aroyo2006BilbaoPrograms,Aroyo2006BilbaoGroups,Aroyo2011CrystallographyServer,Vergniory2019AMaterials}.
Because our DFT calculations in Sec.~\ref{siDFT} were performed using \texttt{Quantum Espresso}~\cite{Giannozzi2009QUANTUMMaterials, Giannozzi2017AdvancedESPRESSO, Giannozzi2020QuantumExascale} and not \texttt{VASP}, we began by implementing custom code, detailed below, to extract the crystal symmetry eigenvalues from our band structure calculations in a format compatible with~\href{https://cryst.ehu.es/cgi-bin/cryst/programs/magnetictopo.pl?tipog=gesp}{\texttt{Check Topological Mat}}.

To construct an input file for~\href{https://cryst.ehu.es/cgi-bin/cryst/programs/magnetictopo.pl?tipog=gesp}{\texttt{Check Topological Mat}} from our DFT band structure calculation in Fig.~\ref{fig:si_dft}a, we first analyzed nonmagnetic SG 164 ($P\bar{3}m1$), the 3D symmetry group $G$ of \ce{EuCd2As2} in the PM state, to determine the set of maximal ${\bf k}$ points (\emph{i.e.} the $\mathcal{T}$-invariant ${\bf k}$ points, as well as high-symmetry points without $\mathcal{T}$ like $K$ and $H$ in Fig.~\ref{fig:si_dft}a, see Refs.~\cite{Bradlyn2017TopologicalChemistry,Elcoro2021MagneticChemistry}).
At each maximal ${\bf k}$ point, there specifically exists an infinite subgroup of symmetries $G_{\bf k}\subseteq G$, termed the little group, for which each symmetry:
\begin{equation}
g \in G_{\bf k},
\end{equation}
leaves ${\bf k}$ invariant up to an integer linear combination ${\bf K}$ of the reciprocal lattice vectors $\mathbf{b}_{1,2,3}$ in Eq.~(\ref{reciplattvecs}):
\begin{equation}
g{\bf k}\text{ mod }{\bf K}={\bf k}.
\label{eq:KmodforGk}
\end{equation}
For each grouping of bands that share the same energy eigenvalue $E_{n\mathbf{k}}$ (up to a numerical tolerance factor of $0.002$ eV), we next obtained the set of $m$ (approximately) degenerate Bloch eigenstates $\{|\psi^{1}_{n\mathbf{k}}\rangle, |\psi^{2}_{n\mathbf{k}}\rangle, \ldots, |\psi^{m}_{n\mathbf{k}}\rangle\}$ in an orthonormal basis.
The $m$ states that share the energy $E_{n\mathbf{k}}$ may either form an enforced degeneracy, or may simply be a set of states for which SOC and orbital splitting is very weak, but in either case transform into themselves or other degenerate states in the set under the action of the little group symmetries $g\in G_{\bf k}$~\cite{Bradley1972TheSolids, Bradlyn2017TopologicalChemistry}.

Using the~\texttt{spglib} software library~\cite{Togo2024Spglib:Search}, we then obtained a finite set of coset symmetry representatives with respect to the infinite group of translations for the full SG $G$, and restricted consideration to those that are also coset representatives of $G_{\bf k}$ using Eq.~(\ref{eq:KmodforGk}).
Of the coset representatives of $G_{\bf k}$ obtained from~\texttt{spglib}, a smaller subset consists only of unitary symmetries $O$, which hence carry eigenvalues that can be numerically obtained.
Specifically, each unitary coset representative $O\in G_{\bf k}$ admits a decomposition of the form~\cite{Seitz1936TheCrystals, Bradley1972TheSolids}:
\begin{equation}
O=\{R|t\},
\label{eq:Obreakdown}
\end{equation}
where $R$ is a point group operation and $t$ is a (possibly fractional) translation.
For the $m$ (approximately) degenerate Bloch states with energy $E_{n\mathbf{k}}$, the action of the unitary little group symmetry $O$ is represented by an $m\times m$ matrix $O^{\mathbf{k}}$:
\begin{equation}
O|\psi^{i}_{n\mathbf{k}}\rangle=\sum_{j=1}^{m}
\left[O^{\mathbf{k}}\right]_{ij}\,|\psi^{j}_{n\mathbf{k}}\rangle,
\end{equation}
where the matrix elements of $O^\mathbf{k}$ are hence given by:
\begin{equation}
\left[O^{\mathbf{k}}\right]_{ij}
=\langle\psi^{j}_{n\mathbf{k}}|\,O\,|\psi^{i}_{n\mathbf{k}}\rangle.
\end{equation}
In the plane-wave spinor basis of our DFT calculations, $O$ in Eq.~(\ref{eq:Obreakdown}) acts by transforming $\mathbf{k}$ vectors under $R$, multiplying by the phase factor $e^{-i(R\mathbf{k})\cdot t}$ from the combined action of $R$ and $t$, and by applying the SU(2) spin rotation that corresponds to the proper rotation piece of $R$.
For each $O$ and (approximate) band multiplet at $E_{n\mathbf{k}}$, we lastly obtain a character (trace) $\chi_{n\mathbf{k}}(O)$ by computing:
\begin{equation}
\chi_{n\mathbf{k}}(O)=\sum_{i=1}^{m}
\langle\psi^{i}_{n\mathbf{k}}|O|\psi^{i}_{n\mathbf{k}}\rangle
=\mathrm{Tr}\ O^{\mathbf{k}}.
\label{eq:traceChars}
\end{equation}
The characters $\chi_{n\mathbf{k}}(O)$ are equal to the sum of the symmetry eigenvalues of $O$ for the $m$ degenerate states at $E_{n\mathbf{k}}$, and across all maximal ${\bf k}$ points and band degeneracies form a set of symmetry data vectors that can be used to determine band connectivity and topology~\cite{Bradlyn2017TopologicalChemistry,Elcoro2021MagneticChemistry}.

In Table~\ref{tab:euccd2as2_irreps}, we reproduce the output from~\href{https://cryst.ehu.es/cgi-bin/cryst/programs/magnetictopo.pl?tipog=gesp}{\texttt{Check Topological Mat}} for the symmetry character trace file generated by our code (the tabulation of each $\chi_{n\mathbf{k}}(O)$ in Eq.~(\ref{eq:traceChars})) applied to the DFT calculation of the PM-state electronic structure of \ce{EuCd2As2} in Sec.~\ref{siDFT}.
The rows of Table~\ref{tab:euccd2as2_irreps} indicate groupings of connected bands above the core electronic shell, listed in the order realized by artificially, independently filling the Bloch states at each ${\bf k}$, rather than using a fixed, horizontal Fermi energy.
This allows energetically isolated bands and gaps to be classified as ``topological band insulators'' that, while metallic in the bulk, still exhibit topological boundary states~\cite{Wieder2018WallpaperInsulator,Wieder2022TopologicalSymmetry}.
In order, the columns in Table~\ref{tab:euccd2as2_irreps} list the irreducible small corepresentations associated to the characters $\chi_{n\mathbf{k}}(O)$ (Eq.~(\ref{eq:traceChars})) at each maximal momentum ${\bf k}$ in Fig.~\ref{fig:si_dft}a, the total number of connected bands in the grouping including spin (dim), the classification of the bands in the language of Topological Quantum Chemistry (type, \emph{e.g.} SEBR, see Ref.~\cite{Vergniory2022AllMaterials}), the symmetry indicators (SIs) $(Z_{2,3},Z_{4})$ in SG 164 ($P\bar{3}m1$) of the energetically isolated bands (top. ind., respectively the $k_{z}=\pi$ weak $\mathbb{Z}_{2}$ invariant and the strong $\mathbb{Z}_{4}$ invariant), the cumulative band dimension realized by individually occupying each Bloch state up to an electronic filling $\nu$ above the core shell (filling $\nu$), and the cumulative topology and SIs $(Z_{2,3},Z_{4})$ of the symmetry-allowed gap at $\nu$ (all type / top. ind.).
Most relevant to the present study of \ce{EuCd2As2}, $Z_{4}\text{ mod }2$ in Table~\ref{tab:euccd2as2_irreps} corresponds to the Fu-Kane $\mathbb{Z}_{2}$ invariant for strong 3D topological insulating (TI) states~\cite{Fu2007TopologicalSymmetry,Fu2007TopologicalDimensions,Hasan2010Colloquium:Insulators}, such that the rows with odd values of $Z_{4}$ in ``all type / top. ind.'' indicate electronic fillings at which topological surface Dirac cones may appear, modulo the effects of surface chemical potentials~\cite{Vergniory2022AllMaterials}.

\begin{table}[t]
\centering
\setlength{\tabcolsep}{4pt}
\renewcommand{\arraystretch}{1.1}
\scriptsize
\resizebox{17cm}{!}{%
\begin{tabular}{ccccccccccc}
\toprule
$\Gamma$ & A & H & K & L & M & dim & type & top.\ ind. & filling $\nu$ & all type / top.\ ind. \\
 & & & & & & & & $(Z_{2,3},Z_{4})$ & & \ \ \ \ \ \ \ \ \ \ \ \ \ $(Z_{2,3},Z_{4})$ \\
\midrule
$\bar{\Gamma}_{8}(2)$ & $\bar{A}_{8}(2)$ & $\bar{H}_{6}(2)$ & $\bar{K}_{6}(2)$ & $\bar{L}_{3}\bar{L}_{4}(2)$ & $\bar{M}_{3}\bar{M}_{4}(2)$ & 2 & TRIVIAL & $(0,0)$ & 2 & TRIVIAL / $(0,0)$ \\
$\bar{\Gamma}_{9}(2)$ & $\bar{A}_{9}(2)$ & $\bar{H}_{6}(2)$ & $\bar{K}_{6}(2)$ & $\bar{L}_{5}\bar{L}_{6}(2)$ & $\bar{M}_{5}\bar{M}_{6}(2)$ & 2 & TRIVIAL & $(0,0)$ & 4 & TRIVIAL / $(0,0)$ \\
$\bar{\Gamma}_{9}(2)$ & $\bar{A}_{9}(2)$ & $\bar{H}_{4}\bar{H}_{5}(2)$ & $\bar{K}_{4}\bar{K}_{5}(2)$ & $\bar{L}_{5}\bar{L}_{6}(2)$ & $\bar{M}_{5}\bar{M}_{6}(2)$ & 2 & FRAGILE & $(0,0)$ & 6 & TRIVIAL / $(0,0)$ \\
$\bar{\Gamma}_{6}\bar{\Gamma}_{7}(2)$ & $\bar{A}_{6}\bar{A}_{7}(2)$ & $\bar{H}_{6}(2)$ & $\bar{K}_{6}(2)$ & $\bar{L}_{5}\bar{L}_{6}(2)$ & $\bar{M}_{5}\bar{M}_{6}(2)$ & 2 & FRAGILE & $(0,0)$ & 8 & TRIVIAL / $(0,0)$ \\
$\bar{\Gamma}_{8}(2)$ & $\bar{A}_{8}(2)$ & $\bar{H}_{6}(2)$ & $\bar{K}_{6}(2)$ & $\bar{L}_{3}\bar{L}_{4}(2)$ & $\bar{M}_{3}\bar{M}_{4}(2)$ & 2 & TRIVIAL & $(0,0)$ & 10 & TRIVIAL / $(0,0)$ \\
$\bar{\Gamma}_{9}(2)$ & $\bar{A}_{9}(2)$ & $\bar{H}_{4}\bar{H}_{5}(2)$ & $\bar{K}_{4}\bar{K}_{5}(2)$ & $\bar{L}_{5}\bar{L}_{6}(2)$ & $\bar{M}_{5}\bar{M}_{6}(2)$ & 2 & FRAGILE & $(0,0)$ & 12 & TRIVIAL / $(0,0)$ \\
$\bar{\Gamma}_{4}\bar{\Gamma}_{5}(2)$ & $\bar{A}_{4}\bar{A}_{5}(2)$ & $\bar{H}_{6}(2)$ & $\bar{K}_{6}(2)$ & $\bar{L}_{5}\bar{L}_{6}(2)$ & $\bar{M}_{3}\bar{M}_{4}(2)$ & 2 & SEBR & $(1,3)$ & 14 & SEBR / $(1,3)$ \\
$\bar{\Gamma}_{6}\bar{\Gamma}_{7}(2)$ & $\bar{A}_{6}\bar{A}_{7}(2)$ & $\bar{H}_{6}(2)$ & $\bar{K}_{6}(2)$ & $\bar{L}_{3}\bar{L}_{4}(2)$ & $\bar{M}_{5}\bar{M}_{6}(2)$ & 2 & SEBR & $(1,1)$ & 16 & TRIVIAL / $(0,0)$ \\
$\bar{\Gamma}_{8}(2)$ & $\bar{A}_{9}(2)$ & $\bar{H}_{4}\bar{H}_{5}(2)$ & $\bar{K}_{4}\bar{K}_{5}(2)$ & $\bar{L}_{3}\bar{L}_{4}(2)$ & $\bar{M}_{5}\bar{M}_{6}(2)$ & 2 & SEBR & $(1,0)$ & 18 & SEBR / $(1,0)$ \\
$\bar{\Gamma}_{9}(2)$ & $\bar{A}_{8}(2)$ & $\bar{H}_{6}(2)$ & $\bar{K}_{6}(2)$ & $\bar{L}_{5}\bar{L}_{6}(2)$ & $\bar{M}_{3}\bar{M}_{4}(2)$ & 2 & SEBR & $(1,0)$ & 20 & TRIVIAL / $(0,0)$ \\
$\bar{\Gamma}_{8}(2)$ & $\bar{A}_{8}(2)$ & $\bar{H}_{6}(2)$ & $\bar{K}_{6}(2)$ & $\bar{L}_{5}\bar{L}_{6}(2)$ & $\bar{M}_{3}\bar{M}_{4}(2)$ & 2 & SEBR & $(1,3)$ & 22 & SEBR / $(1,3)$ \\
$\bar{\Gamma}_{9}(2)$ & $\bar{A}_{9}(2)$ & $\bar{H}_{6}(2)$ & $\bar{K}_{6}(2)$ & $\bar{L}_{3}\bar{L}_{4}(2)$ & $\bar{M}_{5}\bar{M}_{6}(2)$ & 2 & SEBR & $(1,1)$ & 24 & TRIVIAL / $(0,0)$ \\
$\bar{\Gamma}_{4}\bar{\Gamma}_{5}(2)$ & $\bar{A}_{6}\bar{A}_{7}(2)$ & $\bar{H}_{4}\bar{H}_{5}(2)$ & $\bar{K}_{4}\bar{K}_{5}(2)$ & $\bar{L}_{3}\bar{L}_{4}(2)$ & $\bar{M}_{5}\bar{M}_{6}(2)$ & 2 & SEBR & $(1,0)$ & 26 & SEBR / $(1,0)$ \\
$\bar{\Gamma}_{8}(2)$ & $\bar{A}_{9}(2)$ & $\bar{H}_{6}(2)$ & $\bar{K}_{6}(2)$ & $\bar{L}_{5}\bar{L}_{6}(2)$ & $\bar{M}_{3}\bar{M}_{4}(2)$ & 2 & TRIVIAL & $(0,0)$ & 28 & SEBR / $(1,0)$ \\
$\bar{\Gamma}_{6}\bar{\Gamma}_{7}(2)$ & $\bar{A}_{4}\bar{A}_{5}(2)$ & $\bar{H}_{6}(2)$ & $\bar{K}_{6}(2)$ & $\bar{L}_{5}\bar{L}_{6}(2)$ & $\bar{M}_{3}\bar{M}_{4}(2)$ & 2 & SEBR & $(1,0)$ & 30 & TRIVIAL / $(0,0)$ \\
$\bar{\Gamma}_{9}(2)$ & $\bar{A}_{8}(2)$ & $\bar{H}_{4}\bar{H}_{5}(2)$ & $\bar{K}_{4}\bar{K}_{5}(2)$ & $\bar{L}_{3}\bar{L}_{4}(2)$ & $\bar{M}_{5}\bar{M}_{6}(2)$ & 2 & FRAGILE & $(0,0)$ & 32 & TRIVIAL / $(0,0)$ \\
$\bar{\Gamma}_{8}(2)$ & $\bar{A}_{9}(2)$ & $\bar{H}_{4}\bar{H}_{5}(2)$ & $\bar{K}_{4}\bar{K}_{5}(2)$ & $\bar{L}_{3}\bar{L}_{4}(2)$ & $\bar{M}_{3}\bar{M}_{4}(2)$ & 2 & SEBR & $(1,1)$ & 34 & SEBR / $(1,1)$ \\
$\bar{\Gamma}_{8}(2)$ & $\bar{A}_{8}(2)$ & $\bar{H}_{6}(2)$ & $\bar{K}_{6}(2)$ & $\bar{L}_{5}\bar{L}_{6}(2)$ & $\bar{M}_{5}\bar{M}_{6}(2)$ & 2 & SEBR & $(1,2)$ & 36 & SEBR / $(0,3)$ \\
\makecell{$\bar{\Gamma}_{4}\bar{\Gamma}_{5}(2)$\\$\bar{\Gamma}_{9}(2)$} &
\makecell{$\bar{A}_{8}(2)$\\$\bar{A}_{4}\bar{A}_{5}(2)$} &
\makecell{$\bar{H}_{6}(2)$\\$\bar{H}_{6}(2)$} &
\makecell{$\bar{K}_{6}(2)$\\$\bar{K}_{6}(2)$} &
\makecell{$\bar{L}_{3}\bar{L}_{4}(2)$\\$\bar{L}_{3}\bar{L}_{4}(2)$} &
\makecell{$\bar{M}_{5}\bar{M}_{6}(2)$\\$\bar{M}_{3}\bar{M}_{4}(2)$} &
4 & FRAGILE & $(0,0)$ & 40 & SEBR / $(0,3)$ \\
$\bar{\Gamma}_{9}(2)$ & $\bar{A}_{9}(2)$ & $\bar{H}_{4}\bar{H}_{5}(2)$ & $\bar{K}_{4}\bar{K}_{5}(2)$ & $\bar{L}_{5}\bar{L}_{6}(2)$ & $\bar{M}_{5}\bar{M}_{6}(2)$ & 2 & FRAGILE & $(0,0)$ & 42 & SEBR / $(0,3)$ \\
$\bar{\Gamma}_{6}\bar{\Gamma}_{7}(2)$ & $\bar{A}_{6}\bar{A}_{7}(2)$ & $\bar{H}_{6}(2)$ & $\bar{K}_{6}(2)$ & $\bar{L}_{5}\bar{L}_{6}(2)$ & $\bar{M}_{3}\bar{M}_{4}(2)$ & 2 & SEBR & $(0,1)$ & 44 & TRIVIAL / $(0,0)$ \\
$\bar{\Gamma}_{8}(2)$ & $\bar{A}_{9}(2)$ & $\bar{H}_{6}(2)$ & $\bar{K}_{6}(2)$ & $\bar{L}_{5}\bar{L}_{6}(2)$ & $\bar{M}_{5}\bar{M}_{6}(2)$ & 2 & SEBR & $(0,3)$ & 46 & SEBR / $(0,3)$ \\
$\bar{\Gamma}_{9}(2)$ & $\bar{A}_{9}(2)$ & $\bar{H}_{4}\bar{H}_{5}(2)$ & $\bar{K}_{4}\bar{K}_{5}(2)$ & $\bar{L}_{5}\bar{L}_{6}(2)$ & $\bar{M}_{5}\bar{M}_{6}(2)$ & 2 & FRAGILE & $(0,0)$ & 48 & SEBR / $(0,3)$ \\
\makecell{$\bar{\Gamma}_{6}\bar{\Gamma}_{7}(2)$\\$\bar{\Gamma}_{9}(2)$} &
\makecell{$\bar{A}_{8}(2)$\\$\bar{A}_{6}\bar{A}_{7}(2)$} &
\makecell{$\bar{H}_{6}(2)$\\$\bar{H}_{6}(2)$} &
\makecell{$\bar{K}_{6}(2)$\\$\bar{K}_{6}(2)$} &
\makecell{$\bar{L}_{5}\bar{L}_{6}(2)$\\$\bar{L}_{5}\bar{L}_{6}(2)$} &
\makecell{$\bar{M}_{5}\bar{M}_{6}(2)$\\$\bar{M}_{3}\bar{M}_{4}(2)$} &
4 & SEBR & $(1,0)$ & 52 & SEBR / $(1,3)$ \\
\multicolumn{11}{c}{\dashfill}\\
$\bar{\Gamma}_{9}(2)$ & $\bar{A}_{9}(2)$ & $\bar{H}_{4}\bar{H}_{5}(2)$ & $\bar{K}_{4}\bar{K}_{5}(2)$ & $\bar{L}_{5}\bar{L}_{6}(2)$ & $\bar{M}_{5}\bar{M}_{6}(2)$ & 2 & FRAGILE & $(0,0)$ & 54 & SEBR / $(1,3)$ \\
$\bar{\Gamma}_{9}(2)$ & $\bar{A}_{9}(2)$ & $\bar{H}_{6}(2)$ & $\bar{K}_{6}(2)$ & $\bar{L}_{5}\bar{L}_{6}(2)$ & $\bar{M}_{5}\bar{M}_{6}(2)$ & 2 & TRIVIAL & $(0,0)$ & 56 & SEBR / $(1,3)$ \\
$\bar{\Gamma}_{6}\bar{\Gamma}_{7}(2)$ & $\bar{A}_{6}\bar{A}_{7}(2)$ & $\bar{H}_{6}(2)$ & $\bar{K}_{6}(2)$ & $\bar{L}_{5}\bar{L}_{6}(2)$ & $\bar{M}_{5}\bar{M}_{6}(2)$ & 2 & FRAGILE & $(0,0)$ & 58 & SEBR / $(1,3)$ \\
$\bar{\Gamma}_{9}(2)$ & $\bar{A}_{9}(2)$ & $\bar{H}_{6}(2)$ & $\bar{K}_{6}(2)$ & $\bar{L}_{3}\bar{L}_{4}(2)$ & $\bar{M}_{5}\bar{M}_{6}(2)$ & 2 & SEBR & $(1,1)$ & 60 & TRIVIAL / $(0,0)$ \\
\bottomrule
\end{tabular}
}
\caption{{\bf Symmetry data and topological classification from~\href{https://cryst.ehu.es/cgi-bin/cryst/programs/magnetictopo.pl?tipog=gesp}{\texttt{Check Topological Mat}} for the PM state of \ce{EuCd2As2}.}
The output obtained from the~\href{https://cryst.ehu.es/cgi-bin/cryst/programs/magnetictopo.pl?tipog=gesp}{\texttt{Check Topological Mat}} tool on the Bilbao Crystallographic Server~\cite{Aroyo2006BilbaoPrograms,Aroyo2006BilbaoGroups,Aroyo2011CrystallographyServer,Vergniory2019AMaterials,Vergniory2022AllMaterials} for the symmetry characters $\chi_{n\mathbf{k}}(O)$ (trace file, see Eq.~(\ref{eq:traceChars}) and the surrounding text) from our DFT calculation of the electronic structure of \ce{EuCd2As2} in the PM state (SG 164 ($P\bar{3}m1$), see Sec.~\ref{siDFT}).
The rows indicate groupings of connected bands above the core shell, listed in the order realized by independently filling the Bloch states at each ${\bf k}$.
In order, the columns list the irreducible small corepresentations associated to the little group symmetry characters $\chi_{n\mathbf{k}}(O)$ (Eq.~(\ref{eq:traceChars})) at each maximal momentum ${\bf k}$ in Fig.~\ref{fig:si_dft}a, the total number of connected bands in the grouping including spin (dim), the classification of the bands in the language of Topological Quantum Chemistry (type, \emph{e.g.} SEBR, see Ref.~\cite{Vergniory2022AllMaterials}), the symmetry-based indicators (SIs) of stable band topology $(Z_{2,3},Z_{4})$ in SG 164 ($P\bar{3}m1$) for the energetically isolated bands (top. ind., respectively the $k_{z}=\pi$ weak $\mathbb{Z}_{2}$ invariant and the strong $\mathbb{Z}_{4}$ invariant), the cumulative band dimension realized by individually occupying each Bloch state up to an electronic filling $\nu$ above the core shell (filling $\nu$), and the cumulative topology and SIs $(Z_{2,3},Z_{4})$ of the symmetry-allowed gap at $\nu$ (all type / top. ind.).
The columns ``filling $\nu$'' and ``all type / top. ind.'' are hence additive, and are obtained by respectively summing the data in the ``dim'' and ``top. ind.'' columns (including modulo operations for the cumulative SIs).
Counting above the core shell, \ce{EuCd2As2} has $\nu_{F}=51$ valence electrons (Eq.~(\ref{eq:IntrinsicFillingTake2})), and is hence a $\mathcal{T}$-enforced metal in the PM state~\cite{Watanabe2015FillingCrystals,Wieder2016Spin-orbitGroups} (ESFD in the nomenclature of Refs.~\cite{Vergniory2019AMaterials,Vergniory2022AllMaterials}).
The first band gap allowed by symmetry above $\nu_{F}$ occurs at $\nu=\nu_{F}+1=52$ (dashed horizontal line).
The gap at $\nu=52$, as well as all of the symmetry-allowed gaps from $\nu=\nu_{F}-5=46$ to $\nu=\nu_{F} + 7=58$, exhibit the stable topology of a 3D topological insulator (TI), as indicated by their nontrivial Fu-Kane $\mathbb{Z}_{2}$ invariants $Z_{4}\text{ mod }2=1$~\cite{Fu2007TopologicalSymmetry,Fu2007TopologicalDimensions,Hasan2010Colloquium:Insulators}.
We may hence associate the unoccupied $(001)$-surface Dirac fermion in our DFT calculations of \ce{EuCd2As2} in Figs.~\ref{fig:si_dft}b,~\ref{fig:si_DFT_ARPES_dispersions}a, and~\ref{fig:sxzconstenergy} to 3D TI gaps away from $E_{F}$, with the most likely candidates being the unoccupied topological band-insulating gaps at $\nu_{F}<\nu \leq 58$ (see the text following Eq.~(\ref{eq:IntrinsicFillingTake2})).}
\label{tab:euccd2as2_irreps}
\end{table}

In the PM state, \ce{EuCd2As2} carries an intrinsic (valence) electronic filling above the core shell $\nu_{F}$ of:
\begin{equation}
\nu_{F} = 51.
\label{eq:IntrinsicFillingTake2}
\end{equation}
Owing to the requirement from $\mathcal{T}$ symmetry that noninteracting, nonmagnetic insulators have even numbers of occupied bands~\cite{Watanabe2015FillingCrystals,Wieder2016Spin-orbitGroups}, \ce{EuCd2As2} in the PM state is therefore, in the language of Ref.~\cite{Vergniory2022AllMaterials}, an enforced semimetal with a Fermi degeneracy (ESFD) at $\nu=\nu_{F}$, and not a band insulator.
The first gap in the PM state of \ce{EuCd2As2} allowed by symmetry instead occurs at a filling of $\nu=\nu_{F}+1=52$ electrons, as indicated by the dashed horizontal line in Table~\ref{tab:euccd2as2_irreps}.
Remarkably, not only does the gap at $\nu=52$ in \ce{EuCd2As2} carry the cumulative topology of a 3D TI state ($Z_{4}=3$ in ``all type / top. ind.''), but in fact \emph{all} of the symmetry-allowed band gaps realize 3D TI states over a range of $\Delta \nu = 12$ electrons spanning from $\nu=\nu_{F}-5=46$ to $\nu=\nu_{F} + 7=58$.
Hence, though surface chemical potentials may still shift topological surface Dirac cones away from the average energy of nontrivial band-insulating gaps (see the text following Eq.~(\ref{eq:Asurf_eigen})), the huge filling range $\Delta\nu = 12$ of nontrivial 3D TI gaps surrounding $\nu_{F}$ in Table~\ref{tab:euccd2as2_irreps} is much larger than the scale of chemical potential effects, implying that the unoccupied $(001)$-surface Dirac cone in our DFT calculations (Figs.~\ref{fig:si_dft}b,~\ref{fig:si_DFT_ARPES_dispersions}a, and~\ref{fig:sxzconstenergy}) can be associated to a band-insulating 3D TI phase with $\nu\neq \nu_{F}$.
Even further restricting to fillings above the Fermi energy $\nu>\nu_{F}$, it is also unlikely that surface chemical potential effects can shift a Dirac cone by energies equivalent to changing $\nu$ by $7$ electrons, and we may hence more specifically reasonably associate the $(001)$-surface Dirac cone in Figs.~\ref{fig:si_dft}b,~\ref{fig:si_DFT_ARPES_dispersions}a, and~\ref{fig:sxzconstenergy} to the smaller set of unoccupied 3D TI gaps with $\nu_{F}<\nu \leq 58$ in Table~\ref{tab:euccd2as2_irreps}.

In this sense, \ce{EuCd2As2} is reminiscent of the repeat-topological (RTopo) materials introduced in Ref.~\cite{Vergniory2022AllMaterials}, especially the example of Bi$_2$Mg$_3$, which respects the same symmetry group SG 164 ($P\bar{3}m1$) as \ce{EuCd2As2} in the PM state.
RTopo materials were specifically defined in Ref.~\cite{Vergniory2022AllMaterials} as materials that are stable topological band insulators at intrinsic filling $\nu_{F}$, but also host cumulative stable topological band-insulating states at the first band gap below $\nu_{F}$ allowed by symmetry.
In Bi$_2$Mg$_3$, the RTopo classification gives rise to two sets of $(001)$-surface topological Dirac cones, which at intrinsic filling respectively manifest as surface states in the projected gap just above $E_{F}$, and as surface resonance bands below $E_{F}$ that coexist with the surface projections of the bulk valence manifold and are associated to the topological band gap at $\nu<\nu_{F}$~\cite{Vergniory2022AllMaterials,Chang2019RealizationMg3Bi2,Zhou2019ExperimentalEpitaxy}.
If \ce{EuCd2As2} in the PM phase were doped to $\nu=\nu_{F}+1=52$ (the dashed line in Table~\ref{tab:euccd2as2_irreps}), it would also similarly be classified as RTopo.
The unoccupied surface Dirac cone in \ce{EuCd2As2} (Figs.~\ref{fig:si_dft}b,~\ref{fig:si_DFT_ARPES_dispersions}a, and~\ref{fig:sxzconstenergy}) and the associated topological band gaps at $\nu>\nu_{F}$ in Table~\ref{tab:euccd2as2_irreps} can hence be viewed as a generalization of the concept of repeat-topology -- originally introduced to predict topological boundary states accessible in equilibrium ARPES experiments at intrinsic filling -- to topological surface conduction bands that are accessible by doping and pump-probe ARPES experiments like those in the present study.

\newpage

\section{Transient non-equilibrium distributions from transitions induced by \qty{1.2}{eV} and \qty{1.5}{eV} optical pulses}\label{sec:si_rise}
\begin{figure}[H]
    \includegraphics[width=1\linewidth]{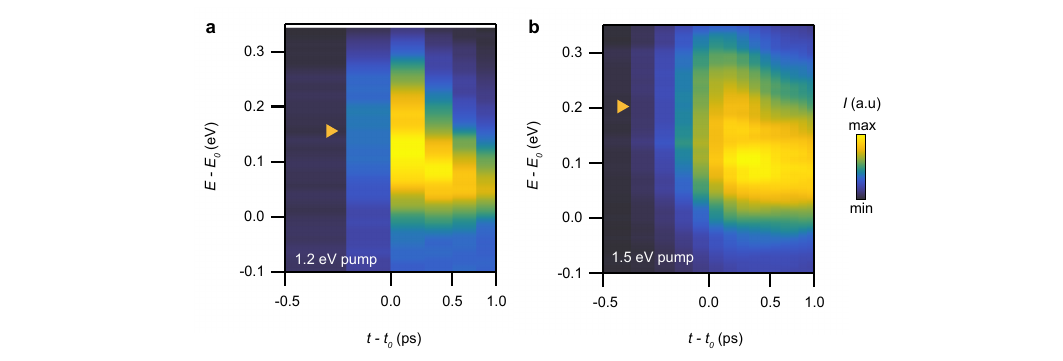}
    \caption{\label{fig:si_rise}\textbf{Rapid intra-band scattering immediately after photo-excitation: }\textbf{(a)} Time trace taken using \qty{1.2}{eV} pump photons with low fluence (\qty{15}{\micro J/cm^2}) at $T=20\,\si{K}$. \textbf{(b)} Time trace taken using \qty{1.5}{eV} pump photons with low fluence (\qty{25}{\micro J/cm^2}) at $T=9.5\,\si{K}$. To improve visualization at initial delays, horizontal axes are scaled logarithmically, using a rigid offset to account for negative values. 
    The orange arrows are guides to the eye indicating the approximate locations of the initial electronic occupation.}
\end{figure}

To confirm that the unoccupied surface states are populated through direct inter-band transitions from photo-excitation by the IR pump, and to establish a timescale for the rapid intra-band scattering by which electrons reach quasi-thermal non-equilibrium distributions in the unoccupied bands, we in this section plot EDCs at initial times over logarithmically spaced temporal delays (Fig.~\ref{fig:si_rise}).
Although the pump pulses (and consequent spread of the initial transient population) are broadband due to their short temporal duration, faint localizations immediately before $t_0$ appear slightly localized at $E-E_0\approx150\,\si{meV}$ and $E-E_0\approx200\,\si{meV}$ for the cases of \qty{1.2}{eV} and \qty{1.5}{eV} photon excitation energies, respectively (panels a and b of Fig.~\ref{fig:si_rise}).

Importantly, the \qty{50}{meV} difference between the locations of the faint localizations (orange arrows) in Fig.~\ref{fig:si_rise}a,b is significantly smaller than the difference between the excitation photon energies, owing to the high velocity of the linearly dispersing bulk band below $E_F$ from which these electrons originate (see Fig.~\ref{fig:overview}b of the main text).
These transient non-equilibrium distributions rapidly reorganize via electron-electron scattering, and by $t_0$, the resulting distribution becomes well-fit by a density of states multiplied by a Fermi-Dirac distribution (see Ref.~\cite{Gierz2013SnapshotsGraphene} and Fig.~\ref{fig:si_muTwaterfall} below). 

\newpage
\section{Estimating the position of the nondispersing bulk conduction band}
\label{bulk_gap_SI}

\begin{figure}[H]

\includegraphics[width=\textwidth]{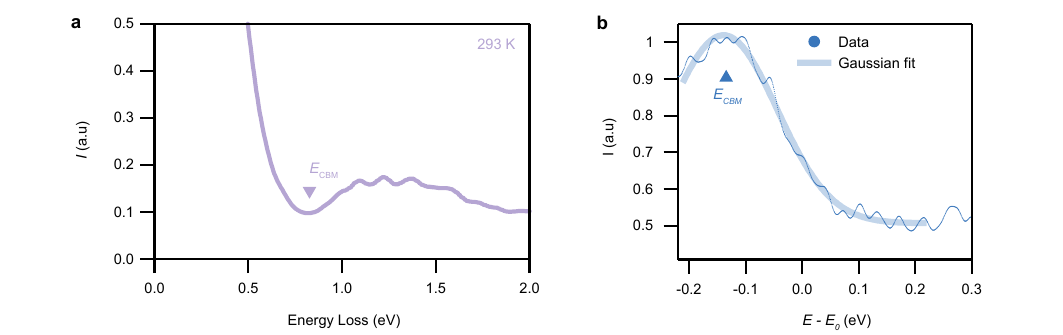}
\caption{\textbf{Estimating the position of the bulk conduction band:}
\textbf{(a)} Momentum-integrated spectrum obtained from electron energy loss spectroscopy (EELS), taken in the paramagnetic state of \ce{EuCd2As2} at \qty{293}{K}. 
The local minimum near $E-E_F\approx\qty{0.7}{eV}$ that gives way to a broad continuum corresponds to the inter-band transition edge. 
The plasmon feature (Fig.~\ref{fig:electron-plasmon_coupling}b of the main text) lies above the vertical range of this plot.
\textbf{(b)} Momentum-integrated Tr-ARPES spectrum (measured relative to $E_0\approx E_F + \qty{0.8}{eV}$, the bottom of the unoccupied surface band), integrated between \qty{7}{ps} and \qty{10}{ps} after photo-excitation by \qty{1.5}{eV} photons of low fluence (\qty{15}{\micro J/cm^2}). We superimpose a single-Gaussian fit, which shows a center value at approximately $E_0-\qty{0.13}{eV}$.}
\label{fig:si_HE-EELS}
\end{figure}

In this section, we use electron energy loss spectroscopy (EELS) measurements to confirm the energy above the Fermi level $E_{F}$ of the bulk conduction band that is nondispersing to leading order at ${\bf k}={\bf 0}$ (the bulk $\Gamma$ point, see Fig.~\ref{fig:si_dft}a).
First, in Fig.~\ref{fig:electron-plasmon_coupling}b of the main text, we plot our EELS measurements at low energies, which show a strong peak at approximately \qty{0.12}{eV} that we attribute to a plasmon. 
Here, in Fig.~\ref{fig:si_HE-EELS}a, we plot EELS measurements at higher energies taken on the \ce{EuCd2As2} samples used in this study.
The higher-energy EELS measurements in Fig.~\ref{fig:si_HE-EELS}a show a broad continuum of energy loss above $E-E_F\approx\qty{0.7}{eV}$, indicative of either a bulk or surface band minimum at that energy. 
Consistent with this observation, our first-principles electronic band structure calculations in Sec.~\ref{theory_SI} also show a weakly dispersing bulk conduction band at $\Gamma$ with approximately the same energy $E_{CBM}$ (Fig.~\ref{fig:si_dft}a).
Furthermore, the value of $E_{CBM}\approx E_F + \qty{0.7}{eV}$ obtained from our EELS measurements is also in agreement with previous optical spectroscopy measurements on samples characterized as semimetallic \cite{Wu2024TheSingularity}, which place a conduction band minimum between \qty{0.7}{eV} and \qty{0.8}{eV} relative to the maximum of the hole-like valence band.
Optical spectroscopy and Tr-ARPES studies of semiconducting samples \cite{Santos-Cottin2023EuCd2As2:Semiconductor} similarly place a local conduction band minimum at \qty{0.8}{eV}.

Next, we additionally analyze our Tr-ARPES data to confirm the position of this bulk conduction band relative to $E_0$, the minimum of the unoccupied surface states.
In Fig.~\ref{fig:si_HE-EELS}b, we show an EDC integrated between \qty{7}{ps} and \qty{10}{ps} after photo-excitation of \ce{EuCd2As2} samples by \qty{1.5}{eV} photons at low fluence.
We then fit the TR-ARPES data in Fig.~\ref{fig:si_HE-EELS}b to a Gaussian profile, finding a center value of approximately \qty{-0.13}{eV}.
This is consistent with our analysis of our measured Tr-ARPES time traces (Sec.~\ref{RateEq_SI}, Fig.~\ref{fig:si_decay_vs_energy} below), in which we assign the bottom of the bulk conduction band at $E_{CBM}=E_0 - \qty{0.12}{eV}$ as the energy level with the longest electron population lifetimes.

\newpage

\section{Extracting and fitting transient electronic temperatures and dispersions from Tr-ARPES data at low and intermediate excitation fluence}
\label{fitting_SI}

\begin{figure}[H]
\includegraphics{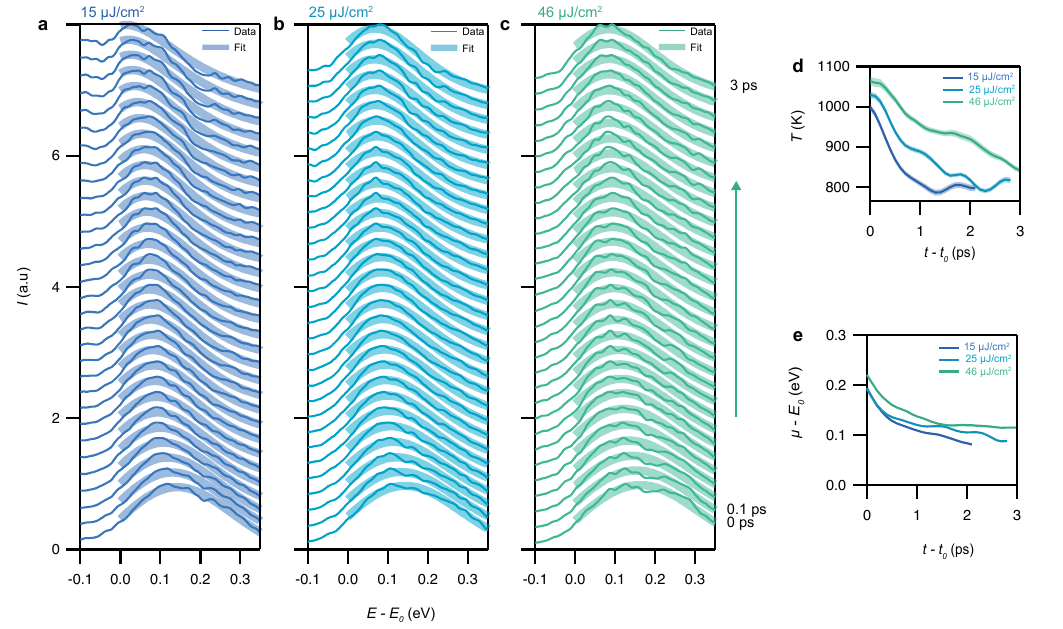}
\caption{\textbf{Waterfall plots for extracting time-dependent chemical potentials and temperatures from ARPES EDCs at low and intermediate excitation fluence:} \textbf{(a-c)} Waterfall plots showing EDCs for the unoccupied surface states and corresponding fitted Fermi-Dirac distributions (Eq.~(\ref{eq:ARPESintensity})) at time delays between \qty{0}{ps} and \qty{3}{ps} and energies measured relative to the bottom of the unoccupied surface state at $E_0\approx E_F+\qty{0.8}{eV}$. 
\textbf{(d, e)} Temperature ($T$) and chemical potential ($\mu$) parameters extracted from the data in (a-c) as functions of time. 
The error bars for $T$ and $\mu$ (respectively the shaded regions in (d, e)) represent a 68\% confidence interval for the fits.
In (e), the error bars are vanishingly small.
The data and fits in (a-e) were taken at low excitation fluence (below \qty{46}{\micro J/cm^2}) or at intermediate fluence (\qty{46}{\micro J/cm^2}, see Sec.~\ref{sec:Clarity}).}
\label{fig:si_muTwaterfall}
\end{figure}

In our data, we observe that for low fluence excitation (below \qty{46}{\micro J/cm^2}, see Sec.~\ref{sec:Clarity}), the momentum-integrated intensity of the measured ARPES spectra are well-approximated at all times by a linear density of states multiplied by a Fermi-Dirac distribution, convolved with a fixed Gaussian energy resolution: 
\begin{equation}
I(E,t)=A_0(t)(E-\alpha)\frac{1}{\mathrm{e}^{(E-\mu(t))/k_BT(t)}+1}*\text{Gauss}(E, \sigma=\qty{10}{meV}).
\label{eq:ARPESintensity}
\end{equation}
In Eq.~(\ref{eq:ARPESintensity}), the parameter $\alpha$ is chosen to permit the presence of a nonzero density of states at the bottom of the unoccupied surface state $E_0\approx E_F+\qty{0.8}{eV}$, which we attribute to a rigid background due to either (or both) scattering from impurities and final state effects. These artifacts are independent of pump fluence and time after photo-excitation, and we therefore in our fits hold $\alpha$ constant at \qty{0.05}{eV}.
We have also in Eq.~(\ref{eq:ARPESintensity}) allowed the prefactor $A_{0}(t)$ to carry an explicit time dependence, because $A_{0}(t)$ scales with the total particle number in the unoccupied states, whose conservation is broken by significant radiative and impurity-mediated decay of electrons towards the equilibrium Fermi level.
In Fig.~\ref{fig:si_muTwaterfall}, we show the results of fitting ARPES EDCs at different times to the function $I(E,t)$ in Eq.~(\ref{eq:ARPESintensity}).

Importantly, in modeling data taken at high excitation fluences (above \qty{46}{\micro J/cm^2}), we no longer employ Eq.~(\ref{eq:ARPESintensity}) over energy ranges near $E_X$, since Eq.~(\ref{eq:ARPESintensity}) does not account for an excitonic population.
Specifically, as described in the main text, we observe that under stronger excitation, the electronic distribution sharpens \qty{100}{meV} above the bottom of the unoccupied surface state $E_{0}$ at increasing delay time after photo-excitation, which we attribute to the formation of excitons. 
In the presence of excitons, the model of Eq.~(\ref{eq:ARPESintensity}) is no longer a reasonable physical description at energies near $E_X$, because the excitons that we resolve in that region (specifically the electrons resulting from breaking excitons with the UV probe) must be ascribed their own statistical distribution.
Further analysis of the high-fluence case and modeling of exciton formation is provided below in Sec.~\ref{RateEq_SI}.

\begin{figure}[t]
\includegraphics{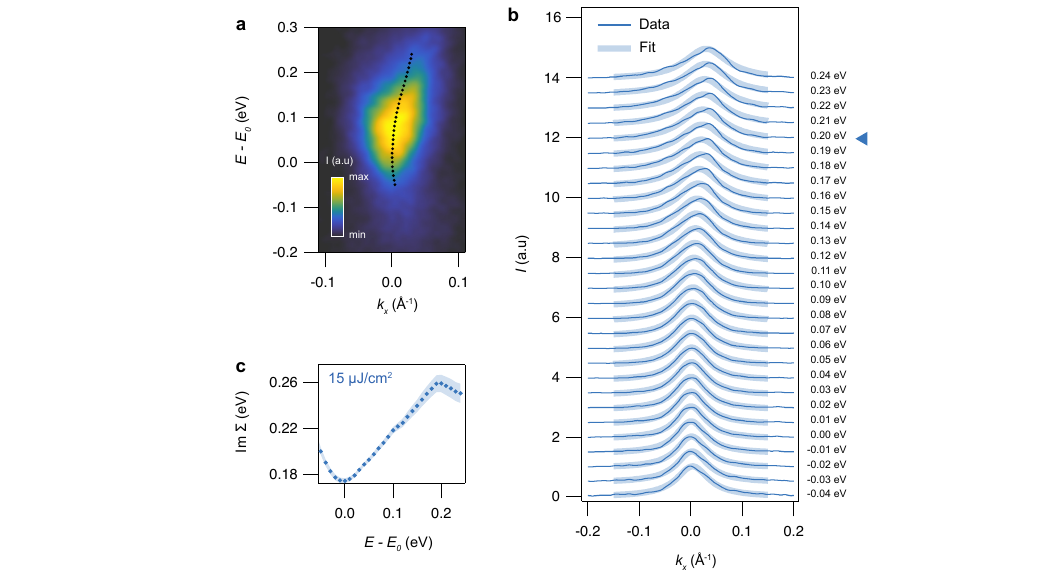}
\caption{\textbf{Waterfall plots for extracting dispersion from Lorentzian fits for low fluence excitation:} \textbf{(a)} Time-integrated ARPES spectrum for $t\in[\qty{0}{ps}, \qty{2}{ps}]$ for low fluence excitation (\qty{15}{\micro J/cm^2}) at \qty{1.5}{eV}. The measurements in (a) were performed at a temperature of \qty{9.5}{K}. 
The black line in (a) tracks the centers of Lorentzian fits (Eq.~(\ref{eq:singleParticleA})) of the momentum distribution curves (MDCs). 
\textbf{(b)} Normalized MDCs at several energy windows, with single Lorentzian fits superimposed. 
\textbf{(c)} Imaginary self energy $\text{Im }\Sigma(E)$ extracted from the half width at half maximum (HWHM) of the Lorentzian fits in (b). 
The error bars (shaded regions) in (b, c) enclose a 95\% confidence interval for the fits.}
\label{fig:si_dispersion_waterfall}
\end{figure}

Returning to the analysis of our ARPES data at low fluence excitation, in a one-band quasi-equilibrium system, the ARPES intensity can be written as \cite{Boschini2024Time-resolvedMaterials, Damascelli2003Angle-resolvedSuperconductors}: 
\begin{equation}
I(k,E)=A(k,E)|M(k,E)|^2f(E),
\end{equation}
where $f(E)$ is the Fermi-Dirac distribution, $M(k,E)$ is a matrix element for the excitation of a bound electron to a plane-wave state, and $A(k,E)$ is the single-particle spectral function, given by: 
\begin{equation}
A(k,E)=-\frac{1}{\pi}\frac{\Sigma''(k,E)}{\left[E-\epsilon(k)-\Sigma'(k,E)\right]^2+\left[\Sigma''(k,E)\right]^2}.
\label{eq:singleParticleA}
\end{equation}
In Eq.~(\ref{eq:singleParticleA}), $\epsilon(k)$ is a single-particle energy eigenvalue, and $\Sigma'(k,E)=\text{Re }\Sigma$ and $\Sigma''(k,E)=\text{Im }\Sigma$ are respectively the real and imaginary parts of the self energy, which are connected through the Kramers-Kronig relations.
If we assume $\Sigma''$ to be independent of $k$ (that is, scattering mechanisms encoded by $\Sigma''$ are assumed to vary slowly with momentum), cuts of $A(k,E)$ taken at constant $E$ have a Lorentzian profile in $k$ centered at $\epsilon(k)+\Sigma'(E)$ with a half width at half maximum (HWHM) proportional to $\Sigma''$.

Under the assumption that $\text{Im }\Sigma$ is $k$-independent, we next use Eq.~(\ref{eq:singleParticleA}) to track the dispersion and imaginary self energy of the unoccupied surface states via Lorentzian fits of the MDCs measured at low excitation fluence. 
Our ARPES spectra were prepared by integrating over a short window starting at $t=t_0$ to account for rapid intra-band scattering and to allow significant spectral weight at all energy levels.
A representative example ARPES spectrum from our data is shown in Fig.~\ref{fig:si_dispersion_waterfall}a.
Since we do not observe band renormalization or other time-dependent photovoltage shifts in our ARPES measurements (see Sec.~\ref{charging_SI} and Fig.~\ref{fig:fluence_dependence}a of the main text), the results of this fitting process are independent of our choice of the integration duration. 
In Fig.~\ref{fig:si_dispersion_waterfall}b, we show the corresponding waterfall plots from the ARPES data in Fig.~\ref{fig:si_dispersion_waterfall}a.
The fits in Fig.~\ref{fig:si_dispersion_waterfall}b show good agreement with individual cuts, although the raw MDCs exhibit some observable asymmetry -- particularly at high energies -- resulting in small deviations from the fits.
Though not shown above, we further find that the deviations from the fits persist if we instead employ a two-Lorentzian model (with the restriction from symmetry that the two peaks of the Lorentzians be at $\pm k_{\text{fit}}$).
From this, we deduce that the asymmetry in the raw MDCs in Fig.~\ref{fig:si_dispersion_waterfall}b can be attributed to $k$-dependent terms in the imaginary part of the self energy $\text{Im } \Sigma$, which we previously neglected for simplicity.

Nevertheless, because our single-Lorentzian fits with $k$-independent $\text{Im } \Sigma$ exhibit significantly higher numerical stability, carry a reduced number of free parameters, and have a lower reliance on extremely precise initial guess values compared to two-Lorentzian fits or more complicated models with $k$-dependent $\Sigma$, we focus our analysis in the main text only on results from single-Lorentzian fits with $k$-independent $\text{Im } \Sigma$ (Fig.~\ref{fig:electron-plasmon_coupling}a of the main text, the raw spectrum and waterfall plot for which are shown in Fig.~\ref{fig:si_dispersion_waterfall}a,b).
In Fig.~\ref{fig:si_dispersion_waterfall}c, we lastly show the $k$-independent imaginary self energy $\text{Im } \Sigma$ extracted from our single-Lorentzian fits.

\newpage
\section{Absence of time-dependent photovoltage shifts and conduction band renormalization}
\label{charging_SI}

\begin{figure}[H]
\centering
\includegraphics{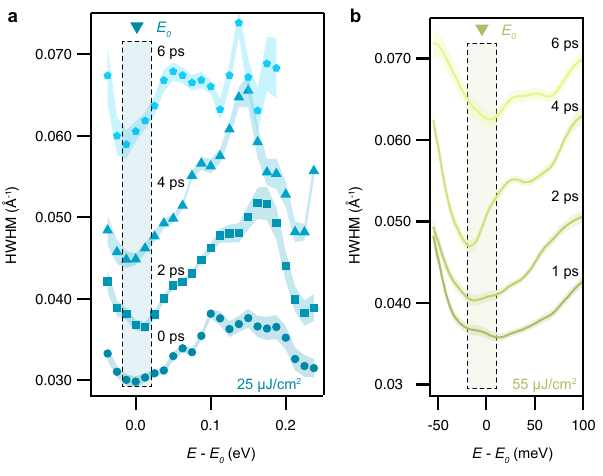}
\caption{\textbf{Absence of photovoltage effects and band gap renormalization: } Plot of the HWHM of MDCs at energies relative to the bottom of the surface spectral feature $E_0\approx E_F+\qty{0.8}{eV}$ for various delay times after excitation by \qty{1.5}{eV} photons at fluences of \textbf{(a)} \qty{25}{\micro J/cm^2} and \textbf{(b)} \qty{55}{\micro J/cm^2}.
Rigid vertical offsets have been added for clarity. 
Across all delay times, we observe no significant shifts in the energies of the local minima of the HWHM (indicated by the dashed rectangular regions in (a, b), which are guides to the eye) relative to the bottom of the surface band at $E_{0}$ for either fluence. 
The fitting process used to extract the HWHM in this figure is identical to that previously used to extract the imaginary self energy $\text{Im }\Sigma(E)$ from ARPES cuts in Sec.~\ref{fitting_SI} (see Fig.~\ref{fig:si_dispersion_waterfall}c and the surrounding text).
The shaded regions in (a, b) are error bars, and represent a 95\% confidence interval for the fits.}
\label{fig:si_charging}
\end{figure}

In wide-gap semiconductors, it is well established that Tr-ARPES measurements can be affected by photovoltage effects that shift the effective Fermi level probed by ARPES~\cite{Schmitt2022FormationTime} and band gap renormalization~\cite{Chen2018BandPhosphorus}. 
If present, these effects could shift the bottom of the surface spectral feature at $E_0\approx E_F+\qty{0.8} {eV}$ (respectively indicated by blue and dark yellow triangles in Fig.~\ref{fig:si_charging}a,b) upwards to the exciton energy $E_X=E_0 + \qty{100}{meV}$ (determined below in Fig.~\ref{fig:si_decay_vs_energy}b and the surrounding text), leading to an apparent long-lived population at the middle of the dispersion extracted at $t\approx\qty{0}{ps}$.

To investigate these effects in our measurements, we track in Fig.~\ref{fig:si_charging} the position of the local minimum of the ARPES linewidth (indicated by the half width at half maximum, HWHM) as a function of energy under low (Fig.~\ref{fig:si_charging}a) and high (Fig.~\ref{fig:si_charging}b) excitation fluence using the fitting process detailed in Sec.~\ref{fitting_SI}.
We specifically noted in Sec.~\ref{fitting_SI} (in particular in the discussion of Fig.~\ref{fig:si_dispersion_waterfall}) that the HWHM of ARPES MDCs are generally proportional to the imaginary self energy $\text{Im }\Sigma(E)$ of the unoccupied states.
From this, we then extracted $\text{Im }\Sigma$ under low and intermediate fluence excitation, when the excitonic population can be neglected.
Under high fluence excitation, the analogous interpretation of ARPES linewidths is only strictly valid at energies where spectral weight contributed by excitonic populations is negligible, that is, far from the exciton energy, which here lies at $E_X=E_0 + \qty{100}{meV}$ (Fig.~\ref{fig:si_decay_vs_energy}b).
However, to facilitate a more direct interpretation of our data, and for completeness, we nevertheless plot the ARPES linewidth HWHM at high excitation fluence in Fig.~\ref{fig:si_charging}b across larger energy ranges that extend to $E_{X}$.

Overall, the data in Fig.~\ref{fig:si_charging} show that the global minimum HWHM appears at approximately $E_0$ at all delay times, irrespective of the excitation fluence.
We attribute this HWHM minimum to the bottom of the unoccupied surface states, and conclude that the minimum of the surface band does not exhibit time-dependent photovoltage shifts.  
This allows to us to exclude a particularly significant trivial (non-excitonic) interpretation of our data, as we have shown that the persistent population interpreted in this work as an exciton is not simply an accumulation of electrons at the bottom of a shifted or renormalized unoccupied band.

\newpage

\section{Rate equation modeling of carrier relaxation and plasmon-driven exciton formation}
\label{RateEq_SI}

To evaluate whether the population dynamics presented in Fig.~\ref{fig:mahan_exciton} of the main text are consistent with the proposed mechanism of plasmon-driven exciton formation, we numerically model the temporal evolution of the population using a rate equation description.
The methodology and results are described in detail below.

\begin{figure}[H]
\centering
\includegraphics[width=\textwidth]{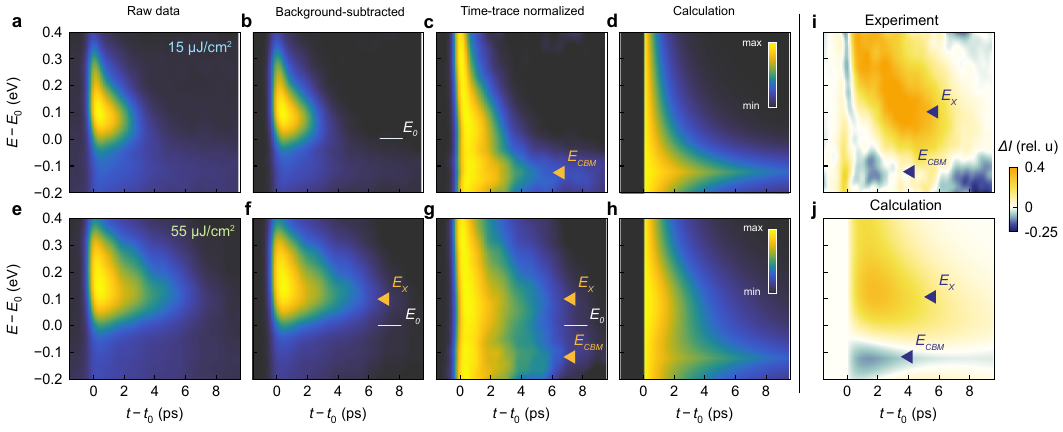}
\caption{\textbf{Theoretical modeling of plasmon-driven exciton formation:}
\textbf{(a)} Measured EDCs ($k_x\in[-0.15, 0.15]\ \qty{}{\per\angstrom}$) as a function of delay time for low fluence (\qty{15}{\micro J/cm^2}) excitation.
\textbf{(b)} Background-subtracted time trace from (a), where the background is evaluated as the averaged EDC over $t-t_0\in[-2, -0.8]\ \qty{}{ps}$ (see Sec.~\ref{sec:dataProcessing} for details of the background subtraction procedure).
\textbf{(c)} Normalized time trace, defined by independently rescaling the time trace at each energy level of the background-subtracted trace in (b) such that the maximum normalized intensity at each energy in (c) reaches a value of 1 at approximately $t_0$ (see Sec.~\ref{sec:dataProcessing}).
\textbf{(d)} Theoretical model of a normalized ARPES time trace at low fluence dynamics ($I_\ell(E,t)$ in Eq.~(\ref{eq:startingI}) with the parameter values in Table~\ref{tab:si_simulation_parameters}), showing qualitative and quantitative agreement with (c).
\textbf{(e-g)} Respectively the measured EDCs, background-subtracted time trace, and normalized time trace for high fluence (\qty{55}{\micro J/cm^2}) excitation, with (f, g) generated using the same procedure as (b, c).
\textbf{(h)} Theoretical model of a normalized ARPES time trace at high fluence dynamics ($I_h(E,t)$ in Eqs.~(\ref{eq:sumHighFluence}),~(\ref{eq:electronicHighFluence}), and~(\ref{eq:plasmonExcitonModel}) using the parameter values in Table~\ref{tab:si_simulation_parameters}), showing qualitative agreement with (g).
\textbf{(i)} Difference between the experimentally obtained background-subtracted and normalized time traces for high (g) and low (c) fluence photo-excitation (\emph{i.e.} the data in (c) subtracted from the data in (g)).
\textbf{(j)} Difference between the theoretically modeled normalized time traces for high (h) and low (d) fluence excitation (\emph{i.e.} (d) subtracted from (h)), showing strong qualitative and quantitative agreement with (i).
The procedures used to both select numerical input parameters and to obtain fitted model parameters for (d, h, j) are detailed in the text following Eq.~(\ref{eq:Iell}).}
\label{fig:simulation}
\end{figure}
 
\subsection{Data Processing}
\label{sec:dataProcessing}

We begin in Fig.~\ref{fig:simulation}a by plotting a momentum-integrated ARPES time trace ($E-E_0$ vs $t-t_0$) of the unoccupied surface and bulk states from Tr-ARPES measurements taken at low excitation fluence (\qty{15}{\micro J/cm^2}).
We next process the EDC at each delay time in Fig.~\ref{fig:simulation}a by subtracting a background that we attribute to dark noise, which specifically includes ``hot pixels'' in the electron spectrometer, parasitic spectral weight from bands below $E_F$, and pump-induced scatter. 
Assuming noise from these sources to be independent of delay relative to $t_0$, we compute the subtracted background by averaging each EDC in Fig.~\ref{fig:simulation}a over a window of negative delays significantly before the arrival of the pump.
In Fig.~\ref{fig:simulation}b, we show the resulting the background-subtracted time trace obtained by applying this procedure to the data in Fig.~\ref{fig:simulation}a.

There are additionally several other complications that we must overcome in modeling photo-excitation in \ce{EuCd2As2}.
First, the unoccupied surface states and the weakly dispersing bulk conduction band below them have significantly different photoemission matrix elements.
This is evident from Fig.~\ref{fig:overview}b of the main text, in which the surface states have significantly higher photoemission intensity than the bulk conduction band, despite both sets of states being expected to be occupied under the non-equilibrium conditions of the experiment.
The significant difference between the matrix elements of the unoccupied surface states and the weakly dispersing bulk conduction band at ${\bf k}={\bf 0}$ makes it challenging to simultaneously visualize dynamics in the two bands in the raw data.
Moreover, completely reproducing the fluence-dependent experimental measurements within a theoretical model of population dynamics solely within the unoccupied surface band would require numerous additional parameters to account for the electronic density of states, thermal band broadening, and the Fermi edge.
Even in a simple two-temperature model, we would still be required to fit the numerous additional parameters.
Finally, in the most general optical pump-probe measurement of a metal or semiconductor, increasing the pump fluence primarily drives a trivial increase in the electronic occupation of bands above the Fermi level~\cite{Chen2018BandPhosphorus}, with other -- less trivial -- observables appearing as secondary effects.

To eliminate contributions from these trivial degrees of freedom, which are not critical to our proposed model of plasmon-driven exciton formation, we next independently normalize the maximum ARPES intensity at each energy level to take a maximum value of 1.
The maximum is necessarily reached at $t_0$, up to small energy- and fluence-dependent variation due to rapid intra- and inter-band scattering (see Sec.~\ref{sec:si_rise}).
This normalization of our EDCs allows a complete description of electron population dynamics for low fluence excitation using only two dynamical parameters, as we will shortly detail below in Sec.~\ref{sec:si_req_model}.
Beginning with the background-subtracted EDCs for low fluence excitation in Fig.~\ref{fig:simulation}b, the Tr-ARPES time trace resulting from this normalization procedure is shown in Fig.~\ref{fig:simulation}c.
The analogous momentum-integrated raw time trace, background-subtracted time trace, and maximum-normalized trace for high fluence excitation are respectively shown in panels e, f, and g of Fig.~\ref{fig:simulation}.

Lastly, we perform a final phase of image processing to visualize the distinction between the unoccupied state occupation dynamics at low and high fluence, 
In Fig.~\ref{fig:simulation}i (Fig.~\ref{fig:mahan_exciton}b of the main text), we subtract the normalized time trace taken under low fluence excitation from the high fluence analog, resulting in a ``normalized difference'' trace.
We interpret the negative (blue) signal at $E_{CBM}$ in Fig.~\ref{fig:simulation}i within the first \qty{2}{ps} after photo-excitation as holes associated with exciton formation, given that the negative signal indicates a brief process of significantly faster decay that occurs only after high fluence excitation.
We then attribute the relative excess of electrons around $E_X\approx E_0+\qty{100}{meV}$ at later delays after high fluence excitation (the broad orange region in Fig.~\ref{fig:simulation}i) to electrons of the exciton, as well as to the overall increase in population lifetime due to the weaker effective electron-boson scattering following high fluence excitation.
It is important to emphasize that the electrons and holes that contribute to an excitonic bound state are required to have identical temporal profiles~\cite{Mori2023Spin-polarizedInsulator}. 
Below, in Eq.~(\ref{eq:rate_diffeq}) and the surrounding text, we will use a rate equation model to demonstrate how the experimentally obtained normalized difference trace in Fig.~\ref{fig:simulation}i is consistent with such electron and hole dynamics.

\subsection{Model fitting and discussion}
\label{sec:si_req_model}

To describe plasmon-driven exciton formation, we introduce a modification of the two-temperature relaxation model~\cite{Allen1987TheoryMetals} commonly employed to analyze Tr-ARPES experiments that probe the relaxation of hot electrons via coupling to a cold bosonic bath~\cite{Bao2022PopulationCd3As2, Gierz2014Non-equilibriumSpectroscopy, Sun2012DynamicsGraphene}.
For sufficiently low energy scales, the dynamics at each fluence can be completely described using two key parameters: an effective electron-boson coupling constant (which acquires a fluence dependence due to screening effects and changes in the initial hot electron bath temperature, see Fig.~\ref{fig:si_muTwaterfall}d), and an overall decay constant representing the slow decay of electrons from the unoccupied bands to states below $E_F$ via impurity scattering and radiative decay.
Beginning with this model, we then introduce plasmon-driven exciton formation as an additional process, superimposed over the thermal decay, that transfers spectral weight from the bulk conduction band towards the higher-energy surface band. 
We attribute this transferred spectral weight to a bath of long-lived quasiparticles, specifically excitons.

To begin constructing out model, we first note that in experimental ARPES measurements, excitons manifest as electrons near the conduction band when photoemission breaks the electron-hole pairs, accompanied by a dearth of electrons in the effective valence band~\cite{Rustagi2018PhotoemissionExcitons, Madeo2020, Karni2022StructureHole, Mori2023Spin-polarizedInsulator}.
In our model, we assume that the quasiparticle population dynamics do not strongly influence those of the weakly correlated electrons and holes in the effective conduction and valence bands, respectively.
We then note that the onset of the ARPES intensity at the instant of photo-excitation in our experiments is approximately probe-limited (\qty{200}{fs}, see the End Matter and Sec.~\ref{sec:si_rise}).
We hence treat the onset of the intensity at photo-excitation as a step function, given that the dynamics described above are significantly slower, and set the observed intensities for both low ($I_\ell(E,t)$) and high ($I_h(E,t)$) excitation fluences before $t_0$ to zero:
\begin{equation}
I_\ell(E,t<t_0) = I_h(E,t<t_0)=0,
\label{eq:Iell}
\end{equation}
where $t_{0} = \qty{0}{ps}$.

Next, we describe the electronic population $I^e_F(E,t)$ at a fluence $F$ (with $F=\ell, h$ respectively for low and high fluences), energy $E$, and time $t>t_{0}$ using the rate equation:
\begin{equation}
\frac{\mathrm{d}I^e_F(E, t)}{\mathrm{d}t}=-\bar{\alpha}_F(E)\ I^e_F(E, t),
\label{eq:rate_diffeq}
\end{equation} 
where $\bar{\alpha}_F(E)$ is the effective energy-dependent scattering rate. 
Next, plasmon-mediated hot carrier decay can be described similarly to optical phonon-mediated decay using a two-temperature model~\cite{Allen1987TheoryMetals}. 
In the two-temperature model, $\bar{\alpha}_F(E)$ in Eq.~(\ref{eq:rate_diffeq}) is given by~\cite{Bao2022PopulationCd3As2, Sun2012DynamicsGraphene}: \begin{equation}
\bar{\alpha}_F(E)=\alpha_F|E-E_{CBM}| + \eta_F.
\label{eq:si:linear_decay_rate}
\end{equation}

\begin{table}[t]
    \centering
    \begin{tabular}{|c|c|}
        \hline
        \textbf{Parameter} & \textbf{Value} \\
        \hline
        Screened electron-plasmon coupling strengths at low fluence ($\alpha_\ell$) & \qty{1.9}{ps^{-1}eV^{-1}}\\
        Screened electron-plasmon coupling strength at high fluence ($\alpha_h$) & \qty{0.95}{ps^{-1}eV^{-1}}\\
        Energy of the conduction band minimum ($E_{CBM}$) & \qty{-0.12}{eV}\\
        FWHM of the conduction band ($\varsigma_{CBM}$) & \qty{0.15}{eV}\\
        Exciton electron energy ($E_X$) & \qty{0.098}{eV} $\pm$ \qty{0.016}{eV}\\
        FWHM of the exciton electron and hole ($\varsigma_X$) & \qty{0.246}{eV} $\pm$ \qty{0.046}{eV}\\
        Exciton spectral weight ($A_X$)& 0.1\\
        Persisting conduction band intensity at low fluence ($\chi_\ell$)& 0.155 \\
        Persisting conduction band intensity at high fluence ($\chi_h$) & 0.048 \\
        Impurity scattering and radiative decay rate for low fluence ($\eta_\ell$) & \qty{0.177}{ps^{-1}} $\pm$ \qty{7.16E-4}{ps^{-1}}\\
        Impurity scattering and radiative decay rate for high fluence ($\eta_h$) & \qty{0.141}{ps^{-1}} $\pm$ \qty{5.3E-4}{ps^{-1}}\\
        Excitonic formation time ($\tau_{Xf}$) & \qty{1.4}{ps} $\pm$ \qty{36.2}{fs}\\
        Excitonic lifetime ($\tau_{Xl}$) & \qty{4.386}{ps} $\pm$ \qty{297}{fs}\\
        
        \hline
    \end{tabular}
\caption{Physical meaning and numerical values for the input parameters for our model of the electronic population dynamics in the unoccupied band structure (Fig.~\ref{fig:simulation}d,h,j and Eqs.~(\ref{eq:startingI}),~(\ref{eq:sumHighFluence}),~(\ref{eq:electronicHighFluence}), and~(\ref{eq:plasmonExcitonModel})). 
The values of $E_{CBM}$ and $E_X$ were respectively obtained from the fits of the experimental data in panels a and e of Fig.~\ref{fig:si_decay_vs_energy}, and are specified relative to the bottom of the unoccupied surface state $E_0\approx E_F+\qty{0.8}{eV}$. 
We provide uncertainties for parameters that we obtained from fits of the normalized time traces in Fig.~\ref{fig:simulation}, where each uncertainty range represents one standard deviation.
The methods that we employed to both select numerical input parameters and to obtain fitted parameters are detailed in the text following Eq.~(\ref{eq:Iell}).}
\label{tab:si_simulation_parameters}
\end{table}

In Table \ref{tab:si_simulation_parameters}, we provide the parameter values that we use to model plasmon-driven exciton formation in \ce{EuCd2As2}.
Due to the large number of parameters necessary to simulate time-dependent ARPES spectra across multiple fluences -- even in a simplified model like the one that we apply here -- we specifically fix several parameters by evaluating individual cuts of the data prior to extracting the critical decay rate and formation time parameters from full time traces, as discussed below.

\begin{figure}[t]
\centering
\includegraphics{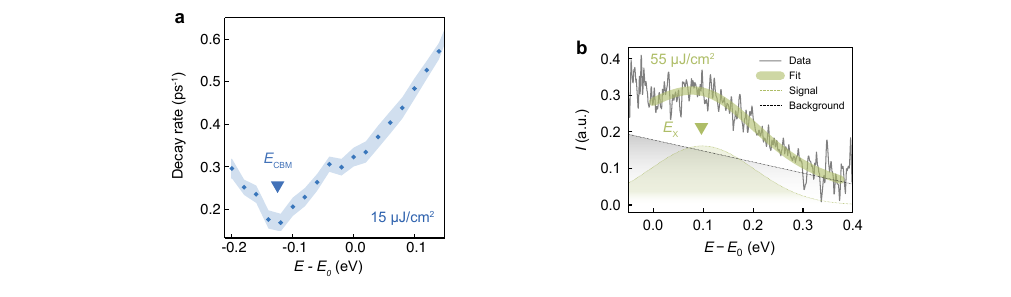}
\caption{\textbf{Extracting additional parameters for theoretical modeling of plasmon-driven exciton formation:}
\textbf{(a)} Decay rate as a function of energy ($E-E_0$), obtained from single exponential fits of the low fluence (\qty{15}{\micro J/cm^2}) Tr-ARPES data in Fig.~\ref{fig:simulation}a. 
Using the magnitude of the slope of the decay rate fit in (a) at energies above $E_{CBM}$ in the vicinity of $E_{0}$, we obtain an estimate for the low-fluence effective electron-plasmon coupling strength $\alpha_\ell$ in Table~\ref{tab:si_simulation_parameters} and Eq.~(\ref{eq:si:linear_decay_rate}). 
In Eq.~(\ref{eq:si:linear_decay_rate}), we then set the value of the parameter $E_{CBM}$ in Table~\ref{tab:si_simulation_parameters} and Eq.~(\ref{eq:si:linear_decay_rate}) to the energy at which the decay rate reaches its minimum value $E_{CBM}=\qty{-0.12}{eV}$.
In (a), the shaded region represents a 68\% confidence interval for the fits.
\textbf{(b)} EDC extracted from the normalized ARPES data at high fluence (\qty{55}{\micro J/cm^2}) in Fig.~\ref{fig:simulation}g, integrated between \qty{5}{ps} and \qty{6}{ps} after $t_0$. 
In (b), the raw data is shown as a black line, and a fit from a single Gaussian with a linear background is superimposed (shaded green line).
The Gaussian fit in (b) has been truncated at $E_0$, below which additional Gaussian functions representing the conduction band and the holes of the exciton would be necessary.
We additionally in (b) plot the decomposition of the fit into a Gaussian (green shaded region) and the linear background (black shaded region).
Using this decomposition, we set the value of $\varsigma_{X}$ in Table~\ref{tab:si_simulation_parameters} to the FWHM of the Gaussian piece of the fit in (b), where $\varsigma_{X}$ is related to the standard deviation $\sigma_{X}$ of the high-fluence electron and hole distributions in Eq.~(\ref{eq:plasmonExcitonModel}) via Eq.~(\ref{eq:RWHMvsSigmaX}).
We then set the value of the exciton energy parameter $E_{X}$ in Table~\ref{tab:si_simulation_parameters} and Eq.~(\ref{eq:plasmonExcitonModel}) to the energy of the maximum of the Gaussian piece of the fit in (b).}
\label{fig:si_decay_vs_energy}
\end{figure}

To begin assigning parameter values in our model of plasmon-driven exciton formation, we first note that the parameters $\alpha_\ell$ and $\alpha_h$ in Eqs.~(\ref{eq:rate_diffeq}) and~(\ref{eq:si:linear_decay_rate}) respectively encode the effective electron-plasmon coupling strengths for the low fluence and high fluence cases.
In our calculations, we set $\alpha_\ell$ to the absolute value of the slope of the fit of the decay rate over energy (Fig.~\ref{fig:si_decay_vs_energy}a) near $E_0$ at low fluence (\qty{15}{\micro J/cm^2}) photo-excitation.
We then estimate $\alpha_h$ to be a factor of 2 lower than $\alpha_\ell$ to qualitatively match the overall slowdown that we observe in Tr-ARPES even at energies far above $E_0$ after high fluence excitation (Fig.~\ref{fig:simulation}e). 
Next, in Eq.~(\ref{eq:si:linear_decay_rate}), $\eta_F$ represents an energy-independent scattering rate that originates from impurity-mediated and radiative processes by which electrons relax from the unoccupied bands to states below $E_F$. 
In our calculations, we leave $\eta_F$ to be evaluated as a fitting parameter.

Lastly, we obtain a value for $E_{CBM}$ in Eq.~(\ref{eq:si:linear_decay_rate}) from Tr-ARPES measurements by assigning the $E_{CBM}$ to the energy at which the population decay rate is minimized (Fig.~\ref{fig:si_decay_vs_energy}a).
In Table~\ref{tab:si_simulation_parameters}, we provide the value of $E_{CBM}$ obtained from the data in Fig.~\ref{fig:si_decay_vs_energy}a as an energy relative to the bottom of the unoccupied surface state $E_0\approx E_F+\qty{0.8}{eV}$. 
Importantly, the value of $E_{CBM}$ obtained through this method lies within the broad energy range over which spectral weight from the bulk conduction band is observed in ARPES measurements (indicated with a blue triangle in Fig.~\ref{fig:si_HE-EELS}b and a yellow triangle in Fig.~\ref{fig:overview}b of the main text).
Furthermore, the value of $E_{CBM}$ extracted from Fig.~\ref{fig:si_decay_vs_energy}a also coincides with the energy of a bulk conduction band in our PM-state DFT calculations that is nondispersing at $\Gamma$ to leading order (Fig.~\ref{fig:si_dft}a).

The presence of only two dynamical parameters for the thermal dynamics at each fluence -- $\alpha_F$ and $\eta_F$ in Eq.~(\ref{eq:si:linear_decay_rate}) (recalling that we above chose $\alpha_{h}=\alpha_\ell /2$) -- highlights the advantage of the time trace normalization process employed in this work.
Despite being equivalent to a two-temperature model within sufficiently low energy ranges~\cite{Sun2012DynamicsGraphene, Bao2022PopulationCd3As2}, Eq.~(\ref{eq:si:linear_decay_rate}) circumvents the numerous additional parameters necessary to perform full, dynamical simulations of the experimental data, including parameters that are difficult to directly measure.
For example, statistical mechanical parameters like the non-equilibrium chemical potential and electronic temperature are strictly necessary for more comprehensive implementations of a two-temperature model~\cite{Bao2022PopulationCd3As2}, and can be extracted from the experimental data (Sec.~\ref{fitting_SI}). 
However, full dynamical simulations would also require knowledge of the photoemission matrix elements, density of states, and broadening for each of the two relevant bands, values for which are much more challenging to theoretically estimate while reliably cross-checking against Tr-ARPES data.

To construct the final ingredient in our modified two-temperature model, we lastly note that as can be seen in Fig.~\ref{fig:simulation}c,g, a small population of electrons persists at a broad energy range surrounding $E_{CBM}$ beyond the measurement window of our experiment, showing negligibly slow decay. 
The appearance of these long-lived electrons is consistent with previously observed populations at the bottom of bulk conduction bands in a range of photo-doped materials~\cite{Sobota2012Ultrafast3, Boschini2024Time-resolvedMaterials}.
In our model, we approximate the long-lived electron population at $E_{CBM}$ to have step-function-like temporal dynamics, with an onset at $t_0$ and an infinite lifetime.
In the experimentally obtained normalized time traces in Fig.~\ref{fig:simulation}c,g, the overall intensity of the long-lived conduction electron population is fluence-dependent, being larger in the case of low fluence excitation.
We approximate the profile of the long-lived conduction electron contribution in EDCs at $t\geq 0$ as:
\begin{equation}
\tilde\chi_F(E)=\chi_F\exp\left(-(E-E_{CBM})^2 / 2\sigma_{CBM}^2\right),    
\label{eq:longElec}
\end{equation}
where $F=\ell, h$, such that Eq.~(\ref{eq:longElec}) takes the form of a Gaussian with height $\chi_F$ centered at $E=E_{CBM}$.
In Eq.~(\ref{eq:longElec}), we set $\chi_F$ to the spectral intensity of the conduction band for the normalized Tr-ARPES trace measured under the corresponding fluence (Fig.~\ref{fig:simulation}c,g), averaged between 8 and \qty{9}{ps} after photo-excitation.
The Gaussian function in Eq.~(\ref{eq:longElec}) has a full width half maximum (FWHM) of $\varsigma_{CBM}$, which is related to the Gaussian width $\sigma_{CBM}$ by:
\begin{equation}
\sigma_{CBM}=\varsigma_{CBM}/(2\sqrt{2\log 2}).
\end{equation}
In Table~\ref{tab:si_simulation_parameters}, we provide the corresponding numerical values of $\chi_F$ and $\varsigma_{CBM}$ used in our calculations.

Having now assembled all of the necessary theoretical machinery, we first model the photoemission intensity for low fluence excitation.
In the low fluence regime, the reabsorption of plasmons by cooled electrons is negligible and the experimentally observed ARPES intensity $I_\ell(E, t)$ originates entirely from electrons (\emph{i.e.} without any excitonic contribution), such that:
\begin{equation}
I_\ell(E, t)=I_\ell^e(E, t).
\label{eq:IellE}
\end{equation}
Next, following the normalization procedure discussed in Sec.~\ref{sec:dataProcessing}, we compute the energy-normalized spectral intensity using Eqs.~(\ref{eq:rate_diffeq}),~(\ref{eq:si:linear_decay_rate}),~(\ref{eq:longElec}), and~(\ref{eq:IellE}): 
\begin{equation}
I_\ell(E, t) = \Theta(t)\left[ (1-\tilde\chi_\ell(E)) \exp\left[-t(\alpha_\ell|E-E_{CBM}|+\eta_\ell)\right] +\tilde\chi_\ell(E)\right],
\label{eq:startingI}
\end{equation}
where $\Theta(t)$ is the Heaviside step function, and where $\alpha_\ell$, $E_{CBM}$, and the parameter $\tilde\chi_\ell$ in $\tilde\chi_\ell(E)$ (Eq.~(\ref{eq:RWHMvsSigmaX})) take the values specified in Table~\ref{tab:si_simulation_parameters}.
We then  we fit the two-dimensional experimental dataset at low fluence in Fig.~\ref{fig:simulation}c to Eq.~(\ref{eq:startingI}) using the \texttt{minimize} function of the \texttt{lmfit} Python package.
From this fitting, we obtain the value $\eta_\ell\approx\qty{0.177}{ps^{-1}}$, which lies in close agreement with the experimentally observed minimum decay rate at $E_{CBM}$ in Fig.~\ref{fig:si_decay_vs_energy}a. 
In Fig.~\ref{fig:simulation}d, we plot the resulting normalized time trace $I_\ell(E,t)$ from our theoretical model (Eq.~(\ref{eq:startingI})) using the parameters specified in the text above and summarized in Table~\ref{tab:si_simulation_parameters}.

For high fluence excitation, the presence of excitons requires us to modify Eq.~(\ref{eq:startingI}) to account for the arrival (along the timescale of exciton formation $\tau_{Xf}$) and decay (along the timescale of the exciton lifetime $\tau_{Xl}$) of electrons at the exciton energy $E_X$~\cite{Mori2023Spin-polarizedInsulator}.
We hence modify Eq.~(\ref{eq:startingI}) by adding terms to account to these effects, as well as a corresponding term of opposite sign at $E_{CBM}$ to account for hole formation, resulting in an intensity $I_h(E, t)$ that can be expressed as:
\begin{equation}
I_h(E, t)=I_h^e(E, t)+I_h^X(E, t),
\label{eq:sumHighFluence}
\end{equation}
where $I_h^e(E, t)$ and $I_h^X(E, t)$ are respectively the electronic and excitonic contributions to the photoemission intensity.

First, we take the function $I_h^X(E, t)$ in Eq.~(\ref{eq:sumHighFluence}) to be positive at energies where the electron of the exciton appears in the spectra, and negative at energies where the hole of the exciton instead appears.
As a simple approximation, we assume that the exciton electron and hole have identical Gaussian profiles in the energy-normalized spectra that carry standard deviations of $\sigma_X$ and are respectively centered at $E_{X}$ and $E_{CBM}$. 
To obtain values for $\sigma_X$ and $E_{X}$, we respectively taking $\sigma_{X}$ to be the standard deviation and $E_{X}$ to be the center value of the Gaussian piece of the fit in Fig.~\ref{fig:si_decay_vs_energy}b of the spectral feature at $E\approx E_0+\qty{100}{meV}$ in the experimentally obtained high fluence normalized trace in Fig.~\ref{fig:simulation}g.
For $\sigma_{X}$, we specifically first extract the FWHM $\varsigma_{X}$ of the Gaussian part of the fit in Fig.~\ref{fig:si_decay_vs_energy}b, which is then related to $\sigma_{X}$ through the relation:
\begin{equation}
\sigma_{X}=\varsigma_{X}/(2\sqrt{2\log{2}}).
\label{eq:RWHMvsSigmaX}
\end{equation}
The values of $\varsigma_{X}$ and $E_{X}$ obtained through this fitting procedure are provided in Table~\ref{tab:si_simulation_parameters}.
It is important to note that the energy-dependent normalization process employed throughout our analysis (see Sec.~\ref{sec:dataProcessing}) leads to a value of $\sigma_X$ that significantly exceeds the width of the excitonic feature in the experimental spectral function that we obtain from ARPES measurements (see for example Fig.~\ref{fig:fluence_dependence}a of the main text at high fluence and later delays).
Conversely, the value of $E_X$ obtained from the Gaussian fit in Fig.~\ref{fig:si_decay_vs_energy}b exhibits good qualitative agreement with the energy of the persistent sharpened spectral weight observed in Tr-ARPES after high fluence excitation (denoted with yellow triangles in Fig.~\ref{fig:simulation}b,c and a green triangle in Fig.~\ref{fig:fluence_dependence}b of the main text).

Using the rate equation model modifications detailed above, we arrive at a model for the photoemission intensity at high fluence excitation in which the electronic contribution $I_h^e(E, t)$ in Eq.~(\ref{eq:sumHighFluence}) takes the same form as the low fluence intensity $I_\ell(E, t)$ in Eq.~(\ref{eq:startingI}):
\begin{equation}
I_h^e(E, t) = \Theta(t) \left[(1-\tilde\chi_h(E))\exp\left[-t(\alpha_h|E-E_{CBM}|+\eta_h)\right]+ \tilde\chi_h(E)\right],
\label{eq:electronicHighFluence}
\end{equation}
whereas the additional excitonic contribution $I_h^X(E, t)$ in Eq.~(\ref{eq:sumHighFluence}) instead takes the form:
\begin{align}
I_h^X(E,t)&=\Theta(t)\bigg[\frac{A_X}{\sigma_X\sqrt{2\pi}}\exp\left(-\frac{(E-E_X)^2}{2\sigma_X^2}\right)\left(1-\mathrm{e}^{-t/\tau_{Xf}}\right)\mathrm{e}^{-t(\tau^{-1}_{Xl}+\eta_h)}\nonumber\\&\qquad\qquad\qquad-\frac{A_X}{\sigma_X\sqrt{2\pi}}\exp\left(-\frac{(E-E_{CBM})^2}{2\sigma_X^2}\right)\left(1-\mathrm{e}^{-t/\tau_{Xf}}\right)\mathrm{e}^{-t(\tau^{-1}_{Xl}+\eta_h)}\bigg].
\label{eq:plasmonExcitonModel}
\end{align}
In Eq.~(\ref{eq:plasmonExcitonModel}), the positive Gaussian term adds electronic spectral weight at $E_{X}$, whereas the negative Gaussian term removes spectral weight at $E_{CBM}$, respectively modeling the contributions of electrons and holes to the measured spectral intensity due to exciton breaking during the photoemission process.

Next, we obtain values for the key dynamical parameters, $\tau_{Xf}$, $\tau_{Xl}$, and $\eta_h$ in Eqs.~(\ref{eq:electronicHighFluence}) and~(\ref{eq:plasmonExcitonModel}) (provided in Table~\ref{tab:si_simulation_parameters}) by fitting the experimental normalized time trace in Fig.~\ref{fig:simulation}g to Eqs.~(\ref{eq:electronicHighFluence}) and~(\ref{eq:plasmonExcitonModel}) using the \texttt{minimize} function of the \texttt{lmfit} Python package. 
To prevent overfitting, we hold all other parameters fixed to their listed values in Table~\ref{tab:si_simulation_parameters} throughout the fitting process. 
We find that the fit for $\tau_{Xf}$, $\tau_{Xl}$, and $\eta_h$ converges with only small residual errors, and that the values obtained from fitting lie well within the parameter guardrail values that we used to ensure convergence to a physically reasonable local minimum.  However, we note that the free parameters $\tau_{Xf}$ and $\tau_{Xl}$ do show a moderately strong correlation of $\rho(\tau_{Xf}, \tau_{Xl})\approx -0.56$ in the fit, and thus their values should be interpreted with caution.

We conclude by plotting $I_h(E,t)$ (Eqs.~(\ref{eq:sumHighFluence}),~(\ref{eq:electronicHighFluence}), and~(\ref{eq:plasmonExcitonModel})) in Fig.~\ref{fig:simulation}h using the parameter values obtained above and listed in Table~\ref{tab:si_simulation_parameters}.
Fig.~\ref{fig:simulation}h exhibits strong qualitative agreement with the experimentally measured high fluence normalized ARPES intensity in Fig.~\ref{fig:simulation}g. 
We next in Fig.~\ref{fig:simulation}j plot the difference between our theoretically modeled normalized time traces at high and low fluence (\emph{i.e.} Fig.~\ref{fig:simulation}d subtracted from Fig.~\ref{fig:simulation}h).
The theoretical normalized difference trace in Fig.~\ref{fig:simulation}j also shows strong qualitative and quantitative agreement with the experimental difference time trace in Fig.~\ref{fig:simulation}i (detailed above in Sec.~\ref{sec:dataProcessing}), even though the parameters in the low and high fluence modeling functions $I_\ell(E,t)$ and $I_h(E,t)$ (respectively Eqs.~(\ref{eq:startingI}) and~(\ref{eq:sumHighFluence})) were obtained from independent fits.

\subsection{Statistical clarity and consistency across different target and reference measurements}
\label{sec:Clarity}

\begin{figure}[H]
\centering
\includegraphics{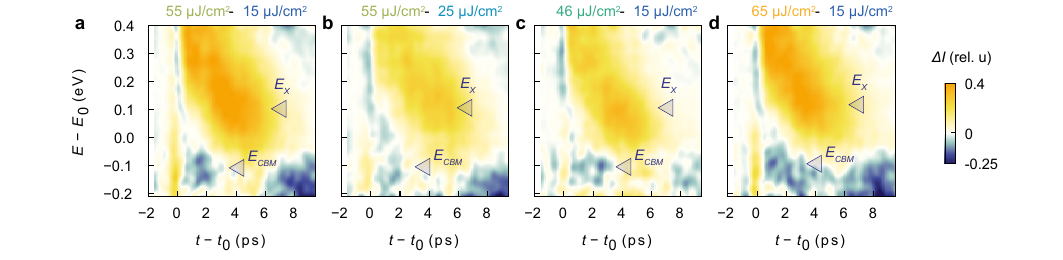}
\caption{\textbf{Normalized difference time traces for a selection of target and reference fluences:} 
In each panel in this figure, we show the normalized difference time trace between Tr-ARPES measurements taken at a target (higher) and reference (lower) fluence, generated using the data processing procedure detailed in Sec.~\ref{sec:dataProcessing} (see panels a, b, c, e, f, g, and i in Fig.~\ref{fig:simulation}).
\textbf{(a)} High fluence: \qty{55}{\micro J/cm^2}; low fluence: \qty{15}{\micro J/cm^2} (reproduced from Fig.~\ref{fig:simulation}i). 
\textbf{(b)} High fluence: \qty{55}{\micro J/cm^2}; low fluence: \qty{25}{\micro J/cm^2}. 
\textbf{(c)} High (intermediate) fluence: \qty{46}{\micro J/cm^2}; low fluence: \qty{15}{\micro J/cm^2}. \textbf{(d)} High fluence: \qty{65}{\micro J/cm^2}; low fluence: \qty{25}{\micro J/cm^2}. 
All panels in this figure are plotted using the same color scale as Fig.~\ref{fig:simulation}i.}
\label{fig:si_normDiff}
\end{figure}

In Fig.~\ref{fig:si_normDiff}, we present a comparison between normalized difference time traces analogous to that shown in Fig.~\ref{fig:simulation}i (and Fig.~\ref{fig:mahan_exciton}b of the main text) for a selection of target (\emph{i.e.} higher) and reference (low) fluence measurements. 
Each panel in Fig.~\ref{fig:si_normDiff} was generated using the same data processing procedure (detailed in Sec.~\ref{sec:dataProcessing}) and plotted using the same color scale as Fig.~\ref{fig:simulation}i.
Across all panels in Fig.~\ref{fig:si_normDiff}, we consistently qualitatively observe two of the key experimental findings in this work: extended population lifetimes at $E_X$ and a dearth of electrons at $E_{CBM}$ within the first \qty{2}{ps}.
For completeness, we in particular show in Fig.~\ref{fig:si_normDiff}c a difference time trace with an intermediate fluence of \qty{46}{\micro J/cm^2} as the higher (target) fluence. 
Previously, in Sec.~\ref{fitting_SI}, we had identified \qty{46}{\micro J/cm^2} as a sufficiently low fluence to extract parameters of the electronic statistical distribution (\emph{i.e.} the chemical potential $\mu$ and temperature $T$) within the first three picosecond after photo-excitation. 
However, consistent with our observation that the crossover from the low fluence regime to the high fluence regime is continuous (discussed in the text surrounding Fig.~\ref{fig:mahan_exciton} of the main text), we also in Fig.~\ref{fig:si_normDiff}c resolve a relative dearth of electrons around $E_{CBM}$ at initial delays after excitation for the intermediate fluence \qty{46}{\micro J/cm^2} when compared to the significantly lower reference fluence of \qty{15}{\micro J/cm^2}.

\subsection{Scattering beyond the rate equation model}
\label{sec:si_beyondREQ}

In this section, we discuss the two features of our experimental data (Fig.~\ref{fig:simulation}i) that are not clearly reproduced in the theoretical model of plasmon-driven exciton formation introduced in Sec.~\ref{sec:si_req_model} (Fig.~\ref{fig:simulation}j, produced from Eqs.~(\ref{eq:startingI}),(\ref{eq:sumHighFluence}),~(\ref{eq:electronicHighFluence}), and~(\ref{eq:plasmonExcitonModel})): a transient negative signal above $E_{X}\approx E_0+\qty{100}{meV}$ immediately before $t_{0}$, and a dearth of electrons below $E_{0}$ at later delay times and high excitation fluence.

\begin{figure}[t]
\centering
\includegraphics{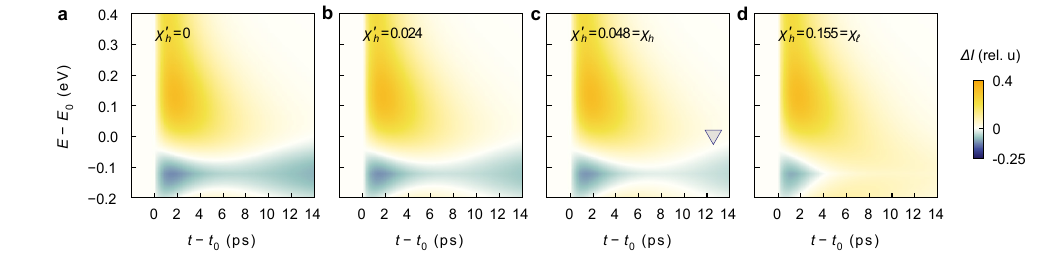}
\caption{\textbf{Theoretical modeling of normalized difference traces for selected alternative values of $\chi_h$:}
In this figure, we recompute the theoretically modeled normalized difference time trace (see Fig.~\ref{fig:simulation}j, Sec.~\ref{sec:si_req_model}, and the text following Eq.~(\ref{eq:plasmonExcitonModel})) using alternative values $\chi_{h}'$ of the high fluence persisting conduction band intensity $\chi_{h}$ in Eq.~(\ref{eq:longElec}).
As detailed in Table~\ref{tab:si_simulation_parameters} and the text surrounding Eq.~(\ref{eq:longElec}),  Fig.~\ref{fig:simulation}j (a reproduction of Fig.~\ref{fig:mahan_exciton}c of the main text) was generated using $\chi_{h}=0.048$.
\textbf{(a)} Theoretical normalized difference time trace using $\chi_h' = 0$. 
\textbf{(b)} Theoretical difference trace  with $\chi_h' = 0.024=\chi_h/2$. 
\textbf{(c)} Theoretical difference trace with $\chi_h' = 0.048=\chi_h$. 
\textbf{(d)} Theoretical difference trace with $\chi_h' = 0.155=\chi_\ell$, where $\chi_\ell$ is the high fluence persisting conduction band intensity (see Table~\ref{tab:si_simulation_parameters} and the text surrounding Eq.~(\ref{eq:longElec})). 
With the exception of $\chi_h$, the calculations in each panel were performed using the same parameter values as those listed in Table~\ref{tab:si_simulation_parameters}. 
All panels in this figure are plotted using the same color scale as Fig.~\ref{fig:simulation}j.
In panel (c), which represents a reproduction of Fig.~\ref{fig:simulation}j (and Fig.~\ref{fig:mahan_exciton}c of the main text) plotted to longer delay times $t-t_{0}$, we have used a purple outlined triangle to draw attention to the reemergence of a negative relative intensity in the conduction band at later delay times.}
\label{fig:si_normDiff_cbh_offset}
\end{figure}

First, in the experimentally obtained normalized difference time trace in Fig.~\ref{fig:simulation}i, we observe a transient negative signal above the exciton energy $E_{X}\approx E_0+\qty{100}{meV}$ immediately before $t_0$, and a transient positive signal below $E_{X}$.
These features are by construction absent in our theoretical model (Fig.~\ref{fig:simulation}j), which instead assumes that the onset of intensity in all relevant electronic states is instantaneous, occurring at exactly $t_0$.
Although a rigorous exploration of the experimentally observed dynamics immediately before $t_0$ requires time resolution finer than that used in our experiment (\qty{200}{fs}), we posit that the transient signals above and below $E_{X}$ just before $t_{0}$ are consequences of the nontrivial fluence dependence of the rapid thermalization that follows optical inter-band excitations by a femtosecond pulse.
Consistent with this, we observe in Fig.~\ref{fig:si_rise} that the optical pump induces an energy-localized population immediately before $t_0$.
In Fig.~\ref{fig:si_muTwaterfall}, we further show that by $t_0$, the transient populations have thermalized into a hot electron distribution spanning multiple unoccupied bands, with the temperature of the thermalized distribution increasing with increasing fluence.
The presence of a faster thermalization process at higher fluences would lead to both a relative dearth of electrons immediately before $t_0$ at energies directly accessed by the optical inter-band transition -- which are primarily above $E_{X}\approx E_0+\qty{100}{meV}$ (Fig.~\ref{fig:si_rise}) -- as well as a relative excess of electrons at lower energies, in agreement with the features present just before $t_{0}$ in Fig.~\ref{fig:simulation}i.

Second, we observe a strong relative dearth of electrons below $E_0$ at later delays for high excitation fluence (\emph{i.e.} a large negative region at $t>t_0 + \qty{6}{ps}$ in Fig.~\ref{fig:simulation}i). 
A similar feature is also present in the corresponding theoretical calculation (Fig.~\ref{fig:simulation}j), though it is significantly more faint.
The discrepancy between the experimental and theoretical signals below $t_{0}$ at later delays in Fig.~\ref{fig:simulation}i,j can largely be attributed to an overestimation of the intensity of the persisting conduction band population in the case of high fluence, which can be seen by contrasting the experimental and theoretical high fluence normalized traces near $E_{CBM}=\qty{-0.12}{eV}$ in panels g and h of Fig.~\ref{fig:simulation}, respectively.
One possible origin of this overestimation is that our model approximates the conduction band lifetime to be infinite, whereas the real material carries a finite conduction band lifetime.
Additionally, more reliable estimates of the conduction band intensity at late delays are challenging to obtain, owing to competition between the extremely faint persisting conduction band population and the noise floor of our experiment.

To further investigate the effects of our assumptions regarding the conduction band intensity contribution, we recompute in Fig.~\ref{fig:si_normDiff_cbh_offset} the theoretical normalized difference time trace in Fig.~\ref{fig:simulation}j for varying values $\chi'_{h}$ of the high fluence persisting conduction band intensity $\chi_{h}$ in Eq.~(\ref{eq:longElec}), while keeping all other model parameters fixed to the values listed in Table~\ref{tab:si_simulation_parameters}.
Previously, to generate Fig.~\ref{fig:simulation}j, $\chi_h$ was set to the average intensity of the conduction band at late delays in the experimental normalized time trace at high fluence excitation (Fig.~\ref{fig:simulation}g). 
In Fig.~\ref{fig:si_normDiff_cbh_offset} we vary $\chi_{h}'$ from 0, through its value in Fig.~\ref{fig:simulation}j $\chi_{h}'=\chi_h=0.048$, to the low fluence persisting conduction band intensity $\chi_\ell$.
Through this, we find that the reemergence of the conduction band at later delays is freely tunable in the theoretical model in Sec.~\ref{sec:si_req_model} via the parameter $\chi_{h}$ in Eq.~(\ref{eq:longElec}).
We further find that when $\chi_{h}'$ is set to the value $\chi_{h}$ obtained by fitting a Gaussian function to the experimental data at long delay times (see the text following Eq.~(\ref{eq:longElec})), a negative signal reappears in the conduction band (indicated by an outlined purple triangle in Fig.~\ref{fig:si_normDiff_cbh_offset}c), in qualitative agreement with the experimental difference trace in Fig.~\ref{fig:simulation}i, albeit appearing at a longer timescale.

Lastly, we note that achieving further quantitative agreement between our experimental measurements and a theoretical model would require a more sophisticated description of the several competing mechanisms for the decay of electrons in the nondispersing bulk conduction band, as well as their fluence dependence, which are beyond the scope of this work.

\newpage

\section{Considering alternative explanations of the Tr-ARPES experimental data}
\label{SI_alternativeExplanations}

In this section, we posit several alternative explanations for our experimental Tr-ARPES observations.
We then discuss why each listed scenario is less likely than the mechanism of plasmon-driven exciton formation emphasized throughout this work.

\begin{itemize}
\item \textbf{The population dynamics follow conventional carrier cooling via bosonic coupling (\emph{i.e.} a two-temperature model) or electron-electron scattering (Fermi liquid dynamics).}
This scenario can be excluded, as it does not capture the existence of a local maximum in population lifetime at $E_{X}\approx E_0+\qty{100}{meV}$, which is evident from the raw data in Fig.~\ref{fig:fluence_dependence} of the main text. 
In this scenario, carrier lifetimes would instead be expected to decrease monotonically with increasing energy. 

\item \textbf{The population dynamics are described by carrier relaxation via bosonic coupling, with the long lifetime at $E_X$ originating from light-induced band renormalization shifting the nondispersing bulk conduction band minimum from $E_{CBM}$ to $E_X$.}
This scenario can be excluded using the data in Fig.~\ref{fig:si_charging}, which demonstrate the absence of significant band renormalization or time-dependent photovoltage shifts due to pump-induced heating. 
Specifically, the bottom of the surface conduction band, indicated in Fig.~\ref{fig:si_charging} as the minimum in the imaginary self energy (MDC HWHM) at $E_{0}$, shows negligible temporal fluctuation.

\item \textbf{The population dynamics follow carrier relaxation via bosonic coupling, with the long lifetime at $E_X$ originating from a population bottleneck at the minimum of an unoccupied (topological) surface band $E_0$~\cite{Zhu2015UltrafastSb2Te3} (which lies close to a surface Dirac crossing, see Figs.~\ref{fig:si_DFT_ARPES_dispersions} and~\ref{fig:sxzconstenergy}).}
Bottlenecks lead to a maximum in population lifetime at the bottom (or Dirac point) of a surface band (here $E_0$), up to broadening effects on the order of tens of meV. 
However, this scenario is not in agreement with our experimental Tr-ARPES measurements of \ce{EuCd2As2}, in which the bottom of the unoccupied band instead remains well-defined at $E_0$ at the same delay times at which a persisting population is observed \qty{100}{meV} higher, at $E_X$ (see for example Fig.~\ref{fig:simulation}e,f,g,i). 

\item \textbf{The long-lived population at $E_X$ is a consequence of Auger processes, resulting from scattering between electrons in the nondispersing bulk conduction band at $E_X - E'$ and higher-energy states at $E_X + E'$, where $E'\approx\qty{200}{meV}$ is the energy separation between the persisting population ($E_X$) and the identified holes in the bulk conduction band ($E_{CBM}$).}
This scenario can be excluded, because the electronic states at $E_X + E'$ are directly probed by our measurements (Fig.~\ref{fig:fluence_dependence} of the main text), and instead show short population lifetimes. 
This indicates that such Auger processes, though not explicitly disallowed, are extremely unlikely at later delays, including past $t_0 + \qty{1}{ps}$. 
Moreover, the states at $E_X + E'$ lie distinctly above the energy accessed by optical inter-band transitions, and therefore never become significantly populated.

\item \textbf{Phonon-mediated scattering from bulk conduction band minima at the $M$ and $L$ points of the Brillouin zone -- which are not probed in this experiment -- could be (nearly-) degenerate with $E_X$, and could therefore act to create the long lifetime at $E_X$~\cite{Jabed2026ControlRenormalization, Fragkos2024Excited1T-ZrTe2}.}
This scenario can be excluded, because such relaxation processes rely on the presence of a cold phonon bath, and are hence lowest order in the fluence. 
A model based on phonon-mediated scattering would therefore fail to explain the emergence of the long-lived population at $E_X$ at only higher fluences (Fig.~\ref{fig:simulation}). 
Such a model would also fail to capture more subtle features of our experimental observations, including the rapid population decrease at $E_{CBM}$ in Fig.~\ref{fig:simulation}i (Fig.~\ref{fig:mahan_exciton}b of the main text).
Lastly, such a model would require discarding the evidence that electronic states are coupled to the bulk plasmon -- which is supported by the experimental data in Fig.~\ref{fig:si_dispersion_waterfall} (Fig.~\ref{fig:electron-plasmon_coupling}a of the main text) -- in favor of electron-phonon coupling, for which there is conversely no explicit evidence.
\end{itemize}